\RequirePackage
		[
			abort,
			l2tabu,
			orthodox,
		]
		{nag}
\RequirePackage{import}

\makeatletter
\def\ece#1#2{\expandafter#1\csname#2\endcsname}%
\def\setproperty#1#2#3{\ece\protected@edef{#1@p#2}{\unexpanded{#3}}}%
\def\getproperty#1#2{%
  \expandafter\ifx\csname#1@p#2\endcsname\relax
  \else \csname#1@p#2\endcsname
  \fi
}%
\def\ifthenelsepropertydefined#1#2#3#4{%
  \expandafter\ifx\csname#1@p#2\endcsname\relax
  #4
  \else#3
  \fi
}%
\def\ifpropertydefined#1#2#3{%
    \ifthenelsepropertydefined{#1}{#2}{#3}{}
}
\def\ifpropertyundefined#1#2#3{%
    \ifthenelsepropertydefined{#1}{#2}{}{#3}
}
\def\raiseifpropertyundefined#1#2#3{%
    \ifpropertyundefined{#2}{#3}{\PackageError{#1}{Property #2 #3 needs to be defined. Put \@backslashchar setproperty{#2}{#3} to your settings file}{Grep for your property :)}}
}
\def\setpropertyifundefined#1#2#3{%
    \ifpropertyundefined{#1}{#2}{%
        \setproperty{#1}{#2}{#3}
    }{}
}
\def\setpropertyifundefinedwithusageinfo#1#2#3{%
    \setpropertyifundefined{#1}{#2}{TODO>> Put \backslash{}setproperty\{#1\}\{#2\}\{<your value here, e.g. ``#3``>\} into content/settings.tex <<}
}

\def\ifthenelseproperty#1#2#3#4{\providetoggle{#1@p#2}\settoggle{#1@p#2}{\getproperty{#1}{#2}}\iftoggle{#1@p#2}{#3}{#4}}
\def\ifthenelsepropertyequal#1#2#3#4#5{\ifthenelse{\equal{\getproperty{#1}{#2}}{#3}}{#4}{#5}}

\makeatother

\makeatletter
\newcommand\RequirePackageWithOption[2][]{%
    \@ifpackageloaded{#2}{%
        \PassOptionsToPackage{#1}{#2}
    }{%
        \RequirePackage[#1]{#2}
    }
}

\newcommand\ImportDefault[2][]{
    \IfFileExists{content/#1#2.tex}{
        \typeout{----- LOAD OWN #1#2 -----}
        \import{content/#1}{#2}
    }{
        \typeout{----- LOAD DEFAULT #1#2 -----}
        \ifthenelsepropertydefined{default}{#2}{%
            \IfFileExists{templates/\getproperty{default}{#2}/#1#2.tex}{
                \import{templates/\getproperty{default}{#2}/#1}{#2}
            }{
                \import{templates/Default/#1}{#2}
            }
        }{%
            \typeout{\getproperty{default}{#1}}
            \ifthenelsepropertydefined{default}{#1}{%
                \IfFileExists{templates/\getproperty{default}{#1}/#1#2.tex}{
                    \import{templates/\getproperty{default}{#1}/#1}{#2}
                }{
                    \import{templates/Default/#1}{#2}
                }
            }{
                \import{templates/Default/#1}{#2}
            }
        }
    }
}

\makeatother

\documentclass
[
  a4paper,
  DIV=calc,
  BCOR=8mm,
  headinclude,
  bibliography=totoc,
  listof=totoc,
  index=totoc,
  british,
  10pt,
  captions=tableheading,
  chapterprefix,
  openright,
]
{scrartcl}

\newif\ifKOMA
\KOMAtrue

\setproperty{default}{titlepage}{Paper}
\setproperty{default}{appendix}{Paper}
\setproperty{default}{content}{Paper}

\setproperty{compilation}{appendix}{true}
\setproperty{compilation}{lof}{false}
\setproperty{compilation}{lot}{false}
\setproperty{compilation}{toc}{false}
\setproperty{compilation}{complementglyphs}{false}
\setproperty{compilation}{final}{true}
\setproperty{compilation}{comment}{footnote}

\usepackage{authblk}
\setproperty{coauthors}{W7Xteaminclude}{True}
\usepackage[]{algorithm2e}

\usepackage{multirow}

\usepackage[]{lineno}

\usepackage[top=2.5cm, bottom=2.5cm, left=2.5cm, right=2.5cm]{geometry}

\usepackage{xstring}
\usepackage{catchfile}

\usepackage
		[
			ngerman,
			main=british
		]
		{babel}

\RequirePackageWithOption
		[
			log-declarations=false 
		]
		{xparse}

\usepackage{xpatch}
\usepackage{shellesc}
\usepackage{ifluatex}

\ifluatex

	\newcommand{\FirstWord}[1]{\luaexec{tex.print(#1)}}
	
\else
\fi

\setpropertyifundefined{color}{main}{black}
\setpropertyifundefined{color}{secondary}{black}

\setpropertyifundefinedwithusageinfo{author}{firstname}{J.R.R.}
\setpropertyifundefinedwithusageinfo{author}{familyname}{Tolkien}
\setpropertyifundefinedwithusageinfo{author}{acadtitle}{Prof.~Dr.~h.\,c.}
\setpropertyifundefinedwithusageinfo{author}{addressstreet}{Way of no return 77}
\setpropertyifundefinedwithusageinfo{author}{addresscity}{12345 Hobbingen}
\AtBeginDocument{%
    \setpropertyifundefined{author}{address}{\getproperty{author}{addressstreet}\\\getproperty{author}{addresscity}}
}
\setpropertyifundefinedwithusageinfo{author}{dateofbirth}{01.01.1970}
\setpropertyifundefinedwithusageinfo{author}{placeofbirth}{Helms Klamm, Rohan}
\setpropertyifundefinedwithusageinfo{author}{nationality}{german}
\setpropertyifundefinedwithusageinfo{author}{mobile}{+47 173 123456}
\setpropertyifundefinedwithusageinfo{author}{phone}{+53 001 123456}
\setpropertyifundefinedwithusageinfo{author}{faxnr}{+53 001 123456}
\setpropertyifundefinedwithusageinfo{author}{email}{minna@barnhelm.com}
\setpropertyifundefinedwithusageinfo{author}{orcid}{1234567}
\setpropertyifundefinedwithusageinfo{author}{signature}{\backslash{}igraph\{path to your signature here>\}}
\setpropertyifundefined{author}{affiliationindices}{1}
\setpropertyifundefined{affiliations}{1}{Max-Planck-Institute for Plasma Physics, 17491 Greifswald, Germany}

\setpropertyifundefined{document}{date}{\today}
\setpropertyifundefinedwithusageinfo{document}{title}{The Lord of the Rings}
\setpropertyifundefinedwithusageinfo{document}{titlehead}{There and back again}  
\setpropertyifundefinedwithusageinfo{document}{subject}{Novel}
\setpropertyifundefinedwithusageinfo{document}{keywords}{ring,mordor}
\setpropertyifundefinedwithusageinfo{document}{location}{Rivendale}
\setpropertyifundefinedwithusageinfo{document}{publishers}{Elrond Verlag}
\setpropertyifundefinedwithusageinfo{document}{dedication}{For my precious}
\setpropertyifundefinedwithusageinfo{document}{logos}{\backslash{}igraph\{<path to your logo here>\}}
\setpropertyifundefinedwithusageinfo{document}{type}{book}
\setpropertyifundefinedwithusageinfo{document}{doi}{145125.1234150}

\setpropertyifundefined{compilation}{externalfolder}{tikz/}
\setpropertyifundefined{compilation}{externalize}{false}
\setpropertyifundefined{compilation}{pdfa}{false}

\setpropertyifundefined{compilation}{final}{false}
\setpropertyifundefined{compilation}{comment}{footnote}

\AtEndPreamble{%
	\ifthenelseproperty{compilation}{pdfa}{\setpropertyifundefined{compilation}{externalize}{false}}{}
}

\setpropertyifundefined{compilation}{minionfonts}{false}
\setpropertyifundefined{compilation}{libertinusfonts}{false}
\setpropertyifundefined{compilation}{complementglyphs}{true}
\setpropertyifundefined{compilation}{fancychapterheadings}{false}
\setpropertyifundefined{compilation}{microtype}{false}
\setpropertyifundefined{compilation}{fontspec}{false}

\setpropertyifundefinedwithusageinfo{letter}{recipientname}{Sauron Mustermann}
\setpropertyifundefinedwithusageinfo{letter}{recipientaddress}{Ring Institute\\Tower 1\\12345 Mordor}
\setpropertyifundefinedwithusageinfo{letter}{subject}{Application for free Ork position}
\setpropertyifundefinedwithusageinfo{letter}{opening}{Dear Madam or Sir,}
\setpropertyifundefined{letter}{closing}{Sincerely yours,}

\setpropertyifundefined{thesis}{faculty}{FACULTY}
\setpropertyifundefined{thesis}{degree}{DEGREE}
\setpropertyifundefined{thesis}{defensedate}{DEFENSEDATE}
\setpropertyifundefined{thesis}{submissionyear}{YEAR OF SUBMISSION}

\makeatletter
\@ifundefined{ifKOMA}{%
    \newif\ifKOMA%
	\KOMAfalse%
}{}

\@ifclassloaded{book}{%
    \setpropertyifundefined{compilation}{titlepage}{true}
    \setpropertyifundefined{compilation}{abstract}{true}
    \setpropertyifundefined{compilation}{toc}{true}
    \setpropertyifundefined{compilation}{bibliography}{true}
    \setpropertyifundefined{compilation}{lof}{true}
    \setpropertyifundefined{compilation}{lot}{true}
    \setpropertyifundefined{compilation}{appendix}{true}
    \setpropertyifundefined{compilation}{affidavit}{true}
    \setpropertyifundefined{compilation}{clsdefineschapter}{true}
}{}
\@ifclassloaded{report}{%
    \setpropertyifundefined{compilation}{titlepage}{true}
    \setpropertyifundefined{compilation}{abstract}{true}
    \setpropertyifundefined{compilation}{toc}{true}
    \setpropertyifundefined{compilation}{bibliography}{true}
    \setpropertyifundefined{compilation}{lof}{true}
    \setpropertyifundefined{compilation}{lot}{true}
    \setpropertyifundefined{compilation}{appendix}{true}
    \setpropertyifundefined{compilation}{affidavit}{true}
    \setpropertyifundefined{compilation}{clsdefineschapter}{true}
}{}
\@ifclassloaded{article}{%
    \setpropertyifundefined{compilation}{titlepage}{true}
    \setpropertyifundefined{compilation}{abstract}{true}
    \setpropertyifundefined{compilation}{toc}{true}
    \setpropertyifundefined{compilation}{bibliography}{true}
}{}
\@ifclassloaded{scrbook}{%
    \setpropertyifundefined{compilation}{frontmatter}{true}
    \setpropertyifundefined{compilation}{backmatter}{true}
    \setpropertyifundefined{compilation}{titlepage}{true}
    \setpropertyifundefined{compilation}{abstract}{true}
    \setpropertyifundefined{compilation}{toc}{true}
    \setpropertyifundefined{compilation}{bibliography}{true}
    \setpropertyifundefined{compilation}{lof}{true}
    \setpropertyifundefined{compilation}{lot}{true}
    \setpropertyifundefined{compilation}{appendix}{true}
    \setpropertyifundefined{compilation}{affidavit}{true}
    \setpropertyifundefined{compilation}{clsdefineschapter}{true}
}{}
\@ifclassloaded{scrreprt}{%
    \setpropertyifundefined{compilation}{titlepage}{true}
    \setpropertyifundefined{compilation}{abstract}{true}
    \setpropertyifundefined{compilation}{toc}{true}
    \setpropertyifundefined{compilation}{bibliography}{true}
    \setpropertyifundefined{compilation}{lof}{true}
    \setpropertyifundefined{compilation}{lot}{true}
    \setpropertyifundefined{compilation}{appendix}{true}
    \setpropertyifundefined{compilation}{affidavit}{true}
    \setpropertyifundefined{compilation}{clsdefineschapter}{true}
}{}
\@ifclassloaded{scrartcl}{%
    \setpropertyifundefined{compilation}{titlepage}{true}
    \setpropertyifundefined{compilation}{abstract}{true}
    \setpropertyifundefined{compilation}{toc}{true}
    \setpropertyifundefined{compilation}{bibliography}{true}
}{}
\@ifclassloaded{moderncv}{%
    \setpropertyifundefined{compilation}{titlepage}{false}
    \setpropertyifundefined{compilation}{abstract}{false}
    \setpropertyifundefined{compilation}{toc}{false}
    \setpropertyifundefined{compilation}{bibliography}{false}
    \setpropertyifundefined{compilation}{glossaries}{false}
}{}
\setpropertyifundefined{compilation}{frontmatter}{false}
\setpropertyifundefined{compilation}{backmatter}{false}
\setpropertyifundefined{compilation}{titlepage}{false}
\setpropertyifundefined{compilation}{abstract}{false}
\setpropertyifundefined{compilation}{toc}{false}
\setpropertyifundefined{compilation}{los}{false}
\setpropertyifundefined{compilation}{bibliography}{false}
\setpropertyifundefined{compilation}{lof}{false}
\setpropertyifundefined{compilation}{lot}{false}
\setpropertyifundefined{compilation}{lol}{false}
\setpropertyifundefined{compilation}{appendix}{false}
\setpropertyifundefined{compilation}{listofpublications}{false}
\setpropertyifundefined{compilation}{acknowledgement}{false}
\setpropertyifundefined{compilation}{affidavit}{false}
\setpropertyifundefined{compilation}{clsdefineschapter}{false}
\setpropertyifundefined{compilation}{acronyms}{false}
\setpropertyifundefined{compilation}{glossaries}{true}

\makeatother

\RequirePackageWithOption[
			fleqn 
		]
		{amsmath}

\usepackage{amsthm}
\usepackage{calc}
\ifluatex
\else
    \usepackage
		[
		]
		{amsfonts}

    \usepackage
		[
		]
		{amssymb}

\fi

\ifluatex
	\usepackage
		[
		]
		{fontspec}

\setproperty{compilation}{fontspec}{true}

	\usepackage
		[
			math-style=ISO,
			bold-style=ISO,
			nabla=upright,
		]
		{unicode-math}

	\AtEndPreamble{
        \ifthenelseproperty{compilation}{minionfonts}{
			\setmathfont{Cambria Math}
			\ifthenelseproperty{compilation}{complementglyphs}{%
				\setmathfont[range=\mathup/{num,latin,Latin,greek,Greek}]{Minion Pro}
				\setmathfont[range=\mathbfup/{num,latin,Latin,greek,Greek}]{MinionPro-Bold}
				\setmathfont[range=\mathit/{num,latin,Latin,greek,Greek}]{MinionPro-It}
				\setmathfont[range=\mathbfit/{num,latin,Latin,greek,Greek}]{MinionPro-BoldIt}
				\setmathfont[range={"0025,"002B,"002C,"002D,"002E,"2212,"2248,"007E,`±,`×,`·,`÷,`¬,`∂,`∆,`∕,`∞,`⌐,`<,`>,`=,`≥,`≤,`−}]{Minion Pro}
				\setmathfont[range={"2211,"221A}]{Latin Modern Math}
			}{}
			\renewcommand{\mathcal}[1]{{\textit{\addfontfeatures{Contextuals=Swash}#1}}}
			\newfontface{\MinionProSwash}{MinionPro-It}
					[
						Contextuals=Swash,
						Ligatures={TeX,Common,Rare,Historic,Contextual,Required}
					]
			\typeout{----- USING MINION FONTS -----}
		}{
		    \ifthenelseproperty{compilation}{libertinusfonts}{
		    	\setmainfont{Libertinus Serif}
		[
			BoldFont          = {* Bold},
  			ItalicFont        = {* Italic},
  			BoldItalicFont    = {* Bold Italic},
			SmallCapsFeatures = {Renderer=Basic},
			Ligatures         = {TeX,Common},
			Scale             = MatchLowercase,
			RawFeature=-lnum;+pnum;+onum;
		]

		    	\setsansfont{Libertinus Sans}
		[
			BoldFont          = {* Bold},
  			ItalicFont        = {* Italic},
  			BoldItalicFont    = {* -Bold Italic},
			SmallCapsFeatures = {Renderer=Basic},
			Ligatures         = {TeX,Common},
			Scale             = MatchLowercase,
			Numbers           = {Lining,Monospaced},
		]
		    	\setmonofont{Libertinus Mono}
		[
			BoldFont          = {* Bold},
  			ItalicFont        = {* Italic},
  			BoldItalicFont    = {* Bold Italic},
			SmallCapsFeatures = {Renderer=Basic},
			Ligatures         = {Common},
			Scale             = MatchLowercase,
			Numbers           = {Lining,Monospaced},
		]
		    	\setmathfont{Cambria Math}
				\ifthenelseproperty{compilation}{complementglyphs}{%
					\setmathfont[range=\mathup/{num,latin,Latin,greek,Greek}]{LibertinusSerif}
					\setmathfont[range=\mathbfup/{num,latin,Latin,greek,Greek}]{LibertinusSerif-Bold}
					\setmathfont[range=\mathit/{num,latin,Latin,greek,Greek}]{LibertinusSerif-Italic}
					\setmathfont[range=\mathbfit/{num,latin,Latin,greek,Greek}]{LibertinusSerif-BoldItalic}
					\setmathfont[range={"0025,"002B,"002C,"002D,"002E,"2212,"2248,"007E,"2208,"221D,"2A85,`±,`×,`·,`÷,`¬,`∂,`∆,`∕,`∞,`⌐,`<,`>,`=,`≥,`≤,`−,`;,}]{Libertinus Math}
					\setmathfont[range={"2211,"221A}]{Latin Modern Math}
					\setmathfont[range=\mathcal/{num,latin,Latin,greek,Greek},Contextuals=Swash]{Cambria Math}
				}{}
		    	\newfontface{\LibertinusSwash}{LibertinusSerif-Italic}
		    			[
		    				Contextuals=Swash,
		    				Ligatures={TeX,Common,Rare,Historic,Contextual,Required}
		    			]
		    	\newfontface{\LibertinusInitials}{LibertinusSerif-Initials}
		    	\typeout{----- USING LIBERTINUS FONTS -----}
		    }{
                \setmathfont{latinmodern-math.otf}
				\ifthenelseproperty{compilation}{complementglyphs}{%
                    \setmathfont[range=\setminus]{Asana Math}
				}{}
            }
        }
	}
\else
	\usepackage
		[
			T1
		]
		{fontenc}
	\usepackage
		[
			utf8
		]
		{inputenc}
	\usepackage{lmodern}
\fi

\usepackage{fontawesome5} 

\ifluatex
    \usepackage[a-3b,pdf17]{pdfx}
    \usepackage{xmpincl}

    \pdfvariable omitcidset=1
	\newcommand{\replaceNBSP}[1]{\luadirect{s,_ = string.gsub("\luatexluaescapestring{#1}", "\luatexluaescapestring{~}", " "); tex.print(-2,s)}}

    \AtBeginDocument{
    	\newwrite\metadatafile
    	\immediate\openout\metadatafile=\jobname.xmpdata
    	\immediate\write\metadatafile{\unexpanded{\Title}{\expanded{\replaceNBSP{\getproperty{document}{title}}}}}
    	\immediate\write\metadatafile{\unexpanded{\Author}{\expanded{\getproperty{author}{firstname} \getproperty{author}{familyname}}}}
    	\immediate\write\metadatafile{\unexpanded{\Subject}{\expanded{\getproperty{document}{subject}}}}
    	\immediate\write\metadatafile{\unexpanded{\Keywords}{\getproperty{document}{keywords}}}
    	\immediate\write\metadatafile{\unexpanded{\PublicationType}{\expanded{\getproperty{document}{type}}}}
    	\immediate\write\metadatafile{\unexpanded{\Doi}{\expanded{\getproperty{document}{doi}}}}
    	\immediate\closeout\metadatafile
    }
\fi

\usepackage
		[
			locale=UK,
			separate-uncertainty,
			retain-zero-exponent=true,
			binary-units, 
		]
		{siunitx}

\DeclareSIUnit{\arbitraryunit}{a.\,u.}

\DeclareSIUnit{\permille}{‰}

\DeclareSIUnit{\sample}{S}

\DeclareSIPrefix{\Femto}{f\kern0.1ex}{-15}

\DeclareSIUnit{\pb}{\pico\barn}
\DeclareSIUnit{\fb}{\femto\barn}
\DeclareSIUnit{\nb}{\nano\barn}

\usepackage{chemfig}

\usepackage{chemmacros}

\usechemmodule{all}

\renewtagform{reaction}[]{(}{)}

\AtBeginEnvironment{reactions}{}
\AtEndEnvironment{reactions}{}

\usepackage{chemformula}

\setchemformula
		{
			arrow-style={
							>=stealth',
							line cap=round,
							thick
						},
		}

\ifthenelseproperty{compilation}{minionfonts}{
	\setchemformula
			{
				arrow-style={
								>=stealth',
								line cap=round,
								thick
							},
				font-spec={[Numbers={Proportional,Lining}]MinionPro}
			}
}{}

\ifthenelseproperty{compilation}{libertinusfonts}{
	\setchemformula
			{
				arrow-style={
								>=stealth',
								line cap=round,
								thick
							},
				font-spec={[Numbers={Proportional,Lining}]LibertinusSerif}
			}
}{}

\RenewChemArrow{->}{\draw[chemarrow,->] (cf_arrow_start) -- (cf_arrow_end);}

\AtEndPreamble{
\usepackage{scrhack}
}

\AtEndPreamble{
	\ifthenelseproperty{compilation}{pdfa}{
	}{
		\usepackage{intopdf}
		
	}
}

\ifthenelseproperty{compilation}{final}{%
    \PassOptionsToPackage{disable}{todonotes}
}{}

\usepackage%
		[
			colorinlistoftodos, 
		]
		{todonotes}

\ifthenelseproperty{compilation}{final}{%
    \PassOptionsToPackage{disable}{todonotes}
}{}

\usepackage%
		[
			colorinlistoftodos, 
		]
		{todonotes}

\newcommand*{\TODO}[2][inline]{%
    \todo[#1]{#2}
}
\newcommand*{\DONE}[2][color=orange!20,inline]{%
    \todo[#1]{\sout{#2}}
}

\AtEndPreamble{%
    \RequirePackage{letltxmacro}

    \LetLtxMacro{\oldmissingfigure}{\missingfigure}
    \renewcommand{\missingfigure}[2][]{%
        \tikzexternaldisable\oldmissingfigure[{#1}]{#2}\tikzexternalenable%
    }

    \LetLtxMacro{\oldtodo}{\todo}
    \renewcommand{\todo}[2][]{%
        \tikzexternaldisable\oldtodo[#1]{#2}\tikzexternalenable%
    }
}

\ifthenelseproperty{compilation}{final}{%
    \PassOptionsToPackage{final}{changes}
}{%
    \ifpropertydefined{compilation}{comment}{%
        \PassOptionsToPackage{commentmarkup=\getproperty{compilation}{comment}}{changes}
    }
}

\usepackage%
		[
		]
		{changes}

\usepackage{seqsplit}

\usepackage{booktabs}

\ifthenelseproperty{compilation}{fontspec}{%
	\AtBeginEnvironment{tabular}{\addfontfeatures{Numbers={Monospaced}}}
    }{}

\usepackage{xcolor} 

\PassOptionsToPackage{dvipsnames,svgnames,x11names,rgb}{xcolor}
\definecolor{darkgreen}{RGB}{0,80,0} 
\definecolor{darkred}{RGB}{80,0,0} 

\usepackage{graphicx}

\usepackage{pgf}  

\usepackage
		[
			format=plain, 
			indention=1em, 
			textfont={color={black},small}, 
			width=0.925\linewidth, 
			hypcap=true, 
		]
		{caption}

\AtBeginDocument
			{
				\captionsetup
						{
                            labelfont={color=\getproperty{color}{main},small,sf,bf}, 
						}
			}

\usepackage{wrapfig}

\usepackage
		[
			hypcap=true, 
			textfont={color={black},small}, 
			width=0.925\textwidth, 
			indention=1em, 
		]
		{subcaption}

\usepackage{tikz}

\usetikzlibrary
			{%
				arrows,
				arrows.meta,
				bending,
				backgrounds,
				calc,
				fit,
				fadings,
				graphs,
				intersections,
				shadings,
				shapes,
				shapes.misc,
				shapes.geometric,
 				patterns,
 				plotmarks,
				chains,  
 				matrix,
 				through,
 				positioning,  
 				scopes,
				spy,
				decorations.shapes,
				decorations.text,
				decorations.pathmorphing,
				decorations.pathreplacing,
				decorations.markings,
				intersections,
			}
\ifluatex
    \usetikzlibrary
			{%
				graphdrawing,
			}
\fi

\makeatletter
\pgfkeys{/pgf/.cd,
  rectangle corner radius/.initial=3pt
}
\newif\ifpgf@rectanglewrc@donecorner@
\def\pgf@rectanglewithroundedcorners@docorner#1#2#3#4{%
  \edef\pgf@marshal{%
    \noexpand\pgfintersectionofpaths
      {%
        \noexpand\pgfpathmoveto{\noexpand\pgfpoint{\the\pgf@xa}{\the\pgf@ya}}%
        \noexpand\pgfpathlineto{\noexpand\pgfpoint{\the\pgf@x}{\the\pgf@y}}%
      }%
      {%
        \noexpand\pgfpathmoveto{\noexpand\pgfpointadd
          {\noexpand\pgfpoint{\the\pgf@xc}{\the\pgf@yc}}%
          {\noexpand\pgfpoint{#1}{#2}}}%
        \noexpand\pgfpatharc{#3}{#4}{\cornerradius}%
      }%
    }%
  \pgf@process{\pgf@marshal\pgfpointintersectionsolution{1}}%
  \pgf@process{\pgftransforminvert\pgfpointtransformed{}}%
  \pgf@rectanglewrc@donecorner@true
}
\pgfdeclareshape{rectangle with rounded corners}
{
  \inheritsavedanchors[from=rectangle] 
  \inheritanchor[from=rectangle]{north}
  \inheritanchor[from=rectangle]{north west}
  \inheritanchor[from=rectangle]{north east}
  \inheritanchor[from=rectangle]{center}
  \inheritanchor[from=rectangle]{west}
  \inheritanchor[from=rectangle]{east}
  \inheritanchor[from=rectangle]{mid}
  \inheritanchor[from=rectangle]{mid west}
  \inheritanchor[from=rectangle]{mid east}
  \inheritanchor[from=rectangle]{base}
  \inheritanchor[from=rectangle]{base west}
  \inheritanchor[from=rectangle]{base east}
  \inheritanchor[from=rectangle]{south}
  \inheritanchor[from=rectangle]{south west}
  \inheritanchor[from=rectangle]{south east}

  \savedmacro\cornerradius{%
    \edef\cornerradius{\pgfkeysvalueof{/pgf/rectangle corner radius}}%
  }

  \backgroundpath{%
    \northeast\advance\pgf@y-\cornerradius\relax
    \pgfpathmoveto{}%
    \pgfpatharc{0}{90}{\cornerradius}%
    \northeast\pgf@ya=\pgf@y\southwest\advance\pgf@x\cornerradius\relax\pgf@y=\pgf@ya
    \pgfpathlineto{}%
    \pgfpatharc{90}{180}{\cornerradius}%
    \southwest\advance\pgf@y\cornerradius\relax
    \pgfpathlineto{}%
    \pgfpatharc{180}{270}{\cornerradius}%
    \northeast\pgf@xa=\pgf@x\advance\pgf@xa-\cornerradius\southwest\pgf@x=\pgf@xa
    \pgfpathlineto{}%
    \pgfpatharc{270}{360}{\cornerradius}%
    \northeast\advance\pgf@y-\cornerradius\relax
    \pgfpathlineto{}%
  }

  \anchor{south west}{%
    \southwest
    \pgfmathparse{\cornerradius*(1-1/sqrt(2))}
    \advance\pgf@x\pgfmathresult pt
    \advance\pgf@y\pgfmathresult pt}

  \anchor{north east}{%
    \northeast
    \pgfmathparse{\cornerradius*(1-1/sqrt(2))}
    \advance\pgf@x-\pgfmathresult pt
    \advance\pgf@y-\pgfmathresult pt}

  \anchor{north west}{%
    \southwest\pgf@xa=\pgf@x
    \northeast\pgf@x=\pgf@xa
    \pgfmathparse{\cornerradius*(1-1/sqrt(2))}
    \advance\pgf@x\pgfmathresult pt
    \advance\pgf@y-\pgfmathresult pt}

  \anchor{south east}{%
    \southwest\pgf@ya=\pgf@y
    \northeast\pgf@y=\pgf@ya
    \pgfmathparse{\cornerradius*(1-1/sqrt(2))}
    \advance\pgf@x-\pgfmathresult pt
    \advance\pgf@y\pgfmathresult pt}

  \anchor{before north east}{\northeast\advance\pgf@y-\cornerradius}
  \anchor{after north east}{\northeast\advance\pgf@x-\cornerradius}
  \anchor{before north west}{\southwest\pgf@xa=\pgf@x\advance\pgf@xa\cornerradius
    \northeast\pgf@x=\pgf@xa}
  \anchor{after north west}{\northeast\pgf@ya=\pgf@y\advance\pgf@ya-\cornerradius
    \southwest\pgf@y=\pgf@ya}
  \anchor{before south west}{\southwest\advance\pgf@y\cornerradius}
  \anchor{after south west}{\southwest\advance\pgf@x\cornerradius}
  \anchor{before south east}{\northeast\pgf@xa=\pgf@x\advance\pgf@xa-\cornerradius
    \southwest\pgf@x=\pgf@xa}
  \anchor{after south east}{\southwest\pgf@ya=\pgf@y\advance\pgf@ya\cornerradius
    \northeast\pgf@y=\pgf@ya}

  \anchorborder{%
    \pgf@xb=\pgf@x
    \pgf@yb=\pgf@y%
    \southwest%
    \pgf@xa=\pgf@x
    \pgf@ya=\pgf@y%
    \northeast%
    \advance\pgf@x by-\pgf@xa%
    \advance\pgf@y by-\pgf@ya%
    \pgf@xc=.5\pgf@x
    \pgf@yc=.5\pgf@y%
    \advance\pgf@xa by\pgf@xc
    \advance\pgf@ya by\pgf@yc%
    \edef\pgf@marshal{%
      \noexpand\pgfpointborderrectangle
      {\noexpand\pgfqpoint{\the\pgf@xb}{\the\pgf@yb}}
      {\noexpand\pgfqpoint{\the\pgf@xc}{\the\pgf@yc}}%
    }%
    \pgf@process{\pgf@marshal}%
    \advance\pgf@x by\pgf@xa%
    \advance\pgf@y by\pgf@ya%
    \pgfextract@process\borderpoint{}%
    \pgf@rectanglewrc@donecorner@false
    \southwest\pgf@xc=\pgf@x\pgf@yc=\pgf@y
    \advance\pgf@xc\cornerradius\relax\advance\pgf@yc\cornerradius\relax 
    \borderpoint
    \ifdim\pgf@x<\pgf@xc\relax\ifdim\pgf@y<\pgf@yc\relax
      \pgf@rectanglewithroundedcorners@docorner{-\cornerradius}{0pt}{180}{270}%
    \fi\fi
    \ifpgf@rectanglewrc@donecorner@\else
      \southwest\pgf@yc=\pgf@y\relax\northeast\pgf@xc=\pgf@x\relax
      \advance\pgf@xc-\cornerradius\relax\advance\pgf@yc\cornerradius\relax
      \borderpoint
      \ifdim\pgf@x>\pgf@xc\relax\ifdim\pgf@y<\pgf@yc\relax
       \pgf@rectanglewithroundedcorners@docorner{0pt}{-\cornerradius}{270}{360}%
      \fi\fi
    \fi
    \ifpgf@rectanglewrc@donecorner@\else
      \northeast\pgf@xc=\pgf@x\relax\pgf@yc=\pgf@y\relax
      \advance\pgf@xc-\cornerradius\relax\advance\pgf@yc-\cornerradius\relax
      \borderpoint
      \ifdim\pgf@x>\pgf@xc\relax\ifdim\pgf@y>\pgf@yc\relax
       \pgf@rectanglewithroundedcorners@docorner{\cornerradius}{0pt}{0}{90}%
      \fi\fi
    \fi
    \ifpgf@rectanglewrc@donecorner@\else
      \northeast\pgf@yc=\pgf@y\relax\southwest\pgf@xc=\pgf@x\relax
      \advance\pgf@xc\cornerradius\relax\advance\pgf@yc-\cornerradius\relax
      \borderpoint
      \ifdim\pgf@x<\pgf@xc\relax\ifdim\pgf@y>\pgf@yc\relax
       \pgf@rectanglewithroundedcorners@docorner{0pt}{\cornerradius}{90}{180}%
      \fi\fi
    \fi
  }
}
\makeatother

\usepackage{pgfplots}
\usepackage{pgfplotstable}

\pgfplotsset{compat=newest}

\usetikzlibrary{pgfplots.groupplots}

\usepgflibrary{decorations.pathmorphing}
\usepgfplotslibrary{fillbetween,colormaps,external,colorbrewer}

\pgfplotsset{aspect ratio/.code args={#1:#2}{%
  \def\axisdefaultwidth{#1 cm}%
  \def\axisdefaultheight{#2 cm}},
}

\pgfplotsset{cycle list/Dark2-8}
\pgfplotsset{colormap/viridis}
\pgfkeys{/pgf/number format/.cd,1000 sep={\,}}

\AtEndPreamble{\ifthenelseproperty{compilation}{externalize}{
		\tikzexternalize[prefix=\getproperty{compilation}{externalfolder}]
		\typeout{----- USING TIKZ EXTERNALIZE -----}
	}{
		\typeout{----- NOT USING TIKZ EXTERNALIZE -----}
	}
}

\usepackage{listings}

\definecolor{javagray}{rgb}{0.55, 0.52, 0.54} 
\definecolor{javared}{rgb}{0.6,0,0} 
\definecolor{javagreen}{rgb}{0.25,0.5,0.35} 
\definecolor{javapurple}{rgb}{0.5,0,0.35} 
\definecolor{javadocblue}{rgb}{0.25,0.35,0.75} 
\definecolor{javaLila}{RGB}{127,0,85}

\let\origthelstnumber\thelstnumber

\makeatletter
\newcommand*\Suppressnumber{%
  \lst@AddToHook{OnNewLine}{%
    \let\thelstnumber\relax%
  }%
}
\newcommand\Reactivatenumber[1]{%
  \global\c@lstnumber#1%
  \global\advance\c@lstnumber\m@ne\relax%
  \lst@AddToHook{OnNewLine}{%
  \let\thelstnumber\origthelstnumber%
  }%
}
\makeatother

\makeatletter
\lst@InputCatcodes
\def\lst@DefEC{%
 \lst@CCECUse \lst@ProcessLetter
^^80^^81^^82^^83^^84^^85^^86^^87^^88^^89^^8a^^8b^^8c^^8d^^8e^^8f%
  ^^90^^91^^92^^93^^94^^95^^96^^97^^98^^99^^9a^^9b^^9c^^9d^^9e^^9f%
  ^^a0^^a1^^a2^^a3^^a4^^a5^^a6^^a7^^a8^^a9^^aa^^ab^^ac^^ad^^ae^^af%
  ^^b0^^b1^^b2^^b3^^b4^^b5^^b6^^b7^^b8^^b9^^ba^^bb^^bc^^bd^^be^^bf%
  ^^c0^^c1^^c2^^c3^^c4^^c5^^c6^^c7^^c8^^c9^^ca^^cb^^cc^^cd^^ce^^cf%
  ^^d0^^d1^^d2^^d3^^d4^^d5^^d6^^d7^^d8^^d9^^da^^db^^dc^^dd^^de^^df%
  ^^e0^^e1^^e2^^e3^^e4^^e5^^e6^^e7^^e8^^e9^^ea^^eb^^ec^^ed^^ee^^ef%
  ^^f0^^f1^^f2^^f3^^f4^^f5^^f6^^f7^^f8^^f9^^fa^^fb^^fc^^fd^^fe^^ff%
	^^^^03b8^^^^03c8^^^^03b7^^^^03bc^^^^03c3^^^^03b1^^^^03a9^^^^03b6%
	^^^^03c9^^^^03b4^^^^03c0%
^^00}
\lst@RestoreCatcodes
\makeatother

\RequirePackageWithOption{hyperref}

\hypersetup
		{
			pdfencoding=unicode,
			pdfpagemode=UseOutlines,
			bookmarksopenlevel=0,
			pdfpagelayout=TwoPageRight,
			hidelinks,
			pdfduplex=DuplexFlipLongEdge,
		}

\usepackage{epigraph}

\usepackage
		[
			math,
			toc,
		]
		{blindtext}

\usepackage
		[
			noabbrev
		]
		{cleveref}

\crefname{line}{line}{lines}
\creflabelformat{line}{#2#1#3}

\crefname{reaction}{reaction}{reactions}
\creflabelformat{reaction}{#2(#1)#3}

\crefname{listing}{code}{codes}

\usepackage
		[
			strict=true, 
			autostyle=true, 
			german=guillemets, 
		]
		{csquotes}

\setquotestyle{german} 

\usepackage{lettrine} 

\setpropertyifundefined{biblatex}{style}{numeric-comp}

\usepackage
		[
			hyperref=true,
			backend=biber, 
			url=true, 
			abbreviate=false,
			isbn=false, 
			doi=true, 
			backref=false, 
			urldate=long, 
			style=\getproperty{biblatex}{style},
			maxbibnames=3, 
			giveninits=true,
			block=none,
            sorting=none,
			defernumbers=true,
		]
		{biblatex}

\makeatletter
\@ifpackageloaded{seqsplit}{
	\DeclareFieldFormat{eid}{Art.\addnbspace\seqsplit{#1}}
}{}
\makeatother

\DeclareSourcemap{
  \maps[datatype=bibtex]{
    \map{
      \step[fieldsource=doi,final]
      \step[fieldset=url,null]
    }
  }
}

\DeclareFieldFormat{journaltitle}{\mkbibemph{#1},} 
\DeclareFieldFormat[inbook,thesis]{title}{\mkbibemph{#1}\addperiod} 
\DeclareFieldFormat[article,misc]{title}{\enquote{#1}} 
\DeclareFieldFormat[article,inproceedings,incollection]{volume}{Vol.~#1} 
\DeclareFieldFormat{edition}{\ifinteger{#1}{\mkbibordedition{#1}\addnbthinspace{}ed.}{#1\isdot}}

\DeclareFieldFormat{eprint:urn}{\textsc{URN}\addcolon\space\ifhyperref{\href{http://www.nbn-resolving.org/#1}{\nolinkurl{#1}}}{\nolinkurl{#1}}}

\DeclareSortingTemplate{ndymdt}{
 	\sort{
		\field{presort}
	}
	\sort[final]{
		\field{sortkey}
	}
	\sort[direction=descending]{
		\field{sortyear}
		\field{year}
		\literal{9999}
	}
	\sort[direction=descending]{
		\field[padside=left,padwidth=2,padchar=0]{month}
		\literal{99}
	}
	\sort[direction=descending]{
		\field[padside=left,padwidth=2,padchar=0]{day}
		\literal{99}
	}
}

\ifluatex
	\def\getAuthorBibCv{\FirstChar{\getproperty{author}{firstname}}.~\getproperty{author}{familyname}}
\else
	\def\getAuthorBibCv{\getproperty{author}{firstname}~\getproperty{author}{familyname}}
\fi

\newboolean{isCoauthorList}
\setboolean{isCoauthorList}{false}

\DefineBibliographyStrings{english}{
    andothers={%
        \ifthenelse{%
            \boolean{isCoauthorList}
        }{%
            et\addabbrvspace al\adddot\addcomma\addspace amongst them \textbf{\textsc{\getAuthorBibCv}}
        }{%
            et\addabbrvspace al\adddot}
    }
}

\DeclareBibliographyCategory{dontbib}
\let\printbibliographyold\printbibliography%
\renewcommand{\printbibliography}[1][]{%
    \typeout{----- printbibliography ------}%
    \printbibliographyold[notcategory=dontbib,#1]
}%
\makeatletter
\newcommand{\tempmaxup}[1]{\def\blx@maxcitenames{99}#1}
\DeclareCiteCommand{\fullcitecontribution}[\tempmaxup]{
    \usebibmacro{prenote}
    \addtocategory{dontbib}{\thefield{entrykey}}
}{
    \textbf{\usebibmacro{maintitle+title}}
    \newline\nopunct\newblock
    \usebibmacro{author}
    \newline\nopunct\newblock
    \usebibmacro{journal+issuetitle}, \usebibmacro{doi+eprint+url} \usebibmacro{addendum+pubstate}
}{
    \multicitedelim
}{
    \usebibmacro{postnote}
}
\makeatother

\ifKOMA
    \usepackage
    		[
    			automark,
    			headsepline, 
    		]
    		{scrlayer-scrpage}
    
    \AtBeginDocument{%
		\addtokomafont{pagehead}{\color{\getproperty{color}{main}}}
		\ifthenelseproperty{compilation}{clsdefineschapter}{%
			\addtokomafont{chapter}{\color{\getproperty{color}{main}}}
		}{}
		\addtokomafont{section}{\color{\getproperty{color}{main}}}
		\addtokomafont{subsection}{\color{\getproperty{color}{main}}}
		\addtokomafont{subsubsection}{\color{\getproperty{color}{main}}}
		\addtokomafont{paragraph}{\color{\getproperty{color}{main}}}

		\addtokomafont{footnoterule}{\color{\getproperty{color}{secondary}}}
		\addtokomafont{headsepline}{\color{\getproperty{color}{secondary}}}

    }
\fi

\usepackage{bookmark}

\bookmarksetup
			{
				open,
				addtohook={
					\ifnum\bookmarkget{level}<1
						\bookmarksetup{bold}
					\fi
				},
			}

\PassOptionsToPackage{automake=immediate}{glossaries-extra}

\usepackage%
		[
			nonumberlist, 
			acronym, 
			toc, 
		]
		{glossaries-extra}

\usepackage{glossary-mcols}

\setabbreviationstyle[acronym]{long-short}

\newglossary[slg]{symbols}{syi}{syg}{List of Symbols} 

\makeglossaries

\newglossarystyle{customGlossaryStyleForListOfSymbols}{%
    \renewenvironment{theglossary}{%
        \begin{longtable}{p{0.12\textwidth}p{\glsdescwidth}p{\glspagelistwidth}}
    }{%
        \end{longtable}
    }
    
    \renewcommand*{\glsgroupheading}[1]{}

}

\newglossarystyle{customListOfSymbols}{
    \setglossarystyle{customGlossaryStyleForListOfSymbols}
    
    \renewenvironment{theglossary}{%
        \begin{longtable}{p{0.12\textwidth}p{\glsdescwidth}p{\glspagelistwidth}}
    }{%
        \end{longtable}
    }
}

\newglossarystyle{customListOfAbbreviations}{%
    \renewenvironment{theglossary}{%
        \begin{longtable}{lp{\glsdescwidth}}
    }{%
        \end{longtable}
    }
    
    \renewcommand*{\glsgroupheading}[1]{}

}

\makeatletter
\newglossarystyle{customIndex}{
    \renewenvironment{theglossary}{%
        \setlength{\parindent}{0pt}
        \setlength{\parskip}{0pt plus 0.3pt}
        \let\item\@idxitem
    }{%
    }
    
    \renewcommand*{\glsgroupheading}[1]{}
    \renewcommand*{\glossaryentryfield}[5]{%
    \item\glstarget{##1}{##2}
        \ifx\relax##4\relax
        \else
            \space(##4)
        \fi
        \dotfill ##3\glspostdescription \space ##5
    }
    \renewcommand*{\glossarysubentryfield}[6]{%
        \ifcase##1\relax
        \item
        \or
            \subitem
        \else
            \subsubitem
        \fi
        \glstarget{##2}{##3}
        \ifx\relax##5\relax
        \else
            \space(##5)
        \fi
        \dotfill ##4\glspostdescription\space ##6
    }
    
    \renewcommand*{\glsgroupheading}[1]{%
        \item\textbf{\glsgetgrouptitle{##1}}\indexspace
    }
}
\makeatother

\newglossarystyle{documentationLabelStyle}{
    \renewenvironment{theglossary}{%
        \begin{longtable}{p{.2\textwidth}p{.3\textwidth}p{.5\textwidth}}
    }{%
        \end{longtable}
    }%

}

\newglossarystyle{documentationSymbolsStyle}{
    \renewenvironment{theglossary}{%
        \begin{longtable}{p{0.12\textwidth}p{.2\textwidth}p{.48\textwidth}p{.2\textwidth}}
    }{%
        \end{longtable}
    }

}

\newglossarystyle{publicationLabelStyle}{
	\renewenvironment{theglossary}{%
		\begin{longtable}{p{.2\textwidth}p{.8\textwidth}}
		}{%
		\end{longtable}
	}%

}

\newglossarystyle{publicationSymbolsStyle}{
}

\usepackage{scrtime}

\usepackage{ragged2e}

\usepackage{marginnote}

\DeclareDocumentCommand{\myMarginnote} 
					{
				 		O{0cm} O{c} m 
				 	}{
						\marginnote
								{
									\ifthispageodd
											{
												\RaggedRight 
											}{
												\RaggedLeft 
											}
									\raisebox{#1}[#1]
											{
												\begin{minipage}[#2]{\marginparwidth}
													\RaggedRight 
													\color{\getMainColor}
													\lineskiplimit=-\maxdimen
													\normalfont\sffamily
													#3\end{minipage}}
											}
					}

\ifluatex
	\usepackage[luatex]{pdflscape}
\else
\fi

\usepackage
		[
			final, 
		]
		{pdfpages}

\pretocmd{\includepdf}{%
    \ifthenelseproperty{compilation}{externalize}{%
        \tikzset{external/optimize=false}%
    }{}
}{}{}

\makeatletter
\@ifpackageloaded{cleveref}{%
    \newcounter{articlenumber}
    \crefname{articlenumber}{article}{articles}
    \Crefname{articlenumber}{Article}{Articles}

}
\makeatother

\usepackage{enumitem}

\setlist[description]{leftmargin=*,style=sameline}

\ifluatex

	\usepackage{luacode}

\usepackage
		[
			draft, 
		]
		{impnattypo}

\usepackage{selnolig}

    \usepackage{pdftexcmds}
    \makeatletter
    \let\pdfstrcmp\pdf@strcmp
    \makeatother
\else
\fi

\def\clap#1{\hbox to 0pt{\hss#1\hss}}

\newlength{\heightOfX}
\AtBeginDocument{\settoheight{\heightOfX}{X}} 

\newlength{\fsize}

\newcommand{\nobreakbefore}
	{%
		\relax
		\ifvmode
		\else
			\ifhmode
				\ifdim\lastskip > 0pt\relax
					\unskip\nobreakspace
				\fi
    		\fi
  		\fi
	}
\let\oldcite\cite
\renewcommand{\cite}{\nobreakbefore\oldcite}
\let\oldref\ref
\renewcommand{\ref}{\nobreakbefore\oldref}

\newcommand{\disabledprotrusion}[1]
		{
			\ifKOMA
                \ifthenelseproperty{compilation}{fontspec}{%
					\begingroup
						\addfontfeatures{Numbers={Lining,Monospaced}}
                }{}
			\fi
            \ifthenelseproperty{compilation}{microtype}{%
				\microtypesetup{protrusion=false}
            }{}
			#1
            \ifthenelseproperty{compilation}{microtype}{%
				\microtypesetup{protrusion=true}
            }{}
			\ifKOMA
                \ifthenelseproperty{compilation}{fontspec}{%
					\endgroup
                }{}
			\fi
		}

\makeatletter
\newcommand{\removeifnextchar}[3]
		{
			\begingroup
			\ltx@LocToksA{\endgroup#2}
			\ltx@LocToksB{\endgroup#3}
			\ltx@ifnextchar{#1}
				{
					\def\next{\the\ltx@LocToksA}
					\afterassignment\next
					\let\scratch= %
				}{
					\the\ltx@LocToksB
				}
		}
\makeatother

\newcommand{\Signature}[2]
		{%
			\vspace{2cm}%
			\noindent%
            \begin{tabular*}{\textwidth}{@{\extracolsep{0pt}} l @{\extracolsep{\fill}} r @{\extracolsep{0pt}}}%
#1, \getproperty{document}{date}	& \rule{0.33\textwidth}{1pt} 	\\
& \raggedleft{\textsc{#2}}
			\end{tabular*}
			\vspace{1cm}
		}

\AtBeginDocument{%
	\newif\ifKOMAandFancyChapterHeadings%
	\KOMAandFancyChapterHeadingsfalse%
    \ifKOMA%
        \ifthenelseproperty{compilation}{fancychapterheadings}{%
            \RedeclareSectionCommand[
                    beforeskip=\dimexpr3.3\baselineskip+1\parskip\relax,
                    innerskip=0pt,
                    afterskip=1.725\baselineskip plus .115\baselineskip minus .192\baselineskip,
                    afterindent=false
                ]%
                {chapter}%
            \KOMAandFancyChapterHeadingstrue%
        }{}%
    \fi%
    \ifKOMAandFancyChapterHeadings%
    \else%
    \fi%
}

\ifKOMA
    \ifthenelseproperty{compilation}{fontspec}{%
		\DeclareTOCStyleEntry[%
							entryformat={%
                                \addfontfeatures{Numbers={Lining,Monospaced}}\bfseries\sffamily\color{\getproperty{color}{main}}
										},
							pagenumberformat={%
											\addfontfeatures{Numbers={Lining,Monospaced}}\bfseries\sffamily
										},
						]
						{default}
						{chapter}
		\DeclareTOCStyleEntry[%
							entryformat={%
											\addfontfeatures{Numbers={Lining,Monospaced}}
										},
							pagenumberformat={%
											\addfontfeatures{Numbers={Lining,Monospaced}}
										},
						]
						{default}
						{section}
		\DeclareTOCStyleEntry[%
							entryformat={%
											\addfontfeatures{Numbers={Lining,Monospaced}}
										},
							pagenumberformat={%
											\addfontfeatures{Numbers={Lining,Monospaced}}
										},
						]
						{default}
						{subsection}
		\DeclareTOCStyleEntry[%
							entryformat={%
											\addfontfeatures{Numbers={Lining,Monospaced}}
										},
							pagenumberformat={%
											\addfontfeatures{Numbers={Lining,Monospaced}}
										},
						]
						{default}
						{subsubsection}
		\DeclareTOCStyleEntry[%
							entryformat={%
											\addfontfeatures{Numbers={Lining,Monospaced}}
										},
							pagenumberformat={%
											\addfontfeatures{Numbers={Lining,Monospaced}}
										},
						]
						{tocline}
						{figure}
		\DeclareTOCStyleEntry[%
							entryformat={%
											\addfontfeatures{Numbers={Lining,Monospaced}}
										},
							pagenumberformat={%
											\addfontfeatures{Numbers={Lining,Monospaced}}
										},
						]
						{tocline}
						{table}
    }{}
\fi

\makeatletter
\newcommand{\headlesschapter}[1]{%
  \begingroup
  \let\@makechapterhead\@gobble 
  \chapter{#1}
  \endgroup
}
\makeatother

\makeatletter
\newcommand\appendgraphicspath[1]{%
  \g@addto@macro\Ginput@path{#1}%
}
\makeatother

\RequirePackage{xkeyval}
\RequirePackage{xstring}
\RequirePackage{xfp}
\RequirePackage{ifthen}
\RequirePackage{tikzscale} 
\makeatletter
\define@key{igraph}{width}{\def\igraph@width{#1}}
\define@key{igraph}{svgwidth}{\def\igraph@svgwidth{#1}}
\define@key{igraph}{angle}{\def\igraph@angle{#1}}
\define@key{igraph}{draft}{\def\igraph@draft{#1}}
\define@key{igraph}{scale}{\def\igraph@scale{#1}}
\define@key{igraph}{height}{\def\igraph@height{#1}}
\define@boolkey{igraph}{scaletext}{\def\igraph@scaletext{#1}}
\newcommand{\igraph}[2][]{%
    \filename@parse{#2}
    \setkeys{igraph}{width=\linewidth} 
    \setkeys{igraph}{svgwidth=\linewidth} 
    \setkeys{igraph}{height=\empty} 
    \setkeys{igraph}{scaletext=false} 
    \setkeys{igraph}{#1}
    \ifthenelse{%
        \equal{\filename@ext}{pgf}%
    }{%
        \typeout{----- INCLUDE pgf @ #2 ------}%
        \let\pgfimageWithoutPath\pgfimage%
        \renewcommand{\pgfimage}[2][]{\typeout{----- INCLUDE pgfimgage @ ##2 ------}\pgfimageWithoutPath[##1]{\filename@area/##2}}%
        \resizebox{\igraph@width}{!}{\input{#2}}%
    }{%
        \ifnum\pdfstrcmp{\filename@ext}{pdf_tex}=0
            \typeout{----- INCLUDE pdf_tex @ #2 ------}%
            \IfSubStr{#1}{height}{%
                \PackageError{igraph}{pdf_tex does not allow the height attribute}{}%
                }{}%
            \ifthenelse{%
                \boolean{\igraph@scaletext}%
            }{%
                \def\svgwidth{\igraph@svgwidth}%
                \resizebox{\igraph@width}{!}{%
                    \import{\filename@area}{\filename@base.\filename@ext}%
                }%
            }{%
                \def\svgwidth{\igraph@width}%
                \import{\filename@area}{\filename@base.\filename@ext}%
            }%
        \else
            \ifthenelse{%
                \equal{\filename@ext}{svg}%
            }{%
                \typeout{----- INCLUDE svg @ #2 ------}%
                \PackageError{igraph}{svg not implemented!}{}%
            }{%
                \ifthenelse{%
                    \equal{\filename@ext}{tikz}%
                }{%
                    \typeout{----- INCLUDE tikz @ #2 ------}%
	                \tikzsetnextfilename{\filename@base}%
                    \begin{minipage}{\igraph@width}%
                        \input{#2}%
                    \end{minipage}%
                }{%
                    \typeout{----- INCLUDE with includegraphics @ #2 ------}%
                    \IfSubStr{#1}{height}{%
                        \includegraphics[#1]{#2}%
                    }{%
                        \includegraphics[width=\igraph@width, #1]{#2}%
                    }%
                }%
            }%
        \fi
    }%
}%
\makeatother

\newlength{\imageh}
\newlength{\imaged}
\newlength{\imagew}
\newcommand{\setimageh}[1]{
 \settoheight{\imageh}{\usebox{#1}}
}
\newcommand{\setimagew}[1]{
 \settowidth{\imagew}{\usebox{#1}}
}
\newcommand{\setimaged}[1]{
 \settodepth{\imaged}{\usebox{#1}}
}
\newsavebox{\imagesavebox}
\newcommand{\setimagedimensions}[1]{
    \savebox{\imagesavebox}{\igraph{#1}}
    \setimageh{\imagesavebox}
    \setimagew{\imagesavebox}
    \setimaged{\imagesavebox}
}

\makeatletter
\newcommand*{\storedata}[2]{%
  \count@=0 %
  \@tfor\@tmp:=#2\do{%
    \advance\count@\@ne
    \expandafter\let\csname data:\the\count@:#1\endcsname\@tmp
  }%
  \expandafter\edef\csname data:0:#1\endcsname{\the\count@}%
}
\newcommand*{\getdata}[2]{%
  \@ifundefined{data:0:#2}{%
    \@latex@error{Undefined data `#2'}\@ehc
  }{%
    \expandafter\@getdata\expandafter{%
      \the\numexpr
        \ifnum\numexpr(#1)<\z@
          \@nameuse{data:0:#2}+1+%
        \fi
        (#1)%
      \relax
    }{#2}{#1}%
  }%
}
\newcommand*{\@getdata}[3]{%
  \ifnum#1<\z@
    \@getdata@error{\the\numexpr(#3)\relax}{#2}%
  \else
    \ifnum#1>\@nameuse{data:0:#2} %
      \@getdata@error{#1}{#2}%
    \else
      \@nameuse{data:#1:#2}%
    \fi
  \fi
}
\newcommand*{\@getdata@error}[2]{%
  \@latex@error{%
    Wrong data selector #1 for `#2',\MessageBreak
    which only contains \@nameuse{data:0:#2} item(s)%
  }\@ehc
}
\makeatother

\storedata{mydata}{{one}{two}{three}}
\storedata{mylongdata}{
  {one} {two} {three} {four} {five} {six} {seven} {eight} {nine} {ten}
  {eleven} {twelve} {thirteen} {fourteen} {fifteen} {sixteen}
  {seventeen} {eighteen} {nineteen} {twenty} {twenty-one} {twenty-two}
  {twenty-three} {twenty-four}
}

\makeatletter
\RequirePackage{forarray}
\RequirePackage{calc}
\RequirePackage{listings}
\newlength\figratiosum
\newlength\figratio
\newlength\figheight
\newlength\figwidth
\newcounter{npaths}
\newcommand{\multigraph}{\begingroup
  \catcode`_=12 \domultigraph}
\newcommand{\catcodeigraph}{\begingroup
  \catcode`_=11 \igraph}
\define@key{multigraph}{width}{\def\multigraph@width{#1}}
\define@key{multigraph}{labels}{\def\multigraph@labels{#1}}
\newcommand{\domultigraph}[3][]{
    \setkeys{multigraph}{width=\linewidth - 1em} 
    \setkeys{multigraph}{labels=\empty} 
    \setkeys{multigraph}{#1}

    \setcounter{npaths}{0}
    \ForEachX{;}{%
        \stepcounter{npaths}
    }{#2}

    \IfSubStr{#1}{labels}{%
        \def\subfigurelabels{}
        \ForEachX{;}{%
            \edef\subfigurelabels{\subfigurelabels"\thislevelitem"}
            \ifnum\thislevelcount=\value{npaths}
            \else
                \edef\subfigurelabels{\subfigurelabels,}
            \fi
        }{\multigraph@labels}
    }{}

    \def\paths{}
    \setlength\figratiosum{0pt}
    \def\figratios{}
    \ForEachX{;}{%
        \edef\paths{\paths"\thislevelitem"}
        \ifnum\thislevelcount=\value{npaths}
        \else
            \edef\paths{\paths,}
        \fi

        \setimagedimensions{\thislevelitem}
        \setlength\figratio{1pt*\ratio{\imagew}{\imageh}}
        \addtolength{\figratiosum}{\figratio}
        \edef\figratios{\figratios"\the\figratio"}
        \ifnum\thislevelcount=\value{npaths}
        \else
            \edef\figratios{\figratios,}
        \fi
    }{#2}

    \storedata{subcaptions}{#3}

    \setlength{\figheight}{1pt*\ratio{\multigraph@width}{\figratiosum}}

    \addtocounter{npaths}{-1}  
    \foreach \i in {0,...,\value{npaths}}{%
        \pgfmathparse{{\figratios}[\i]}
        \setlength\figwidth{\figheight}
        \pgfmathparse{\figwidth * \pgfmathresult}
        \setlength{\figwidth}{\pgfmathresult pt}
        \begin{subfigure}{\figwidth}
            \centering
            \pgfmathparse{{\paths}[\i]}
            \typeout{----- Width for multigraph figure \pgfmathresult : \the\figwidth ------}
            \catcodeigraph{\pgfmathresult}\endgroup
            \subcaption{\getdata{\i+1}{subcaptions}}
            \IfSubStr{#1}{labels}{%
                \pgfmathparse{{\subfigurelabels}[\i]}
                \label{\pgfmathresult}
            }{}
        \end{subfigure}
    }
    \endgroup
}
\makeatother

\newcommand{\makeup}[1]{%
    \ensuremath{
        \ifluatex
            \symup{#1}
        \else
            \mathrm{#1}
        \fi
    }
}
\newcommand{\makebf}[1]{%
    \ensuremath{
        \ifluatex
            \symbf{#1}
        \else
            \mathbf{#1}
        \fi
    }
}

\newcommand{\eg}{e.\,g.\xspace}

\newcommand{\ie}{i.\,e.\xspace}

\newcommand{\vect}[1]{\ensuremath{\vec{#1}}} 
\newcommand{\matr}[1]{\ensuremath{\makebf{#1}}}

\makeatletter
\newcommand{\Grad}[2][\@nil]{%
    \def\tmp{#1}%
    \ifx\tmp\@nnil
    	\ensuremath{\vect{\nabla} #2}
    \else
    	\ensuremath{\vect{\nabla}_{\!\!#1} #2}
    \fi}
\makeatother

\newcommand{\dx}[3][\empty]
		{
			\if{#1}\equal{\empty}
				\frac{\mathrm{d}#2}{\mathrm{d}#3}
			\else
				\frac{\mathrm{d}^{#1}#2}{\mathrm{d}#3^{#1}}
		}
\newcommand{\pdx}[3][\empty]
		{
			\if{#1}\equal{\empty}
				\frac{\partial#3}{\partial#2}
			\else
				\frac{\partial^{#1}#3}{\partial#2^{#1}}
		}
\makeatletter
\providecommand*{\diff}{\@ifnextchar^{\DIfF}{\DIfF^{}}}
\def\DIfF^#1{\mathop{\mathrm{\mathstrut d}}\nolimits^{#1}\gobblespace}
\def\gobblespace{\futurelet\diffarg\opspace}
\def\opspace
		{
			\let\DiffSpace\!
			\ifx\diffarg(
				\let\DiffSpace\relax
			\else
				\ifx\diffarg[%
					\let\DiffSpace\relax
				\else
					\ifx\diffarg\{%
						\let\DiffSpace\relax
					\fi
				\fi
			\fi
			\DiffSpace
		}
\makeatother

\newcommand{\orderof}[1]{\ensuremath{\mathcal{O}(#1)}}

\DeclareMathOperator*{\argmax}{arg\,max}

\newcommand{\tightoverset}[2]{\mathop{#2}\limits^{\vbox to -.5ex{\kern-0.75ex\hbox{$\! #1$}\vss}}} 

\DeclareSIUnit[number-unit-product = \,]{\permille}{\textperthousand}

\newacronym{statistics}{Statistics}{\nopostdesc}
\newacronym[parent=statistics,plural=MCs]{MC}{MC}{Monte Carlo}
\newacronym[parent=statistics,plural=MCMCs]{MCMC}{MCMC}{Markov chain Monte Carlo}
\newacronym[parent=statistics]{MAP}{MAP}{maximum a posteriori}
\newacronym[parent=statistics,plural=LAs,longplural={Laplace approximations}]{LA}{LA}{Laplace approximation}
\newacronym[parent=statistics,first=kernel density estimate (KDE) \cite{Parzen1962,Silverman1986},plural=KDEs,firstplural=kernel density estimates (KDEs) \cite{Parzen1962,Silverman1986},longplural={kernel density estimates}]{KDE}{KDE}{kernel density estimate}
\newacronym[parent=statistics,plural=PDFs,longplural={probability density functions}]{PDF}{PDF}{probability density function}
\newacronym[parent=statistics,plural=GPs,longplural={Gaussian processes}]{GP}{GP}{Gaussian process}
\newacronym[parent=statistics,plural=LGIs,longplural={linear Gaussian inversions}]{LGI}{LGI}{linear Gaussian inversion}
\newacronym[parent=statistics,plural=LSs,longplural={least-squares}]{LS}{LS}{least-square}
\newacronym[parent=statistics]{MaLi}{Max-$\mathcal{L}$}{maximum likelihood}
\newacronym[parent=statistics,plural=iid,longplural={independent and identically distributed}]{iid}{iid}{independent and identically distributed}
\newacronym[parent=statistics]{FWHM}{FWHM}{full-width half-maximum}
\newacronym[parent=statistics]{HDI}{HDI}{highest density interval}
\newacronym[parent=statistics]{MHA}{MHA}{Metropolis-Hastings algorithm}
\newacronym[parent=statistics]{KS}{KS}{Kolmogorov–Smirnov}
\newglossaryentry{PearsonCorrelationSample}{%
	parent=statistics,
	name=\ensuremath{r_{\text{xy}}},
	type=symbols,
	sort=correlation,
	description={Pearson sample correlation coefficient},
	symbol={},
}
\newacronym[parent=statistics,plural=HPs,longplural={hyper-parameters}]{HP}{HP}{hyper-parameter}
\newacronym[parent=statistics,plural=PCA]{PCA}{PCA}{principal component analysis}

\newacronym{dynamicalSystems}{Dynamical systems}{\nopostdesc}
\newacronym[parent=dynamicalSystems]{KAM}{KAM}{Kolmogorov–Arnold–Moser}

\newacronym{methods}{Methods}{\nopostdesc}
\newacronym[parent=methods]{FP}{FP}{Function parametrization}
\newacronym[parent=methods]{MISO}{MISO}{multiple-input single-output}
\newacronym[parent=methods]{MIMO}{MIMO}{multiple-input multiple-output}

\newacronym{functionalAnalysis}{Functional analysis}{\nopostdesc}
\newacronym[parent=functionalAnalysis,plural=FCs,longplural={Fourier coefficients}]{FC}{FC}{Fourier coefficient}

\newacronym{differentialEquations}{Differential Equations}{\nopostdesc}
\newacronym[parent=differentialEquations,plural=ODEs,longplural={Ordinary Differential Equations}]{ODE}{ODE}{Ordinary Differential Equation}
\newacronym[parent=differentialEquations,plural=PDEs,longplural={Partial Differential Equations}]{PDE}{PDE}{Partial Differential Equation}

\newacronym{equations}{Equations}{\nopostdesc}
\newacronym[parent=equations]{LHS}{LHS}{left-hand side}
\newacronym[parent=equations]{RHS}{LHS}{right-hand side}
\newacronym{GeneralPhysics}{General Physics}{\nopostdesc}
\newacronym[parent=GeneralPhysics]{IR}{IR}{Infrared}
\newacronym[parent=GeneralPhysics]{EM}{EM}{electro-magnetic}

\newacronym{Organisational}{Organisational}{\nopostdesc}
\newacronym[parent=Organisational]{OP}{OP}{operation phase}

\newacronym[parent=Organisational]{KKS}{KKS}{\emph{Kraftwerk-Kennzeichensystem} Reference Designation System for Power Plants}
\newacronym[parent=Organisational]{PLM}{PLM}{Project Lifecycle Management}

\newacronym{plasmaTheory}{Plasma Theory}{\nopostdesc}
\newacronym[parent=plasmaTheory,plural=tokamaks]{tokamak}{tokamak}{тороидальная камера в магнитных катушках}
\newacronym[parent=plasmaTheory]{MHD}{MHD}{magnetohydrodynamic}
\newacronym[parent=plasmaTheory]{LCFS}{LCFS}{last closed flux surface}
\newacronym[parent=plasmaTheory]{LCMS}{LCMS}{last closed magnetic surface}
\newacronym[parent=plasmaTheory]{ISS95}{ISS95}{international stellarator scaling 1995}
\newacronym[parent=plasmaTheory]{ISS04}{ISS04}{international stellarator scaling 2004}
\newacronym[parent=plasmaTheory]{NTM}{NTM}{neoclassical tearing modes}
\newacronym[parent=plasmaTheory]{HPP}{HPP}{heat pulse propagation}

\newacronym{plasmaPhysics}{Plasma Physics}{\nopostdesc}

\newacronym[parent=plasmaPhysics]{Omode}{O~mode}{ordinary mode}
\newacronym[parent=plasmaPhysics]{Xmode}{X~mode}{extraordinary mode}
\newacronym[parent=plasmaPhysics]{EBW}{EBW}{electron Bernstein wave}
\newacronym[parent=plasmaPhysics]{TM}{TM}{transverse magnetic}
\newacronym[parent=plasmaPhysics]{TEM}{TEM}{transverse electromagnetic}
\newacronym[parent=plasmaPhysics,plural=EEDFs,firstplural=electron energy distribution functions (EEDFs)]{EEDF}{EEDF}{electron energy distribution function}
\newacronym[parent=plasmaPhysics]{ECE}{ECE}{electron cyclotron emission}

\newacronym[parent=plasmaPhysics]{LFS}{LFS}{low field side}
\newacronym[parent=plasmaPhysics]{HFS}{HFS}{high field side}
\newacronym[parent=plasmaPhysics]{ELM}{ELM}{edge localized mode}
\newacronym[parent=plasmaPhysics]{MARFE}{MARFE}{multifaceted asymmetric radiation from the edge}
\newacronym[parent=plasmaPhysics]{Serpens}{Serpens}{self-regulated plasma edge beneath the last-closed-flux-surface}
\newacronym[parent=plasmaPhysics,plural=SOLs,firstplural=scrape-off layers]{SOL}{SOL}{scrape-off layer}
\newacronym[parent=plasmaPhysics]{ECRH}{ECRH}{electron cyclotron resonance heating}
\newacronym[parent=plasmaPhysics]{ECCD}{ECCD}{electron cyclotron current drive}
\newacronym[parent=plasmaPhysics]{NBCD}{NBCD}{neutral beam current drive}
\newacronym[parent=plasmaPhysics]{ICRH}{ICRH}{ion cyclotron radiation heating}
\newacronym[parent=plasmaPhysics]{FL}{FL}{field line}
\newacronym[parent=plasmaPhysics]{PFR}{PFR}{private flux region}

\newacronym[parent=plasmaPhysics,firstplural=reversed-field pinches]{RFP}{RFP}{reversed-field pinch}
\newacronym[parent=plasmaPhysics]{FRC}{FRC}{field-reversed configuration}

\newacronym[parent=plasmaPhysics]{see}{\textsc{SEE}}{secondary electron emission}
\newacronym[parent=plasmaPhysics]{SE}{\textsc{SE}}{sheath edge}
\newacronym[parent=plasmaPhysics]{SL}{SL}{strike line}
\newacronym[parent=plasmaPhysics]{PWI}{PWI}{plasma wall interaction}

\newacronym{fusionReactorComponent}{Nuclear Fusion Machine Components}{\nopostdesc}
\newacronym[parent=fusionReactorComponent,plural=PFCs,firstplural=plasma facing components (PFCs)]{PFC}{PFC}{plasma facing component}
\newacronym[parent=fusionReactorComponent,]{VV}{VV}{vaccum vessel}
\newacronym[parent=fusionReactorComponent,plural=IVCs,firstplural=in-vessel components (IVCs)]{IVC}{IVC}{in-vessel component}
\newacronym[parent=fusionReactorComponent,plural=TDUs,firstplural=test divertor units (TDUs)]{TDU}{TDU}{test divertor unit}
\newacronym[parent=fusionReactorComponent,plural=HHFDs, firstplural=high heatflux divertors (HHFDs)]{HHFD}{HHFD}{high heatflux divertor}
\newacronym[parent=fusionReactorComponent,]{TDUSE}{TDUSE}{test divertor unit scraper element}
\newacronym[parent=fusionReactorComponent,]{PG}{PG}{Pumping Gap}
\newacronym[parent=fusionReactorComponent,]{TE}{TE}{Target Element}

\newacronym[parent=fusionReactorComponent]{NBI}{NBI}{neutral beam injection}
\newacronym[parent=fusionReactorComponent,plural=PIs,firstplural=pellet injections (PIs)]{PI}{PI}{pellet injection}

\newacronym{plasmaTheoryCode}{Plasma Theory Codes}{\nopostdesc}
\newacronym[parent=plasmaTheoryCode]{FLD}{FLD}{field line diffusion}
\newacronym[parent=plasmaTheoryCode]{FLT}{FLT}{field line tracing}
\newacronym[parent=plasmaTheoryCode]{LOWENDEL}{LOWENDEL}{``loads on Wendelstein''}
\newacronym[parent=plasmaTheoryCode]{VMEC}{VMEC}{variational moments equilibrium code}
\newacronym[parent=plasmaTheoryCode]{EXTENDER}{EXTENDER}{virtual casing principle code}
\newacronym[parent=plasmaTheoryCode]{PIES}{PIES}{Princeton iterative equilibrium solver}
\newacronym[parent=plasmaTheoryCode]{SPEC}{SPEC}{stepped-pressure equilibrium code}
\newacronym[parent=plasmaTheoryCode]{EMC3E}{EMC3-Eirene}{Edge Monte Carlo 3D code coupled to the neutral transport code EIRENE}
\newacronym[parent=plasmaTheoryCode]{TRAVIS}{TRAVIS}{tracing visualized}
\newacronym[parent=plasmaTheoryCode]{SPECE}{SPECE}{solver for plasma electron cyclotron emission} 
\newacronym[parent=plasmaTheoryCode]{EFIT}{EFIT}{equilibrium fitting}
\newacronym[parent=plasmaTheoryCode]{Theo}{THEODOR}{Thermal Energy onto Divertor}
\newacronym[parent=plasmaTheoryCode]{STRAHL}{STRAHL}{Radial impurity transport and emission code}
\newacronym[parent=plasmaTheoryCode]{ROSE}{ROSE}{ROSE Optimises Stellarator Equilibria}
\newacronym[parent=plasmaTheoryCode]{V3FIT}{V3FIT}{three-dimensional equilibrium reconstruction}
\newacronym[parent=plasmaTheoryCode]{SIMPLE}{SIMPLE}{symplectic integration methods for particle loss estimation}

\newacronym[parent=plasmaTheoryCode]{IDA}{IDA}{integrated data analysis}
\newacronym[parent=plasmaTheoryCode]{BSM}{BSM}{Bayesian scientific modeling}
\newacronym[parent=plasmaTheoryCode]{SBI}{SBI}{simulation-based inference}

\newacronym{database}{Databases}{\nopostdesc}
\newacronym[parent=database]{ADAS}{ADAS}{Atomic Data and Analysis Structure}

\newacronym{plasmaDiagnostic}{Plasma Diagnostics}{\nopostdesc}
\newacronym[parent=plasmaDiagnostic]{MPM}{MPM}{Multi-purpose manipulator}
\newacronym[parent=plasmaDiagnostic]{TS}{TS}{Thomson scattering}
\newacronym[parent=plasmaDiagnostic]{CTS}{CTS}{collective Thomson scattering}
\newacronym[parent=plasmaDiagnostic]{XICS}{XICS}{X-ray imaging crystal spectroscopy}
\newacronym[parent=plasmaDiagnostic]{CXRS}{CXRS}{charge exchange recombination spectroscopy}
\newacronym[parent=plasmaDiagnostic]{LiBES}{LiBES}{lithium beam emission spectroscopy}
\newacronym[parent=plasmaDiagnostic]{DCI}{DCI}{deuterium cyanide laser interferometry}
\newacronym[parent=plasmaDiagnostic,plural=RFMs,firstplural=reflectometers (RFMs)]{RFM}{RFM}{reflectometer}
\newacronym[parent=plasmaDiagnostic]{LP}{LP}{Langmuir probe}
\newacronym[parent=plasmaDiagnostic,sort=Ha-Camera]{HAC}{$H_\alpha$-C}{$H_\alpha$-Camera}
\newacronym[parent=plasmaDiagnostic,sort=Ha-Spectrometer]{HAS}{$H_\alpha$-S}{$H_\alpha$-Spectrometer}
\newacronym[parent=plasmaDiagnostic]{IRC}{IRC}{Infrared-Camera}
\newacronym[parent=plasmaDiagnostic]{HB}{HeBS}{Helium beam spectroscopy}
\newacronym[parent=plasmaDiagnostic]{Hexos}{HEXOS}{High efficiency extreme ultraviolet overview spectrometer}
\newacronym[parent=plasmaDiagnostic]{DI}{DI}{dispersion interferometer}

\newacronym{gasDiagnostic}{Gas Diagnostics}{\nopostdesc}
\newacronym[parent=gasDiagnostic]{DRGA}{DRGA}{Diagnostic residual gas analyser}
\newacronym[parent=gasDiagnostic]{LIF}{LIF}{Laser induced fluorescence}

\newacronym{fusionApproach}{Nuclear Fusion Approaches}{\nopostdesc}
\newacronym[parent=fusionApproach]{MCF}{MCF}{magnetic confinement fusion}
\newacronym[parent=fusionApproach]{ICF}{ICF}{inertial confinement fusion}
\newacronym[parent=fusionApproach]{MIF}{MIF}{magneto-inertial fusion}

\newacronym{fusionExperiment}{Nuclear Fusion Experiments}{\nopostdesc}

\newacronym[parent=fusionExperiment,sort=W7]{W7}{\protect\mbox{W7}}{Wendelstein~\protect\mbox{7}}

\newacronym[parent=fusionExperiment,sort=W7-A]{W7A}{\protect\mbox{W7-A}}{Wendelstein~\protect\mbox{7-A}}

\newacronym[parent=fusionExperiment,sort=W7-AS]{W7AS}{\protect\mbox{W7-AS}}{Wendelstein~\protect\mbox{7-AS}}

\newacronym[parent=fusionExperiment,sort=W7-X]{W7X}{\protect\mbox{W7-X}}{Wendelstein~\protect\mbox{7-X}}

\newacronym[parent=fusionExperiment]{NCSX}{NCSX}{National Compact Stellarator Experiment}

\newacronym[parent=fusionExperiment]{CNT}{CNT}{Columbia Non-neutral Torus}

\newacronym[parent=fusionExperiment]{CTH}{CTH}{Compact Toroidal Hybrid}

\newacronym[parent=fusionExperiment,sort=LHD]{LHD}{\protect\mbox{LHD}}{large helical device}

\newacronym[parent=fusionExperiment]{WEGA}{WEGA}{Wendelstein Experiment in Greifswald für die Ausbildung} 

\newacronym[parent=fusionExperiment]{HSX}{HSX}{Helically Symmetric Experiment}

\newacronym[parent=fusionExperiment]{JET}{JET}{Joint European Torus}

\newacronym[parent=fusionExperiment]{MAST}{MAST}{Mega-Ampere Spherical Tokamak}

\newacronym[parent=fusionExperiment]{AUG}{AUG}{ASDEX Upgrade}

\newacronym[parent=fusionExperiment]{ASDEX}{ASDEX}{axialsymmetrisches Divertor-Experiment}

\newacronym[parent=fusionExperiment]{EAST}{EAST}{experimental advanced superconducting tokamak}

\newacronym[parent=fusionExperiment]{ITER}{ITER}{lat.~\textit{the way} (formerly International Thermonuclear Experimental Reactor)}

\newacronym[parent=fusionExperiment]{AlcCM}{Alcator C-Mod}{Alto Campo Toro C-Mod} 

\newacronym[parent=fusionExperiment,sort=D]{D3D}{\mbox{DIII-D}}{Doublet III-D}

\newacronym[parent=fusionExperiment,sort=JT60]{JT60}{\mbox{JT-60}}{Japan Torus-60}
\newacronym[parent=fusionExperiment,sort=JT60SA]{JT60SA}{\mbox{JT-60SA}}{Japan Torus-60 Super Upgrade}

\newacronym[parent=fusionExperiment,sort=KSTAR]{KSTAR}{\mbox{KSTAR}}{Korea Superconducting Tokamak Advanced Research}

\newacronym[parent=fusionExperiment,sort=WEST]{WEST}{\mbox{WEST}}{Tungsten Environment in Steady-state Tokamak (formerly Tore Supra)}

\newacronym[parent=fusionExperiment,sort=TCV]{TCV}{\mbox{TCV}}{Tokamak à configuration variable}

\newacronym[parent=fusionExperiment,sort=TEXT]{TEXT}{\mbox{TEXT}}{Texas Tokamak}
\newacronym[parent=fusionExperiment,sort=TEXTU]{TEXTU}{\mbox{TEXT-U}}{Texas Tokamak - Upgrade}

\newacronym[parent=fusionExperiment,sort=Textor]{Textor}{\mbox{Textor}}{Tokamak Experiment for Technology Oriented Research}

\newacronym[parent=fusionExperiment,sort=Helias]{Helias}{\mbox{Helias}}{helical advanced stellarator}

\newacronym{stellaratorSymmetry}{Stellarator Symmetries}{\nopostdesc}

\newacronym[parent=stellaratorSymmetry,sort=QO]{QO}{QO}{quasi-omnigeneous}
\newacronym[parent=stellaratorSymmetry,sort=QI]{QI}{QI}{quasi-isodynamic}
\newacronym[parent=stellaratorSymmetry,sort=QS]{QS}{QS}{quasi-symmetry}
\newacronym[parent=stellaratorSymmetry,sort=QA]{QA}{QA}{quasi-axisymmetric}
\newacronym[parent=stellaratorSymmetry,sort=QH]{QH}{QH}{quasi-helical}
\newacronym[parent=stellaratorSymmetry,sort=QP]{QP}{QP}{quasi-poloidal}

\newacronym{plasmaLab}{Plasma Physics Laboratories}{\nopostdesc}

\newacronym[parent=plasmaLab]{PPPL}{PPPL}{Princeton Plasma Physics Laporatory}

\newacronym[parent=plasmaLab]{IPP}{IPP}{Max-Planck Institute for Plasma Physics}

\newacronym[sort=W7-X]{magneticConfig}{\mbox{W7-X} Magnetic Configurations}{\nopostdesc}
\newacronym[parent=magneticConfig]{EIM}{\textsc{EIM}}{'Standard' magnetic configuration in the centre of accessible space}
\newacronym[parent=magneticConfig]{DBM}{\textsc{DBM}}{Low iota magnetic configuration}

\newacronym{nuclearReaction}{Nuclear Reactions}{\nopostdesc}
\newacronym[parent=nuclearReaction]{DT}{DT}{Deuterium-Tritium}

\newglossaryentry{physics}{%
	name={General physics quantitites},
	type=symbols,
	description={\nopostdesc},
	symbol={},
}
\newglossaryentry{n}{%
	parent=physics,
	name=\ensuremath{n},
	type=symbols,
	sort=density,
	description={Particle density, $n = N / V$},
	symbol={\si{\per\cubic\meter}}
}
\newglossaryentry{T}{%
	parent=physics,
	name=\ensuremath{T},
	type=symbols,
	sort=temperature,
	description={Temperature},
	symbol={\si{\eV}}
}
\newglossaryentry{energy}{%
	parent=physics,
	name=\ensuremath{W},
	type=symbols,
	sort=energy,
	description={Plasma kinetic energy},
	symbol={\si{\kg\square\meter\per\square\second}}
}
\newglossaryentry{cs}{%
	parent=physics,
	name=\ensuremath{\sigma},
	type=symbols,
	sort=cross-section,
	description={Reaction cross-section},
	symbol={\si{\meter\square}}
}
\newglossaryentry{v}{%
	parent=physics,
	name=\ensuremath{\vect{v}},
	type=symbols,
	sort=velocity,
	description={Particle velocity},
	symbol={\si{\meter\per\second}}
}
\newglossaryentry{me}{%
	parent=physics,
	name=\ensuremath{m_{\text{e}}},
	type=symbols,
	sort= mass ,
	description={Electron mass},
	symbol={\si{\kilo\gram}},
}
\newglossaryentry{mp}{%
	parent=physics,
	name=\ensuremath{m_{\text{p}}},
	type=symbols,
	sort= mass ,
	description={Proton mass},
	symbol={\si{\kilo\gram}},
}

\newglossaryentry{B}{%
	parent=physics,
	name=\ensuremath{\vect{B}},
	type=symbols,
	sort=electromagnetism,
	description={Magnetic field},
	symbol={\si{\tesla} = \si{\kg\per\ampere\per\square\second}}
}
\newglossaryentry{E}{%
	parent=physics,
	name=\ensuremath{\vect{E}},
	type=symbols,
	sort=electromagnetism,
	description={Electric field},
	symbol={\si{\kg\meter\per\ampere\per\cubic\second}}
}
\newglossaryentry{chargedensity}{%
	parent=physics,
	name=\ensuremath{\rho},
	type=symbols,
	sort=electromagnetism,
	description={Electric charge density},
	symbol={\si{\ampere\second\per\cubic\meter}}
}
\newglossaryentry{currentdensity}{%
	parent=physics,
	name=\ensuremath{\vect{j}},
	type=symbols,
	sort=electromagnetism,
	description={Electric current density},
	symbol={\si{\ampere\per\square\meter}}
}
\newglossaryentry{resistance}{%
	parent=physics,
	name=\ensuremath{R},
	type=symbols,
	sort=impedance,
	description={Electric resistance},
	symbol={\si{\ohm}}
}
\newglossaryentry{inductance}{%
	parent=physics,
	name=\ensuremath{L},
	type=symbols,
	sort=inductance,
	description={Inductance},
	symbol={\si{\henry} = \si{\kg\square\meter\per\square\ampere\per\square\second}}
}
\newglossaryentry{larmor}{%
	parent=physics,
	name=\ensuremath{\rho},
	type=symbols,
	sort={Larmor radius},
	description={Larmor radius},
	symbol={\si{\meter}}
}

\newglossaryentry{plasmaphysics}{%
	name={Plasma physics quantitites},
	type=symbols,
	description={\nopostdesc},
	symbol={},
}

\newglossaryentry{vperp}{%
	parent=plasmaphysics,
	name=\ensuremath{v_{\perp}},
	type=symbols,
	sort=velocity,
	description={Field-perpendicular velocity},
	symbol={\si{\meter\per\second}}
}
\newglossaryentry{flowvelocity}{%
	parent=plasmaphysics,
	name=\ensuremath{\vect{v_{\text{f}}}},
	type=symbols,
	sort=velocity,
	description={Plasma flow velocity},
	symbol={\si{\meter\per\second}}
}
\newglossaryentry{p}{%
	parent=plasmaphysics,
	name=\ensuremath{p},
	type=symbols,
	sort=pressure,
	description={Plasma pressure},
	symbol={\si{\kg\per\meter\per\square\second}}
}
\newglossaryentry{V}{%
	parent=plasmaphysics,
	name=\ensuremath{V},
	type=symbols,
	sort=volume,
	description={Plasma volume},
	symbol={\si{\cubic\meter}}
}
\newglossaryentry{viscosity}{%
	parent=plasmaphysics,
	name=\ensuremath{\matr{\pi}},
	type=symbols,
	sort=viscosity,
	description={Plasma viscosity tensor},
	symbol={\si{\kg\per\meter\per\second}}
}
\newglossaryentry{Te}{%
	parent=plasmaphysics,
	name=\ensuremath{T_{\mathrm{e}}},
	type=symbols,
	sort=temperature,
	description={Electron temperature},
	symbol={\si{\eV}}
}
\newglossaryentry{Ti}{%
	parent=plasmaphysics,
	name=\ensuremath{T_{\mathrm{i}}},
	type=symbols,
	sort=temperature,
	description={Ion temperature},
	symbol={\si{\eV}}
}
\newglossaryentry{Tn}{%
	parent=plasmaphysics,
	name=\ensuremath{T_{\mathrm{n}}},
	type=symbols,
	sort=temperature,
	description={Neutral temperature},
	symbol={\si{\eV}}
}
\newglossaryentry{ni}{%
	parent=plasmaphysics,
	name=\ensuremath{n_{\mathrm{i}}},
	type=symbols,
	sort=density,
	description={Ion density},
	symbol={\si{\per\cubic\meter}}
}
\newglossaryentry{ne}{%
	parent=plasmaphysics,
	name=\ensuremath{n_{\mathrm{e}}},
	type=symbols,
	sort=density,
	description={Electron density},
	symbol={\si{\per\cubic\meter}}
}
\newglossaryentry{nn}{%
	parent=plasmaphysics,
	name=\ensuremath{n_{\mathrm{n}}},
	type=symbols,
	sort=density,
	description={Neutral density},
	symbol={\si{\per\cubic\meter}}
}

\newglossaryentry{nimp}{%
	parent=plasmaphysics,
	name=\ensuremath{n_{\mathrm{imp}}},
	type=symbols,
	sort=density,
	description={Impurity density},
	symbol={\si{\per\cubic\meter}}
}

\newglossaryentry{te}{%
	parent=plasmaphysics,
	name=\ensuremath{\tau_{\mathrm{E}}},
	type=symbols,
	sort=time,
	description={Energy confinement time},
	symbol={\si{\second}}
}
\newglossaryentry{radialShift}{%
	parent=plasmaphysics,
	name=\ensuremath{\makeup{\Delta} R},
	type=symbols,
	sort=confinement,
	description={Radial shift of the magnetic axis},
	symbol={\si{\meter}}
}
\newglossaryentry{plasmaBeta}{%
	parent=plasmaphysics,
	name=\ensuremath{\beta},
	type=symbols,
	sort=confinement,
	description={Plasma beta},
	symbol={}
}
\newglossaryentry{Itor}{%
	parent=plasmaphysics,
	name=\ensuremath{I_{\mathrm{tor}}},
	type=symbols,
	sort=confinement,
	description={Toroidal plasma current},
	symbol={\si{\ampere}}
}
\newglossaryentry{Ibs}{%
	parent=plasmaphysics,
	name=\ensuremath{I_{\mathrm{bs}}},
	type=symbols,
	sort=confinement,
	description={Bootstrap current},
	symbol={\si{\ampere}}
}
\newglossaryentry{Ips}{%
	parent=plasmaphysics,
	name=\ensuremath{I_{\mathrm{ps}}},
	type=symbols,
	sort=confinement,
	description={Pfirsch-Schlueter current},
	symbol={\si{\ampere}}
}
\newcommand{\quer}[1]{\mathrel{\hbox{-}\mkern-6.55mu #1}}
\newglossaryentry{ibar}{%
	parent=plasmaphysics,
	name=\ensuremath{\quer{\iota}},
	type=symbols,
	sort=confinement,
	description={Rotational transform},
	symbol={}
}

\newglossaryentry{shear}{%
	parent=plasmaphysics,
	name=\ensuremath{s},
	type=symbols,
	sort=confinement,
	description={Shear, radial derivative of \ensuremath{\quer{\iota}}},
	symbol={}
}

\newglossaryentry{ibarCF}{%
parent=plasmaphysics,
name=\ensuremath{\quer{\iota}_{\mathrm{CF}}},
type=symbols,
sort=confinement,
description={Current free rotational transform},
symbol={}
}

\newglossaryentry{Vp}{%
	parent=plasmaphysics,
	name=\ensuremath{V_\mathrm{p}},
	type=symbols,
	sort=potential,
	description={Plasma potential},
	symbol={\si{\volt}}
}
\newglossaryentry{Vf}{%
	parent=plasmaphysics,
	name=\ensuremath{V_\mathrm{f}},
	type=symbols,
	sort=potential,
	description={Floating potential},
	symbol={\si{\volt}}
}
\newglossaryentry{Isat}{%
	parent=plasmaphysics,
	name=\ensuremath{I_\text{sat}},
	type=symbols,
	sort=current,
	description={Ion saturation current},
	symbol={\si{\ampere}}
}
\newglossaryentry{esat}{%
	parent=plasmaphysics,
	name=\ensuremath{I_{e,\text{sat}}},
	type=symbols,
	sort=current,
	description={Electron saturation current},
	symbol={\si{\ampere}}
}
\newglossaryentry{jsat}{%
	parent=plasmaphysics,
	name=\ensuremath{j_\text{sat}},
	type=symbols,
	sort=current,
	description={Ion saturation current density},
	symbol={\si{\ampere\per\square\meter}}
}
\newglossaryentry{csound}{%
	parent=plasmaphysics,
	name=\ensuremath{c_\mathrm{s}},
	type=symbols,
	sort=velocity,
	description={Ion sound speed},
	symbol={\si{\meter\per\second}}
}
\newglossaryentry{Vbias}{%
	parent=plasmaphysics,
	name=\ensuremath{V_\mathrm{bias}},
	type=symbols,
	sort=potential,
	description={Bias voltage},
	symbol={\si{\volt}}
}
\newglossaryentry{r}{%
	parent=plasmaphysics,
	name=\ensuremath{r},
	type=symbols,
	sort=radius,
	description={Minor radius},
	symbol={\si{\meter}}
}
\newglossaryentry{ra}{%
	parent=plasmaphysics,
	name=\ensuremath{r_a},
	type=symbols,
	sort=radius,
	description={Minor radius of the last closed flux surface},
	symbol={\si{\meter}}
}
\newglossaryentry{reff}{%
	parent=plasmaphysics,
	name=\ensuremath{r_{\mathrm{eff}}},
	type=symbols,
	sort=radius,
	description={Effective minor radius},
	symbol={\si{\meter}}
}
\newglossaryentry{R}{%
	parent=plasmaphysics,
	name=\ensuremath{R},
	type=symbols,
	sort=radius,
	description={Major radius},
	symbol={\si{\meter}}
}
\newglossaryentry{normminrad}{%
	parent=plasmaphysics,
	name=\ensuremath{\varrho},
	type=symbols,
	sort=radius,
	description={Normalised minor radius},
	symbol={}
}
\newglossaryentry{aspectRatio}{%
	parent=plasmaphysics,
	name=\ensuremath{\epsilon},
	type=symbols,
	sort=radius,
	description={Aspect ratio},
	symbol={\si{}}
}
\newglossaryentry{Nfp}{%
	parent=plasmaphysics,
	name=\ensuremath{N_{\text{fp}}},
	type=symbols,
	sort=number,
	description={Number of field periods},
	symbol={\si{}},
}
\newglossaryentry{phiEdge}{%
	parent=plasmaphysics,
	name=\ensuremath{\phi_{\mathrm{edge}}},
	type=symbols,
	sort=magnetic flux,
	description={Total enclosed magnetic toroidal flux},
	symbol={\si{\volt\second}}
}

\newglossaryentry{Pconv}{%
	parent=plasmaphysics,
	name=\ensuremath{P_{\mathrm{conv}}},
	type=symbols,
	sort=power,
	description={Convective power},
	symbol={\si{\watt}}
}
\newglossaryentry{Prad}{%
	parent=plasmaphysics,
	name=\ensuremath{P_{\mathrm{rad}}},
	type=symbols,
	sort=power,
	description={Radiated power},
	symbol={\si{\watt}}
}
\newglossaryentry{Pheat}{%
	parent=plasmaphysics,
	name=\ensuremath{P_{\mathrm{heat}}},
	type=symbols,
	sort=power,
	description={Total heating power},
	symbol={\si{\watt}}
}
\newglossaryentry{Lc}{%
	parent=plasmaphysics,
	name=\ensuremath{L_{\mathrm{c}}},
	type=symbols,
	sort=length,
	description={Connection length},
	symbol={\si{\meter}}
}
\newglossaryentry{heatflux}{%
	parent=plasmaphysics,
	name=\ensuremath{q},
	type=symbols,
	sort=power flux,
	description={Heat flux or heat load},
	symbol={\si{\MW\per\square\meter}}
}
\newglossaryentry{Zav}{%
	parent=plasmaphysics,
	name=\ensuremath{Z_{\mathrm{av}}},
	type=symbols,
	sort=ion charge,
	description={Average ion charge},
	symbol={\si{}},
}
\newglossaryentry{Zeff}{%
	parent=plasmaphysics,
	name=\ensuremath{Z_{\mathrm{eff}}},
	type=symbols,
	sort=ion charge,
	description={Effective ion charge},
	symbol={\si{}},
}
\newglossaryentry{iongyro}{%
	parent=plasmaphysics,
	name=\ensuremath{\rho_{\text{i}}},
	type=symbols,
	sort=radius,
	description={Ion gyro (Lamor) radius},
	symbol={\si{\meter}},
}
\newglossaryentry{mi}{%
	parent=plasmaphysics,
	name=\ensuremath{m_{\text{i}}},
	type=symbols,
	sort=mass,
	description={Ion mass},
	symbol={\si{\kilo\gram}},
}
\newglossaryentry{ndl}{%
	parent=plasmaphysics,
	name=\ensuremath{n\text{d}\ell},
	type=symbols,
	sort=density,
	description={Line integrated density},
	symbol={\si{\per\square\meter}},
}
\newglossaryentry{Wdia}{%
	parent=plasmaphysics,
	name=\ensuremath{W_\text{dia}},
	type=symbols,
	sort=energy,
	description={Diamagnetic energy},
	symbol={\si{\joule}},
}
\newglossaryentry{Qsci}{%
	parent=plasmaphysics,
	name=\ensuremath{Q_{\text{sci}}},
	type=symbols,
	sort=gain,
	description={Fusion energy scientific gain factor},
	symbol={\si{}}
}
\newglossaryentry{Qeng}{%
	parent=plasmaphysics,
	name=\ensuremath{Q_{\text{eng}}},
	type=symbols,
	sort=gain,
	description={Fusion energy engineering gain factor},
	symbol={\si{}}
}

\newglossaryentry{epseff}{%
	parent=plasmaphysics,
	name=\ensuremath{\epsilon_\text{eff}},
	type=symbols,
	sort=effective ripple,
	description={Effective ripple},
	symbol={\si{}},
}

\newglossaryentry{Halpha}{%
	parent=plasmaphysics,
	name=\ensuremath{\text{H}_{α}\xspace},
	type=symbols,
	sort= Wavelengths ,
	description={Hydrogen $\alpha$ transition line, Balmer series transition $n=3 \rightarrow n=2$, \SI{656.5}{\nano\meter}},
	symbol=\si{\nano\meter},
}
\newglossaryentry{sxb}{%
	parent=plasmaphysics,
	name=\ensuremath{\text{S/XB}},
	type=symbols,
	sort= Coefficients,
	description={Ratio of ionisation, excitation and branching ratio. Inverse photons per neutral.},
	symbol=,
}
\newglossaryentry{particleFlux}{%
	parent=plasmaphysics,
	name=\ensuremath{\Gamma},
	type=symbols,
	sort= Particle flux,
	description={Particle flux},
	symbol=\si{\per\second\per\square\meter},
}
\newglossaryentry{stf}{%
	parent=plasmaphysics,
	name=\ensuremath{\gamma_{\text{s}}},
	type=symbols,
	sort=Sheath transmission coefficient,
	description={Sheath transmission coefficient},
	symbol={\si{}},
}
\newglossaryentry{frad}{%
	parent=plasmaphysics,
	name=\ensuremath{f_{\text{rad}}},
	type=symbols,
	sort=Fraction,
	description={Radiated power fraction},
	symbol={\si{}},
}
\newglossaryentry{mfpi}{%
	parent=plasmaphysics,
	name=\ensuremath{\lambda_\text{mfp,i}},
	type=symbols,
	sort=Length,
	description={Mean free path length of ionisation},
	symbol={\si{\meter}},
}
\newglossaryentry{frec}{%
	parent=plasmaphysics,
	name=\ensuremath{f_\text{rec}},
	type=symbols,
	sort=Fraction,
	description={Fraction of ions recycled as neutrals at PFCs},
	symbol={\si{}},
}
\newglossaryentry{Pecrh}{%
	parent=plasmaphysics,
	name=\ensuremath{P_\text{ECRH}},
	type=symbols,
	sort=Power,
	description={ECRH Power},
	symbol={\si{\watt}},
}

\newcommand{\plasmaBeta}{\gls{plasmaBeta}\xspace}
\newcommand{\averagePlasmaBeta}{\ensuremath{\langle\gls{plasmaBeta}\rangle}\xspace}
\newcommand{\Itor}{\gls{Itor}\xspace}

\newcommand{\Te}{\gls{Te}\xspace}
\newcommand{\Ti}{\gls{Ti}\xspace}

\newacronym{engineering}{Engineering}{\nopostdesc}
\newacronym[parent=engineering]{MPCDF}{MPCDF}{Max Planck computing and data facility}

\newacronym[parent=engineering]{CAD}{CAD}{computer-aided design}
\newacronym[parent=engineering]{SOA}{SOA}{service-oriented architecture}
\newacronym[parent=engineering]{ESB}{ESB}{enterprise service bus}
\newacronym[parent=engineering]{HPC}{HPC}{high-performance computing}
\newacronym[parent=engineering,plural=GPUs,longplural={graphical processing units}]{GPU}{GPU}{graphical processing unit}
\newacronym[parent=engineering,plural=TPUs,longplural={tensor processing units}]{TPU}{TPU}{tensor processing unit}

\newacronym[parent=engineering]{DAQ}{DAQ}{data acquisition}
\newacronym[parent=engineering]{ADC}{ADC}{analog-to-digital converter}
\newacronym[parent=engineering]{LO}{LO}{local oscillator}
\newacronym[parent=engineering]{PLL}{PLL}{phase locked loop}
\newacronym[parent=engineering]{CPU}{CPU}{central processing unit}
\newacronym[parent=engineering]{OAMP}{OAmp}{operational amplifier}
\newacronym[parent=engineering]{DAMP}{DAmp}{differential amplifier}
\newacronym[parent=engineering]{PCB}{PCB}{printed circuit board}
\newacronym[parent=engineering]{CMRR}{CMRR}{common mode rejection ratio}
\newacronym[parent=engineering]{IC}{IC}{integrated circuit}
\newacronym[parent=engineering]{CCD}{CCD}{charge-coupled device}
\newacronym[parent=engineering]{PMT}{PMT}{photo-multiplyer tube}
\newacronym[parent=engineering]{ROI}{ROI}{region of interest}
\newacronym[parent=engineering,plural=LOSs,longplural={lines of sight}]{LOS}{LOS}{line of sight}
\newacronym[parent=engineering]{EDICAM}{EDICAM}{Event Detection Intelligent Camera}

\newacronym[parent=engineering]{CFC}{CFC}{carbon fibre reinforced carbon}

\newacronym{machineLearning}{Machine Learning}{\nopostdesc}
\newacronym[parent=machineLearning]{ML}{ML}{machine learning}
\newacronym[parent=machineLearning]{RL}{RL}{reinforcement learning}
\newacronym[parent=machineLearning,plural=NNs,firstplural=artificial neural networks (NNs)]{NN}{NN}{artificial neural network}
\newacronym[parent=machineLearning]{EI}{EI}{expected improvement}

\newacronym[parent=machineLearning,plural=FF-FCs,firstplural=feed-forward fully-connected neural networks (FF-FC)]{FFFC}{FF-FC}{feed-forward fully-connected neural network}
\newacronym[parent=machineLearning,plural=MLPs,firstplural=multilayer perceptrons (MLP)]{MLP}{MLP}{multilayer perceptron}
\newacronym[parent=machineLearning,plural=CNNs,firstplural=convolutional neural networks (CNNs)]{CNN}{CNN}{convolutional neural network}
\newacronym[parent=machineLearning,plural=DCNNs,firstplural=deep convolutional neural networks (DCNNs)]{DCNN}{DCNN}{deep convolutional neural network}
\newacronym[parent=machineLearning,plural=IRNNs,firstplural=Inception-ResNets (IRNNs)]{IRNN}{IRNN}{Inception-ResNet}
\newacronym[parent=machineLearning,plural=DINNs,firstplural=deep inception neural networks (DINNs)]{DINN}{DINN}{deep inception neural network}
\newacronym[parent=machineLearning,plural=GANs,firstplural=generative adversarial neural networks (GANs)]{GAN}{GAN}{generative adversarial neural network}
\newacronym[parent=machineLearning,plural=DQNs,firstplural=deep Q-networks]{DQN}{DQN}{deep Q-network}
\newacronym[parent=machineLearning,plural=AEs,firstplural=autoencoders]{AE}{AE}{autoencoder}
\newacronym[parent=machineLearning,plural=VAEs,firstplural=variational autoencoders]{VAE}{VAE}{variational autoencoder}
\newacronym[parent=machineLearning,plural=PiNNs,firstplural=physics informed neural networks]{PiNN}{PiNN}{physics informed neural network}
\newacronym[parent=machineLearning,plural=DeepONets,firstplural=deep operator networks]{DeepONet}{DeepONet}{deep operator network}

\newacronym[parent=machineLearning]{MNIST}{MNIST}{MNIST} 

\newacronym[parent=machineLearning]{ReLU}{ReLU\xspace}{rectified linear unit}
\newacronym[parent=machineLearning]{SeLU}{SeLU\xspace}{scaled exponential linear unit}
\newacronym[parent=machineLearning]{SiLU}{SiLU\xspace}{sigmoid linear unit}

\newacronym[parent=machineLearning]{rmse}{rmse\xspace}{root-mean-square error}
\newacronym[parent=machineLearning]{mse}{mse\xspace}{mean-squared error}
\newacronym[parent=machineLearning]{msd}{msd\xspace}{mean-squared deviation}
\newacronym[parent=machineLearning]{nmse}{nmse\xspace}{normalized mean-squared error}
\newacronym[parent=machineLearning]{rmsd}{rmsd\xspace}{root-mean-squared deviation}
\newacronym[parent=machineLearning]{nrmsd}{nrmsd\xspace}{normalized root-mean-squared deviation}
\newacronym[parent=machineLearning]{nrmse}{nrmse\xspace}{normalized root-mean-squared error}
\newacronym[parent=machineLearning]{mae}{mae\xspace}{mean absolute error}
\newacronym[parent=machineLearning]{mape}{mape\xspace}{mean absolute percentage error}

\newacronym[parent=machineLearning]{TPE}{TPE}{tree-structured Parzen estimator}
\newacronym[parent=machineLearning]{NAS}{NAS}{neural architecture search}
\newacronym[parent=machineLearning]{NES}{NES}{neural ensemble search}
\newacronym[parent=machineLearning]{SMBO}{SMBO}{sequential model-based optimization}
\newacronym[parent=machineLearning]{BFGS}{BFGS}{Broyden–Fletcher–Goldfarb–Shanno}
\newacronym[parent=machineLearning]{LBFGS}{L-BFGS}{limited-memory Broyden–Fletcher–Goldfarb–Shanno}
\newacronym[parent=machineLearning]{AD}{AD}{automatic differentiation}

\newglossaryentry{ml}{%
    name={Machine learning quantitites},
    type=symbols,
    description={\nopostdesc},
    symbol={},
}
\newglossaryentry{lossfunction}{%
    name=\ensuremath{\mathcal{L}},
	type=symbols,
	parent=ml,
	sort=loss,
	description={Loss function},
    symbol={\si{}}
}

\setproperty{document}{title}{Accelerated Bayesian inference of plasma profiles with self-consistent MHD equilibria at W7-X via neural networks}
\setproperty{author}{firstname}{Andrea}
\setproperty{author}{familyname}{Merlo}
\setproperty{coauthor1}{firstname}{Andrea}
\setproperty{coauthor1}{familyname}{Pavone}
\setproperty{coauthor1}{affiliationindices}{1}
\setproperty{coauthor2}{firstname}{Daniel}
\setproperty{coauthor2}{familyname}{Böckenhoff}
\setproperty{coauthor2}{affiliationindices}{1}
\setproperty{coauthor3}{firstname}{Ekkehard}
\setproperty{coauthor3}{familyname}{Pasch}
\setproperty{coauthor3}{affiliationindices}{1}
\setproperty{coauthor4}{firstname}{Golo}
\setproperty{coauthor4}{familyname}{Fuchert}
\setproperty{coauthor4}{affiliationindices}{1}
\setproperty{coauthor5}{firstname}{Kai Jakob}
\setproperty{coauthor5}{familyname}{Brunner}
\setproperty{coauthor5}{affiliationindices}{1}
\setproperty{coauthor6}{firstname}{Kian}
\setproperty{coauthor6}{familyname}{Rahbarnia}
\setproperty{coauthor6}{affiliationindices}{1}
\setproperty{coauthor7}{firstname}{Jonathan}
\setproperty{coauthor7}{familyname}{Schilling}
\setproperty{coauthor7}{affiliationindices}{1}
\setproperty{coauthor8}{firstname}{Udo}
\setproperty{coauthor8}{familyname}{Höfel}
\setproperty{coauthor8}{affiliationindices}{1}
\setproperty{coauthor9}{firstname}{Sehyun}
\setproperty{coauthor9}{familyname}{Kwak}
\setproperty{coauthor9}{affiliationindices}{1}
\setproperty{coauthor10}{firstname}{Jakob}
\setproperty{coauthor10}{familyname}{Svensson}
\setproperty{coauthor10}{affiliationindices}{1}
\setproperty{coauthor11}{firstname}{Thomas Sunn}
\setproperty{coauthor11}{familyname}{Pedersen}
\setproperty{coauthor11}{affiliationindices}{2}
\setproperty{coauthors}{W7Xteaminclude}{True}
\setproperty{affiliations}{1}{Max-Planck-Institute for Plasma Physics, 17491 Greifswald, Germany}
\setproperty{affiliations}{2}{Type One Energy Group, Madison, WI USA 53703}

\newcommand{\Phiedge}{\ensuremath{\Phi_{\text{edge}}}\xspace}
\newcommand{\Wkin}{\ensuremath{W_{\text{kin}}}\xspace}

\newcommand{\neZero}{\ensuremath{h_{n_e,0}}\xspace}
\newcommand{\neOne}{\ensuremath{h_{n_e,1}}\xspace}
\newcommand{\neTwo}{\ensuremath{h_{n_e,2}}\xspace}
\newcommand{\neThree}{\ensuremath{h_{n_e,3}}\xspace}

\newcommand{\teZero}{\ensuremath{h_{T_e,0}}\xspace}
\newcommand{\teOne}{\ensuremath{h_{T_e,1}}\xspace}
\newcommand{\teTwo}{\ensuremath{h_{T_e,2}}\xspace}
\newcommand{\teThree}{\ensuremath{h_{T_e,3}}\xspace}

\newcommand{\itorOne}{\ensuremath{h_{\Itor,1}}\xspace}
\newcommand{\itorTwo}{\ensuremath{h_{\Itor,2}}\xspace}
\newcommand{\itorThree}{\ensuremath{h_{\Itor,3}}\xspace}
\newcommand{\itorFour}{\ensuremath{h_{\Itor,4}}\xspace}

\newcommand{\expId}[1]{\textit{W7X#1}}

\definecolor{lightNullColor}{HTML}{d1e5f0}
\definecolor{nullColor}{HTML}{67a9cf}
\definecolor{darkNullColor}{HTML}{2166ac}
\definecolor{lightFiniteColor}{HTML}{fddbc7}
\definecolor{finiteColor}{HTML}{ef8a62}
\definecolor{darkFiniteColor}{HTML}{b2182b}

\DeclareSIUnit{\nothing}{\relax}

\DeclareSIUnit{\arbitraryunit}{a.u.}

\let\pgfimage=\includegraphics
\graphicspath{{./}}

\definechangesauthor[name={Andrea Merlo}, color=red]{AM}
\definechangesauthor[name={Daniel Böckenhoff}, color=blue]{DB}
\definechangesauthor[name={Andrea Pavone}, color=orange]{AP}

\makeatletter
\let\blx@rerun@biber\relax
\makeatother

\begin{document}

    \typeout{----- BEGIN DOCUMENT -----}
    \typeout{}
    \typeout{--------------------------------}
    \typeout{----- Document properties: -----}
    \typeout{--------------------------------}
    \typeout{Font size: \the\fsize}
    \typeout{Text width: \the\textwidth}
    \typeout{--------------------------------}
    \typeout{--------------------------------}

    \makeatletter
\@ifundefined{mainmatter}{}{\mainmatter}  
\makeatother
\ifthenelseproperty{compilation}{frontmatter}{%
    \frontmatter
}{}

\ifthenelseproperty{compilation}{titlepage}{%
    \ifKOMA
    \pagestyle{empty}%
    \title{\getproperty{document}{title}%
    \ifpropertydefined{document}{status}{%
    \thanks{
        This is the Accepted Manuscript version of an article accepted for publication in \getproperty{document}{journal}. \getproperty{document}{editor} is not responsible for any errors or omissions in this version of the manuscript or any version derived from it. This Accepted Manuscript is published under a \getproperty{document}{license} licence. The Version of Record is available online at \url{\getproperty{document}{doi}}
    }%
    }}%
    \author[\getproperty{author}{affiliationindices}]{\getproperty{author}{firstname} \getproperty{author}{familyname}}
    \affil[1]{\getproperty{affiliations}{1}}
    
    \ifpropertydefined{coauthor1}{familyname}{
    \author[\getproperty{coauthor1}{affiliationindices}]{\getproperty{coauthor1}{firstname} \getproperty{coauthor1}{familyname}}}{}

    \ifpropertydefined{coauthor2}{familyname}{
    \author[\getproperty{coauthor2}{affiliationindices}]{\getproperty{coauthor2}{firstname} \getproperty{coauthor2}{familyname}}}{}

    \ifpropertydefined{coauthor3}{familyname}{
    \author[\getproperty{coauthor3}{affiliationindices}]{\getproperty{coauthor3}{firstname} \getproperty{coauthor3}{familyname}}}{}

    \ifpropertydefined{coauthor4}{familyname}{
    \author[\getproperty{coauthor4}{affiliationindices}]{\getproperty{coauthor4}{firstname} \getproperty{coauthor4}{familyname}}}{}

    \ifpropertydefined{coauthor5}{familyname}{
    \author[\getproperty{coauthor5}{affiliationindices}]{\getproperty{coauthor5}{firstname} \getproperty{coauthor5}{familyname}}}{}

    \ifpropertydefined{coauthor6}{familyname}{
    \author[\getproperty{coauthor6}{affiliationindices}]{\getproperty{coauthor6}{firstname} \getproperty{coauthor6}{familyname}}}{}

    \ifpropertydefined{coauthor7}{familyname}{
    \author[\getproperty{coauthor7}{affiliationindices}]{\getproperty{coauthor7}{firstname} \getproperty{coauthor7}{familyname}}}{}

    \ifpropertydefined{coauthor8}{familyname}{
    \author[\getproperty{coauthor8}{affiliationindices}]{\getproperty{coauthor8}{firstname} \getproperty{coauthor8}{familyname}}}{}

    \ifpropertydefined{coauthor9}{familyname}{
    \author[\getproperty{coauthor9}{affiliationindices}]{\getproperty{coauthor9}{firstname} \getproperty{coauthor9}{familyname}}}{}

    \ifpropertydefined{coauthor10}{familyname}{
    \author[\getproperty{coauthor10}{affiliationindices}]{\getproperty{coauthor10}{firstname} \getproperty{coauthor10}{familyname}}}{}

    \ifpropertydefined{coauthor11}{familyname}{
    \author[\getproperty{coauthor11}{affiliationindices}]{\getproperty{coauthor11}{firstname} \getproperty{coauthor11}{familyname}}}{}

    \ifpropertydefined{coauthors}{W7Xteaminclude}{
    \author[ ]{the W7-X team}}{}
    
    \ifpropertydefined{affiliations}{2}{
        \affil[2]{\getproperty{affiliations}{2}}
    }{}
    \ifpropertydefined{affiliations}{3}{
        \affil[3]{\getproperty{affiliations}{3}}
    }{}
    \ifpropertydefined{affiliations}{4}{
        \affil[4]{\getproperty{affiliations}{4}}
    }{}
    \ifpropertydefined{affiliations}{5}{
        \affil[5]{\getproperty{affiliations}{5}}
    }{}
    \ifpropertydefined{affiliations}{*}{
        \affil[*]{\getproperty{affiliations}{*}}
    }{}
        
    \date{\today}%
    \maketitle%
    \pagestyle{scrheadings}
\else
	\pagestyle{empty}
	\title{\getproperty{document}{title}}
	\author{\getproperty{author}{firstname} \getproperty{author}{familyname}}
	\date{\getproperty{document}{date}}
	\maketitle
\fi

}{}

\ifthenelseproperty{compilation}{abstract}{%
	\section*{Abstract}

High-\plasmaBeta operations require a fast and robust inference of plasma parameters with a self-consistent \gls{MHD} equilibrium.
Precalculated \gls{MHD} equilibria are usually employed at \gls{W7X} due to the high computational cost.
To address this,
we couple a physics-regularized \gls{NN} model that approximates the ideal-\gls{MHD} equilibrium with the Bayesian modeling framework Minerva.
We show the fast and robust inference of plasma profiles
(electron temperature and density)
with a self-consistent \gls{MHD} equilibrium approximated by the \gls{NN} model.
We investigate the robustness of the inference across diverse synthetic \gls{W7X} plasma scenarios.
The inferred plasma parameters and their uncertainties are compatible with the parameters inferred using the \gls{VMEC},
and the inference time is reduced by more than two orders of magnitude.
This work suggests that \gls{MHD} self-consistent inferences of plasma parameters can be performed between shots.
\glsresetall%

}{}

\ifthenelseproperty{compilation}{glossaries}{
	\glsresetall
}{}

\ifthenelseproperty{compilation}{toc}{%
    \disabledprotrusion{\tableofcontents}
}{}

\ifthenelseproperty{compilation}{frontmatter}{
    \mainmatter
}{}

\makeatletter
\@ifundefined{mainmatter}{}{\mainmatter}
\makeatother






\section{Introduction}\label{sec:introduction}

Bayesian inference is the statistical identification of model parameters that are consistent with given observed data using Bayes theorem.
In a magnetic confinement experiment,
the model parameters describe the plasma state,
which often includes,
but is not limited to the plasma profiles (\eg, temperature and density) and an ideal-\gls{MHD} equilibrium.
Diagnostics raw measurements provide the observed data.


In case of a high \plasmaBeta (the ratio between the plasma kinetic to the magnetic pressure) scenario or a large internal plasma current,
the equilibrium may differ from the vacuum equilibrium.
To correctly infer the plasma state,
a self-consistent, finite-beta ideal-\gls{MHD} equilibrium is required.
\gls{W7X} is an optimized stellarator that features a stiff equilibrium and a low bootstrap current~\cite{Grieger1993,Wolf2008}\added[id=AM,comment=compared to non-optimized stllrtr]{ (the expected bootstrap current at high-\plasmaBeta in the standard configuration is $\simeq\SI{80}{\kilo\ampere}$~\cite{Geiger2014})},
therefore,
equilibrium changes caused by the plasma internal current are predicted to be \replaced[id=AM,comment=DB: Lets discuss. Yes-bs current is RELATIVELY small. But still sometimes at ~100kA. That seems to me to be relevant but I have not done calculations. So far we have not reached these regimes so that might be relevant in the future. AM: I am not super happy of the current solution. At least it is a relative comparison.]{small with respects to an equivalent tokamak.}{small.}
Finite-beta effects,
however,
have been experimentally observed,
and they can still modify the vacuum equilibrium~\cite{Bozhenkov2020}.

In tokamaks,
data analysis workflows regularly include the reconstruction of a self-consistent ideal-\gls{MHD} equilibrium~\cite{Lao1985,Lao1990,Braams1991}.
However,
this is not the case with stellarators.
The inference of \num{3}D ideal-\gls{MHD} equilibria have been already demonstrated in several non-axisymmetric experiments, namely
the \gls{CTH}~\cite{Ma2015,Pandya2021},
the \gls{NCSX}~\cite{Pomphrey2005},
the \gls{HSX}~\cite{Chlechowitz2015,Schmitt2013,Schmitt2014},
\gls{W7AS}~\cite{Callaghan1999},
and \gls{W7X}~\cite{Schilling2018,Andreeva2019,Bozhenkov2020}.
Moreover,
multiple frameworks exist that allow the reconstruction of \num{3}D ideal-\gls{MHD} equilibria:
V3FIT~\cite{Hanson2009,Howell2020},
STELLOPT~\cite{Lazerson2015},
and Minerva~\cite{Svensson2007,Svensson2010}.
However,
because of the computational cost of calculating \num{3}D ideal-\gls{MHD} equilibria,
equilibrium reconstruction is not routinely used in Bayesian inference procedures and data analysis pipelines.
Vacuum fields,
or precalculated finite-beta approximations are used instead~\cite{Andreeva2019}.


Data-driven approaches have been proposed to reduce the computational cost of constructing stellarator equilibria.
Function parametrization was used at \gls{W7AS}~\cite{Callaghan1999,Callaghan2000} and at \gls{W7X}~\cite{Sengupta2004,Sengupta2007,Geiger2010}.
More recently,
\cite{Merlo2023} proposed a physics-regularized \gls{NN} to approximate the ideal-\gls{MHD} solution operator in \gls{W7X} configurations.
Adopting \gls{ML} models to approximate experimental equilibria can drastically accelerate sample-intensive applications (\eg, Bayesian inference) that require the computation of a \gls{MHD} equilibrium.

However,
how does the adoption of \gls{MHD} equilibria approximated by \gls{NN} models affect the inferred plasma parameters?
In this paper,
we show the fast (\orderof{\SI{e2}{\second}}) and robust inference of plasma profiles (electron temperature and density) based on the \gls{TS}~\cite{Pasch2016,Bozhenkov2017} and single channel \gls{DI}~\cite{Knauer2016} diagnostics with a self-consistent \gls{MHD} equilibrium.
The self-consistent \gls{MHD} equilibrium is approximated by the physics-regularized \gls{NN} model proposed in~\cite{Merlo2023}.
The \gls{NN} model allows the efficient exploration of the full posterior distribution of plasma profiles with self-consistent \gls{MHD} equilibria,
yielding an estimation of the profile uncertainties.
Specifically,
we:

\begin{itemize}
    \item couple the physics-regularized \gls{NN} model from~\cite{Merlo2023} in the Bayesian framework Minerva (\cref{sec:methods});
    \item investigate the robustness of the inferred plasma parameters across diverse \gls{W7X} synthetic plasma scenarios (\cref{sec:simple,sec:full});
    \item compare the reconstructed experimental plasma parameters and their uncertainties (estimated with \gls{MCMC}) with three different \gls{MHD} equilibrium models (\cref{sec:experimental}): a fixed, finite-beta \gls{VMEC} equilibrium; a self-consistent \gls{VMEC} equilibrium; and a self-consistent \gls{NN} equilibrium;
\end{itemize}

\section{Methods}\label{sec:methods}

\subsection{Bayesian Analysis}\label{sec:bayesian}

In a model-based simulation,
a mechanistic model,
the model parameters $H$,
and observed data $D$ characterize the phenomenon under study.
In Bayesian analysis,
the model parameters and observations are fully captured by their joint probability distribution:

\begin{gather}\label{eq:joint}
    P(H, D) = P(D | H) P(H) \ ,
\end{gather}

where the \textit{prior} distribution $P(H)$ describes the prior knowledge on the model parameters before taking any observations into account
(\eg, plasma density should be positive),
and the conditional distribution $P(D \vert H)$ denotes the probability of the observed data $D$ given the model parameters $H$
(when seen as a function of $H$ for given $D$,
$P(D \vert H)$ is it also known as the \textit{likelihood} function).

A parametric distribution function (\eg, Gaussian) usually represents $P(D \vert H)$.
The likelihood function is a function that captures the physical model under consideration,
and it is often referred to as the \textit{forward} model.
Experimentally derived uncertainties inform the distribution variance.

Bayesian \textit{inference} updates the model parameters prior distribution to their posterior distribution given the observed data using Bayes formula:

\begin{gather}\label{eq:posterior}
    P(H \vert D) = \frac{P(H, D)}{P(D)} = \frac{P(D \vert H) P(H)}{P(D)} \ ,
\end{gather}

where the \textit{posterior} distribution $P(H \vert D)$ describes the probability of the model parameters given the observed data,
and $P(D)$ is the \textit{evidence},
a normalization constant that describes the probability of the observed data given all possible values of the model parameters:

\begin{gather}\label{eq:evidence}
    P(D) = \int P(D \vert H) P(H) d H \ .
\end{gather}

The posterior distribution of the model parameters describes the degree of belief we have in the parameter values,
which is based on prior knowledge and updated by the observed data.
It encapsulates all the information and uncertainties we have on the model parameters.

\subsection{Model Graph}\label{sec:model-graph}

The Bayesian modeling framework Minerva~\cite{Svensson2007,Svensson2010} is used in this work.
In Minerva,
a graph describes the relationship between model parameters and observed data,
which are both represented as the graph's nodes.
The graph's edges represent the probabilistic relationships between the modeled quantities.
At \gls{W7X},
several diagnostics are already implemented in Minerva,
making it the obvious choice to validate the accelerated inference of plasma profiles with the \gls{NN} model.

\TODO{To reflect whether to include Phiedge here or not since it is fixed for the real inference.}

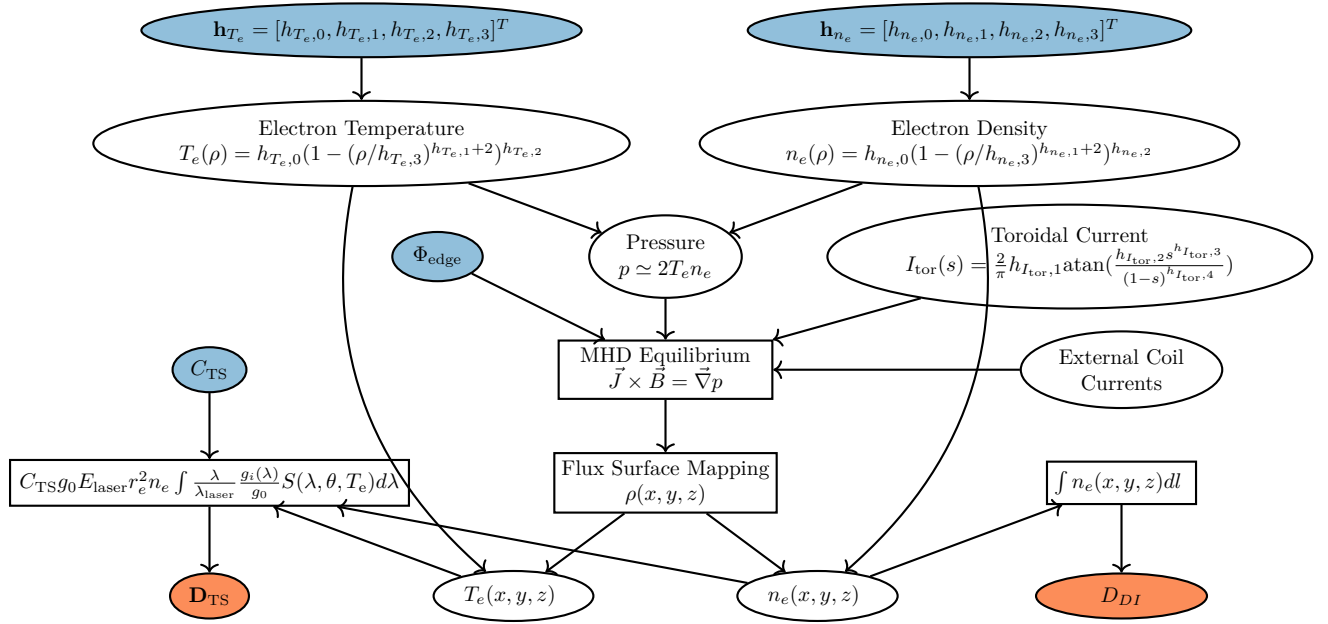
\begin{figure}[h!]
    \centering
\definecolor{free}{HTML}{91bfdb}
\definecolor{obs}{HTML}{fc8d59}

\def\x0{0}
\def\y0{0}
\def\stepx{4}
\def\stepy{1.5}

\begin{tikzpicture}[thick,scale=1, every node/.style={scale=0.8}]
  \node[ellipse,draw,fill=free] (mte) at (\x0 - \stepx,\y0 + \stepy) {
  $\mathbf{h}_{T_e} = [\teZero, \teOne, \teTwo, \teThree]^T$
  };
  \node[ellipse,draw,fill=free] (mne) at (\x0 + \stepx,\y0 + \stepy) {
  $\mathbf{h}_{n_e} = [\neZero, \neOne, \neTwo, \neThree]^T$
  };
  \node[ellipse,draw,minimum width=15em,align=center] (te) at (\x0 - \stepx,\y0) {
    Electron Temperature\\%
    $T_e (\rho) = \teZero (1 - (\rho / \teThree)^{\teOne + 2})^{\teTwo}$
  };
  \node[ellipse,draw,minimum width=15em,align=center] (ne) at (\x0 + \stepx,\y0) {
    Electron Density\\%
    $n_e (\rho) = \neZero (1 - (\rho / \neThree)^{\neOne + 2})^{\neTwo}$
  };
  \node[ellipse,draw,minimum width=15em,align=center] (itor) at (\x0 + 1.33 * \stepx,\y0 - \stepy) {
    Toroidal Current\\%
    $\Itor (s) = \frac{2}{\pi} \itorOne \text{atan}(\frac{\itorTwo s^{\itorThree}}{(1 - s)^{\itorFour}})$
  };
  \node[ellipse,draw,minimum height=2em,align=center] (pressure) at (\x0,\y0 - \stepy) {
    Pressure\\%
    $p \simeq 2 T_e n_e $
  };
  \node[ellipse,draw,fill=free] (phiedge) at (\x0 - 0.75 * \stepx,\y0 - \stepy) {\Phiedge};
  \node[ellipse,draw,text width=6em,align=center] (extcur) at (\x0 + 1.5 * \stepx,\y0 - 2 * \stepy) {External Coil Currents};
  \node[rectangle,draw,minimum width=10em,minimum height=2em,align=center] (mhd) at (\x0,\y0 - 2 * \stepy) {
    MHD Equilibrium\\%
    $\vec{J} \times \vec{B} = \vec{\nabla} p$
  };
  \node[rectangle,draw,minimum width=10em,minimum height=2em,align=center] (mapping) at (\x0,\y0 - 3 * \stepy) {
    Flux Surface Mapping\\%
    $\rho(x, y, z)$
  };
  \node[ellipse,draw,minimum height=2em,align=center] (nerho) at (\x0 + 0.5 * \stepx,\y0 - 4 * \stepy) {
    $n_e(x, y, z)$
  };
  \node[ellipse,draw,minimum height=2em,align=center] (terho) at (\x0 - 0.5 * \stepx,\y0 - 4 * \stepy) {
    $T_e(x, y, z)$
  };
  \node[rectangle,draw,minimum width=6em,minimum height=2em,align=center] (los) at (\x0 + 1.5 * \stepx,\y0 - 3 * \stepy) {
    $\int n_e(x, y, z) dl$
  };
  \node[ellipse,draw,fill=obs,minimum width=8em,minimum height=2em,align=center] (if) at (\x0 + 1.5 * \stepx,\y0 - 4 * \stepy) {$D_{DI}$};
  \node[ellipse,draw,fill=free] (cts) at (\x0 - 1.5 * \stepx,\y0 - 2 * \stepy) {$C_{\text{TS}}$};
  \node[rectangle,draw,text width=18em,minimum height=2em,align=center] (nato) at (\x0 - 1.5 * \stepx,\y0 - 3 * \stepy) {
    $C_{\text{TS}} g_0 E_{\text{laser}} r_e^2 n_e \int \frac{\lambda}{\lambda_{\text{laser}}} \frac{g_i(\lambda)}{g_0} S(\lambda, \theta, \Te) d\lambda$
  };
  \node[ellipse,draw,fill=obs] (ts) at (\x0 - 1.5 * \stepx,\y0 - 4 * \stepy) {$\mathbf{D}_{\text{TS}}$};
  \draw [thick,->] (mte) -- (te);
  \draw [thick,->] (mne) -- (ne);
  \draw [thick,->] (te) -- (pressure);
  \draw [thick,->] (ne) -- (pressure);
  \draw [thick,->] (phiedge) -- (mhd);
  \draw [thick,->] (pressure) -- (mhd);
  \draw [thick,->] (itor) -- (mhd);
  \draw [thick,->] (extcur) -- (mhd);
  \draw [thick,->] (mhd) -- (mapping);
  \draw [thick,->] (ne) to[bend left] (nerho);
  \draw [thick,->] (te) to[bend right] (terho);
  \draw [thick,->] (mapping) -- (nerho);
  \draw [thick,->] (mapping) -- (terho);
  \draw [thick,->] (nerho) -- (los);
  \draw [thick,->] (los) -- (if);
  \draw [thick,->] (cts) -- (nato);
  \draw [thick,->] (nerho) -- (nato);
  \draw [thick,->] (terho) -- (nato);
  \draw [thick,->] (nato) -- (ts);
\end{tikzpicture}
    \caption{%
        A simplified sketch of the Minerva graph.
        Blue ellipses represent free parameters,
        orange ellipses represent observed quantities,
        white ellipses represent model fixed parameters or model internal quantities,
        and white boxes represent physical forward models.
    }%
    \label{fig:minerva-graph}
\end{figure}

\Cref{fig:minerva-graph} shows a simplified picture of the Minerva graph used in this work.
Blue denotes the model free parameters,
and orange denotes the observed quantities.
A \gls{MHD} equilibrium maps the \num{1}D electron temperature and density profiles to a \num{3}D position of magnetic flux surfaces in space.
The \gls{TS}~\cite{Pasch2016,Bozhenkov2017} and the single channel \gls{DI}~\cite{Knauer2016} diagnostics are used to constrain the plasma state:
the \gls{TS} provides information on the electron density and temperature,
while the \gls{DI} measures the line-integrated electron density along a single \gls{LOS}.

\subsubsection{Prior distributions and forward models}\label{sec:forward}

Modified \textit{two power} profiles parametrize the electron density and temperature:

\begin{align}\label{eq:profiles}
    n_e (\rho) = \neZero (1 - (\rho / \neThree)^{\neOne + 2})^{\neTwo} , \\
    T_e (\rho) = \teZero (1 - (\rho / \teThree)^{\teOne + 2})^{\teTwo} ,
\end{align}

where $\rho = \sqrt{\Phi / \Phiedge}$ is the square root of the normalized toroidal flux.
The additional (with respect the standard two power parametrization) parameters \teThree and \neThree allow modeling not-null temperatures and densities beyond the \gls{LCFS},
which are usually observed at \gls{W7X}~\cite{Killer2019a}.
Uniform distributions represent the prior on the plasma profile parameters (see~\cref{sec:prior-distributions}).

Flux surfaces
$\rho(x, y, z)$
computed from a free-boundary ideal-\gls{MHD} equilibrium map the \num{1}D profiles
(\eg, $n_e(\rho)$)
to the \num{3}D real space geometry
($n_e(x, y, z)$).
The set of \gls{W7X} coil currents fixed to their setpoint values,
the net toroidal magnetic flux \Phiedge represented by a uniform prior distribution,
and the \num{1}D pressure and toroidal current profiles parametrize the free-boundary ideal-\gls{MHD} equilibrium.

In this work,
we assume $T_i \simeq T_e$,
$\gls{Zeff} \simeq 1$,
and $n_i \simeq n_e = n$.
As a result,
the pressure is simply given by $p \simeq 2 n_e T_e$.
The purpose of this paper is to investigate the use of \gls{NN} equilibria in a Bayesian inference framework,
rather than providing a highly accurate description of \gls{W7X}
(\eg,
by including a \gls{XICS} to infer the ion temperature,
or spectrometers to infer the plasma effective ion charge \gls{Zeff}).
Therefore,
the relaxation of these assumptions is not in the scope of this paper.

The \gls{NN} model sees only the total pressure $p$ and not $T_e$ or $T_i$.
As a result,
for the scope of this work,
it is not critical how the pressure profile is constructed.
The assumptions only limit the set of experimental scenarios that can be used to test the inference procedure.
Indeed,
the experimental scenarios considered in~\cref{sec:experimental} feature $T_i \simeq T_e$ within experimental uncertainties.
However,
in~\cref{sec:full},
we use synthetic data to investigate the robustness of the inference procedure across diverse,
randomly sampled (within the model parameter prior distributions) \gls{W7X} scenarios.

The toroidal current profile is parametrized as:

\begin{gather}\label{eq:itor}
    \Itor (s) = \frac{2}{\pi} \itorOne \text{atan}(\frac{\itorTwo s^{\itorThree}}{(1 - s)^{\itorFour}}) ,
\end{gather}

where $s = \rho^2$.
However,
given the lack of diagnostics to constrain its profile~\cite{Geiger2010},
the profile shape is fixed:
$\itorTwo = \itorThree = \itorFour = \num{1}$.
Moreover,
given that $\Itor (s = 1) = \itorOne$,
$\itorOne$ is held fixed to the net toroidal current as measured by the Rogowski coil~\cite{Endler2015}.

This work considers two \gls{MHD} equilibrium models:
\gls{VMEC}~\cite{Hirshman1983} and a physics-regularized \gls{NN}~\cite{Merlo2023}.
The \gls{NN} model has been integrated into Minerva with the ONNX Runtime framework~\cite{ONNX2018}.
In the proposed Minerva graph,
the equilibrium model is employed only to provide the self-consistent flux surface mapping.
In this regard,
\cite{Merlo2023} reports that the \gls{NN} model achieves an average flux surface error of $\simeq\SI{0.6}{\milli\meter}$.
\added[id=AM]{
    The \gls{NN} model was trained on a large set of \gls{W7X} magnetic configurations,
    including the reference \gls{W7X} configurations~\cite{Andreeva2002},
    pressure profiles with \averagePlasmaBeta up to \SI{5}{\percent},
    and toroidal current profiles with \Itor up to \SI{20}{\kilo\ampere}.
    \gls{W7X} experimental conditions in previous operational campaigns are well within the training data set~\cite{Klinger2016,Rolf2019}.
}
For further information on the model architecture,
training,
and evaluation,
please refer to~\cite{Merlo2023}.

The \gls{TS} observations constrain the electron density and temperature.
The \gls{TS} system's forward model maps $n_e(x, y, z)$ and $T_e(x, y, z)$ to the number of scattered photons on a set of \textit{volumes} along the \gls{TS} laser beam path.
Polychromator filters are used to measure the scattered spectrum by selecting distinct spectral intervals.
The time integral of the signal of each channel over the length of the scattered laser pulse corresponds to the number of the scattered photons~\cite{Bozhenkov2017}.
Thomson scattering scatters laser photons off of electrons,
and the Doppler effect broadens the Thomson scattered spectra due to the thermal motions of the electrons.
The electron temperature determines the width of the Thomson scattered spectrum,
and electron density determines its amplitude.
The predicted \gls{TS} signal of the $i$-th channel is~\cite{Bozhenkov2017}:

\begin{gather}\label{eq:ts}
    D_{\text{TS}}^i = C_{\text{TS}} g_0 E_{\text{laser}} r_e^2 n_e \int \frac{\lambda}{\lambda_{\text{laser}}} \frac{g_i(\lambda)}{g_0} S(\lambda, \theta, \Te) d\lambda \ ,
\end{gather}

where the integration is over the scattered spectrum wavelength $\lambda$.
$g_0$ is the absolute sensitivity of a reference channel,
$E_{\text{laser}}$ is the laser pulse energy,
$r_e$ is the electron radius,
$\lambda_{\text{laser}}$ is the laser wavelength,
$g_i(\lambda)$ is the absolute sensitivity of the $i$-th channel,
$\theta$ is the scattering angle,
and $S(\lambda, \theta, \Te)$ is the spectral density function.
The $\frac{g_i(\lambda)}{g_0}$ factor represents the relative calibration of each channel,
and the $C_{\text{TS}}$ is the global calibration factor.
For a more detailed description of the \gls{TS} system,
please refer to~\cite{Pasch2016,Bozhenkov2017}.

The forward model of the \gls{DI} system maps the \num{3}D electron density to the line-integrated value along the \gls{DI} \gls{LOS}.
The predicted \gls{DI} signal is:

\begin{gather}\label{eq:if}
    D_{\text{DI}} = \int n_e(x, y, z) dl ,
\end{gather}

where the integration path $\int dl$ is along the \gls{DI} \gls{LOS}.
For a more detailed description of the \gls{DI} system,
please refer to~\cite{Knauer2016}.

The misalignment of the laser system is a common issue that affects \gls{TS} systems.
In a \gls{TS} system,
observation optics with \glspl{LOS} that cross the laser beam path capture the scattered light.
A misaligned laser system can cause the laser beam path to be outside the nominal scattering volumes;
in such a case,
the inferred density will be lower than the actual plasma density.

At \gls{W7X},
the \gls{TS} system is believed to be affected by a systematic shift of the laser beam~\cite{Nelde2021,Fuchert2022}.
Laser misalignment is expected to be the dominant source of error in the electron density profiles,
affecting both the overall scale and shape of the profile.

In this work,
we introduce two additional parameters to account for laser misalignment errors:
the global \gls{TS} calibration factor $C_{\text{TS}}$,
and a systematic uncertainty $\sigma_{\text{laser}}$ that is added to the nominal \gls{TS} uncertainty.
$C_{\text{TS}}$ is a model free parameter,
and it is constrained during inference by the line-integrated density of the \gls{DI} (\ie, the \gls{TS} is cross-calibrated during inference).
The systematic uncertainty due to laser misalignment is held fixed,
and the standard deviation of the $i$-th \gls{TS} channel is modeled as:

\begin{align}
    (\sigma_{\text{TS}}^i)^2 = (\tilde{\sigma}_{\text{TS}}^i)^2 + (\sigma_{\text{laser}}^i)^2 , \\
    \sigma_{\text{laser}}^i = \max(\sigma_0, \alpha D_{\text{TS}}^i) \ ,
\end{align}

where $\tilde{\sigma}_{\text{TS}}^i$ is the nominal \gls{TS} uncertainty.
$\alpha$ and $\sigma_0$ are experimentally estimated (see~\cref{sec:misalignment}).
Ongoing efforts at \gls{W7X} are in place to account for the \gls{TS} laser misalignment.
A comprehensive modeling and exhaustive analysis of this error is beyond the scope of this paper.

\subsection{Inference}\label{sec:inference}

The inference procedure is divided in two stages.
Firstly,
in the so called \gls{MAP} optimization,
the model free parameters that maximizes the left-hand side of~\cref{eq:posterior},
\ie,
the mode of the posterior distribution,
are derived:

\begin{gather}\label{eq:h-map}
    H_{\text{MAP}} = \argmax_{H} P(H \vert D) \ .
\end{gather}

The Hooke and Jeeves optimizer~\cite{Hooke1961} drives the optimization.
Secondly,
the \gls{MCMC} \gls{MHA}~\cite{Metropolis1953,Hastings1970} numerically approximates the full posterior distribution.
The model parameter uncertainties can be derived from the \gls{MCMC} samples.
The inference is carried out on a single core of an Intel(R) Xeon(R) Gold 6136 CPU @ \SI{3}{\giga\hertz}.

We are interested in generating samples from the posterior distribution of the model parameters because they allow propagating their uncertainties into any subsequent quantities.
The computation of the samples is not straightforward:
either a closed form of the posterior is derived
(either analytically,
possible only in very seldom cases,
or by approximating it with a parametric form as it is done in variational inference),
or a sampling algorithm,
such as an \gls{MCMC} method,
is used.
\gls{MCMC} methods have desirable mathematical convergence properties,
and they do not require assumptions on the parametric form of the posterior distribution.
However,
they do not scale well to problems where the computation of the likelihood is inefficient because they require several thousands of evaluations of the likelihood function.
Replacing the \gls{VMEC} code with the \gls{NN} model results in orders of magnitude faster likelihood evaluations,
therefore,
it brings the Bayesian inference approach to a scale that was not possible before.

\section{Results}\label{sec:results}

This section investigates the inference of plasma profiles for both synthetic and experimental \gls{W7X} data:
\cref{sec:simple} visualizes the simple reconstruction of two parameters on simulated diagnostic data;
\cref{sec:full} investigates the robustness of the inference procedure across diverse \gls{W7X} plasma scenarios;
\cref{sec:experimental} compares the reconstruction of plasma profiles and their uncertainties with three different \gls{MHD} equilibrium models:
a fixed, finite-beta \gls{VMEC} equilibrium;
a self-consistent \gls{VMEC} equilibrium;
and a self-consistent \gls{NN} equilibrium.

\subsection{\num{2}D inference on synthetic data}\label{sec:simple}

To introduce the inference procedure,
and to visualize the free parameter values during inference,
a simplified \num{2}D inference on synthetic data is reported.
Apart for the electron density scale factor \neZero and the total toroidal flux enclosed by the plasma \Phiedge,
all other model parameters are fixed.

In this section,
diagnostic data are simulated with this Minerva graph rather than taken from the experiment.
Observations are modeled using normal or truncated normal distributions.
The predicted data from the forward model determine the distribution means,
and the experimentally informed uncertainties determine the distribution standard deviations.
The synthetic diagnostic data are then generated by setting the free parameter values to their target one (see paragraph below),
running the forward model with \gls{VMEC} as the equilibrium solver,
setting the observation means to the predicted diagnostic data,
and finally drawing samples from the observation distributions.
We would like to make two important remarks:
the generated synthetic data are obtained using an accurate equilibrium provided by \gls{VMEC},
while the inference process uses equilibria obtained from the \gls{NN} model;
the synthetic data are sampled from distributions whose sigma is informed by experimental uncertainties,
ensuring that they are not noise-free.

A \gls{W7X} plasma scenario resembling the post-pellet phase with an enhanced confinement~\cite{Bozhenkov2020} is considered (\expId{20181016.037} at $t=\SI{1.7}{\second}$):
central density $n_e(\rho = 0) = \SI{8.0e19}{\meter\tothe{-3}}$,
density peaking factor $n_e(\rho = 0) / n_e(\rho = 0.8) \simeq \num{2.77}$,
central temperature $T_e(\rho = 0) = \SI{3.0}{\kilo\eV}$,
flat temperature profile till $\rho \simeq \num{0.2}$,
net toroidal current $\Itor = \SI{1.3}{\kilo\ampere}$,
and $\Phiedge = \SI{1.945}{\volt\second}$ as from the vacuum equilibrium.
For simplicity,
$\neThree = \teThree = \num{1.0}$.
The resulting volume average plasma beta is $\averagePlasmaBeta = \SI{1.31}{\percent}$.

\begin{figure}[h!]
    \centering
    \igraph[width=0.5\textwidth]{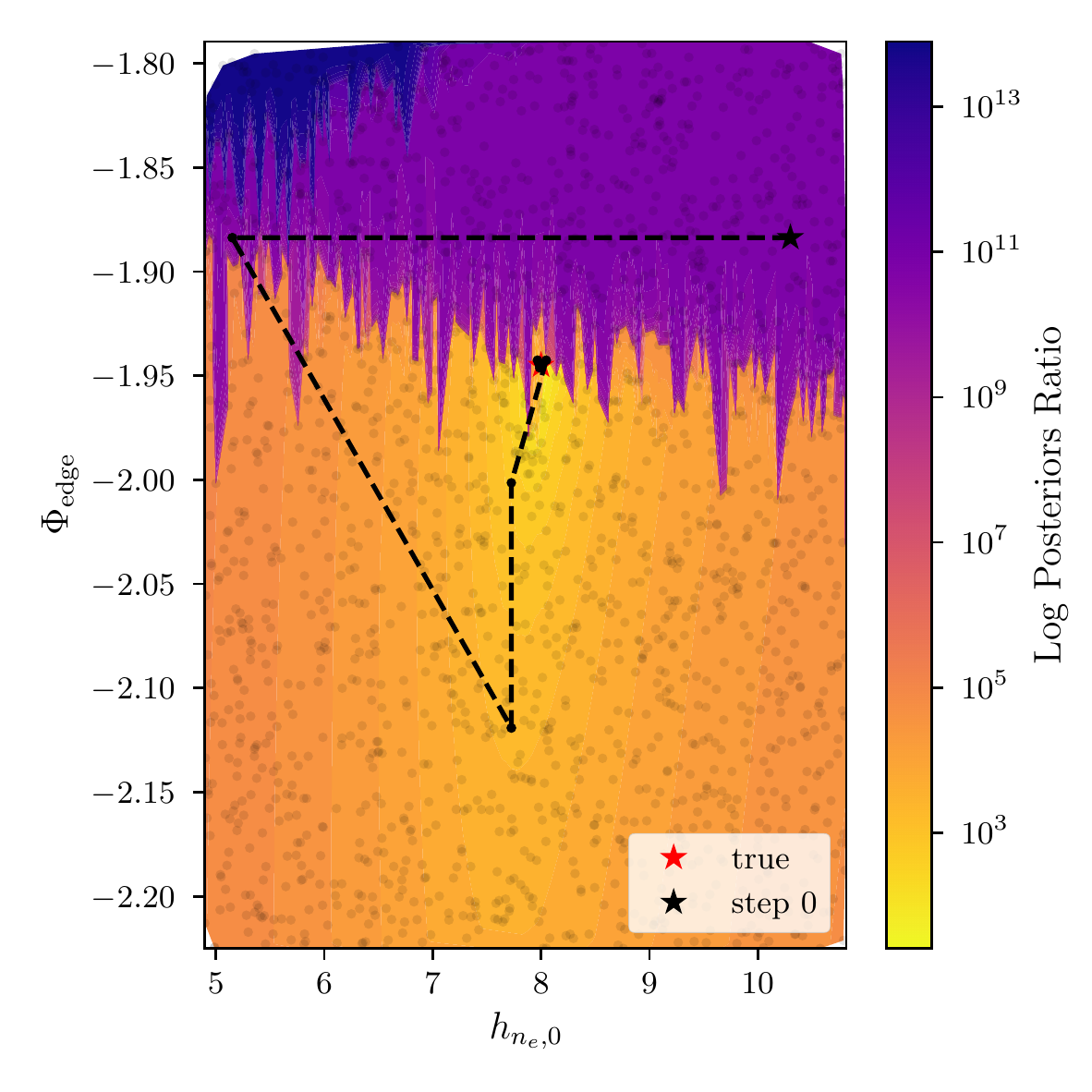}%
    \caption{%
        Trajectory of the free parameters during inference.
        In this case,
        the free parameters are the electron density scale factor \neZero and the net toroidal flux enclosed by the plasma \Phiedge.
        The reconstruction process starts from the black star,
        which represents the initial guess,
        and converges to the red star,
        which represents the true state.
        The dashed black line indicates the order of parameter values encountered during inference.
        The synthetic observations are generated using the Minerva graph and \gls{VMEC} as equilibrium model,
        while the inference is performed using the \gls{NN} as equilibrium model.
        The logarithm of the ratio of the sample posterior probability to the \gls{MAP} posterior probability is also shown.
        The light dark dots represent \num{10000} random samples used to interpolate the posterior distribution.
    }%
    \label{fig:simple}
\end{figure}

In this simplified \num{2}D inference,
the \gls{MAP} optimization quickly converges to the true values (\cref{fig:simple}):
after five iterations,
the inferred values (black dots) are already close to the target ones (red star).
The inference initial guess (black star) is sampled from the parameter prior distributions.
The relative difference between the inferred and true values is less than \SI{1}{\percent} (\cref{tab:simple}).

\Cref{fig:simple} also shows the contours of the full posterior distribution normalized to the posterior distribution of the \gls{MAP} values in the $(\neZero, \Phiedge)$ space:
a dark color indicates that a region of parameter space is less likely than the \gls{MAP} values,
on the other hand,
a light color indicates that a region of parameter space is equally likely as the \gls{MAP} values.
\num{10000} random samples are used to interpolate the posterior distribution.
\Cref{fig:simple} does also highlight the advantage of a fast \gls{MHD} equilibrium model:
sampling \num{10000} parameter values,
which requires computing \num{10000} self-consistent \gls{MHD} equilibria,
took only about \SI{20}{\minute}
(in a serial fashion on a single core).
Hence,
it is now possible to quickly explore the entire posterior distribution.

\begin{table}[h!]
    \caption{
        Reconstructed free parameter values for the \num{2}D inference on synthetic data.
        See~\cref{sec:forward} for the description of each parameter.
    }%
    \label{tab:simple}
    \centering
    \begin{tabular}{lcccc}
\toprule
           Parameter &  true &  initial &  inferred &  difference [\%] \\
\midrule
         $h_{n_e,0}$ &  8.00 &    10.30 &      7.98 &             0.19 \\
$\Phi_{\text{edge}}$ & -1.95 &    -1.88 &     -1.94 &             0.04 \\
\bottomrule
\end{tabular}

\end{table}

\subsection{Inference robustness on synthetic data}\label{sec:full}

Is the \gls{NN} model's approximation of the ideal-\gls{MHD} equilibrium robust across diverse plasma scenarios?
The \gls{NN} model was trained on a large set of \gls{W7X} equilibria for various vacuum magnetic configurations,
plasma pressure and toroidal current profiles.
In this section,
we investigate if the inference procedure is robust across plasma scenarios
(\eg, high- and low-\plasmaBeta, flat or peaked pressure profiles).

Using synthetic data provides the advantage of being able to conveniently scan the parameter space in different trials:
for each inference trial,
the \textit{true} state and the \textit{initial guess} are sampled from the model parameter prior distributions.
The plasma state is represented by the vector $\vec{h} = (\teZero, \teOne, \teTwo, \neZero, \neOne, \neTwo)^T$,
while $\teThree = \neThree = \num{1}$ are fixed.
Similarly to~\cref{sec:simple},
synthetic data are generated by injecting the true plasma state into the graph,
and by setting the observed data to samples drawn from the observation distributions,
using \gls{VMEC} as equilibrium model.
The free parameters are then initialized to their initial guesses,
and the posterior distribution is maximized for \num{100} iterations.
The \gls{NN} model approximates the \gls{MHD} equilibrium during inference.

The inference procedure is robust across plasma state and initial guess (\cref{fig:full-te-am0,fig:full-ne-am0,fig:full-te-am1,fig:full-ne-am1,fig:full-ne-am2,fig:full-te-am2}).
\Cref{fig:full-te-am0,fig:full-ne-am0,fig:full-te-am1,fig:full-ne-am1,fig:full-ne-am2,fig:full-te-am2} show the trajectory of the free parameters
(normalized to their true values)
as a function of the \gls{MAP} iterations.
Because the free parameters are normalized to their true values,
a final value of \num{1} (red dashed line) represents a successfully converged inference.
The trajectory means (solid line) and $\SI{5}{\percent}-\SI{95}{\percent}$ quantiles (dashed lines) are estimated with \num{25} independent inference trials.
The profile scale factors (\neZero and \teZero) are accurately inferred,
whereas the distributions of the profile shape parameters show a non-finite spread in their estimated values.
Yet,
the inferred values are on average close to the target values.
As a representative case,
\cref{tab:full} shows the true,
initial,
and inferred parameter values in case of the median trial
(\ie, the trial that has the median average percentage inference error across all trials).

\begin{table}[h!]
    \caption{
        Target,
        initial guess,
        and inferred parameter values for the median trial.
        See~\cref{sec:forward} for the description of each parameter.
    }%
    \label{tab:full}
    \centering
    \begin{tabular}{lcccc}
\toprule
  Parameter &  true &  initial &  inferred &  difference [\%] \\
\midrule
$h_{T_e,0}$ &  3.87 &     4.37 &      3.87 &             0.20 \\
$h_{T_e,1}$ &  2.70 &     2.17 &      2.87 &             6.44 \\
$h_{T_e,2}$ &  3.19 &     0.20 &      3.38 &             5.93 \\
$h_{n_e,0}$ & 13.92 &    12.20 &     13.92 &             0.03 \\
$h_{n_e,1}$ &  4.93 &     2.95 &      4.96 &             0.62 \\
$h_{n_e,2}$ &  6.73 &     2.90 &      6.72 &             0.22 \\
\bottomrule
\end{tabular}

\end{table}

\begin{figure}[h!]
    \centering
    \multigraph[labels={fig:full-te-am0}{fig:full-ne-am0}]{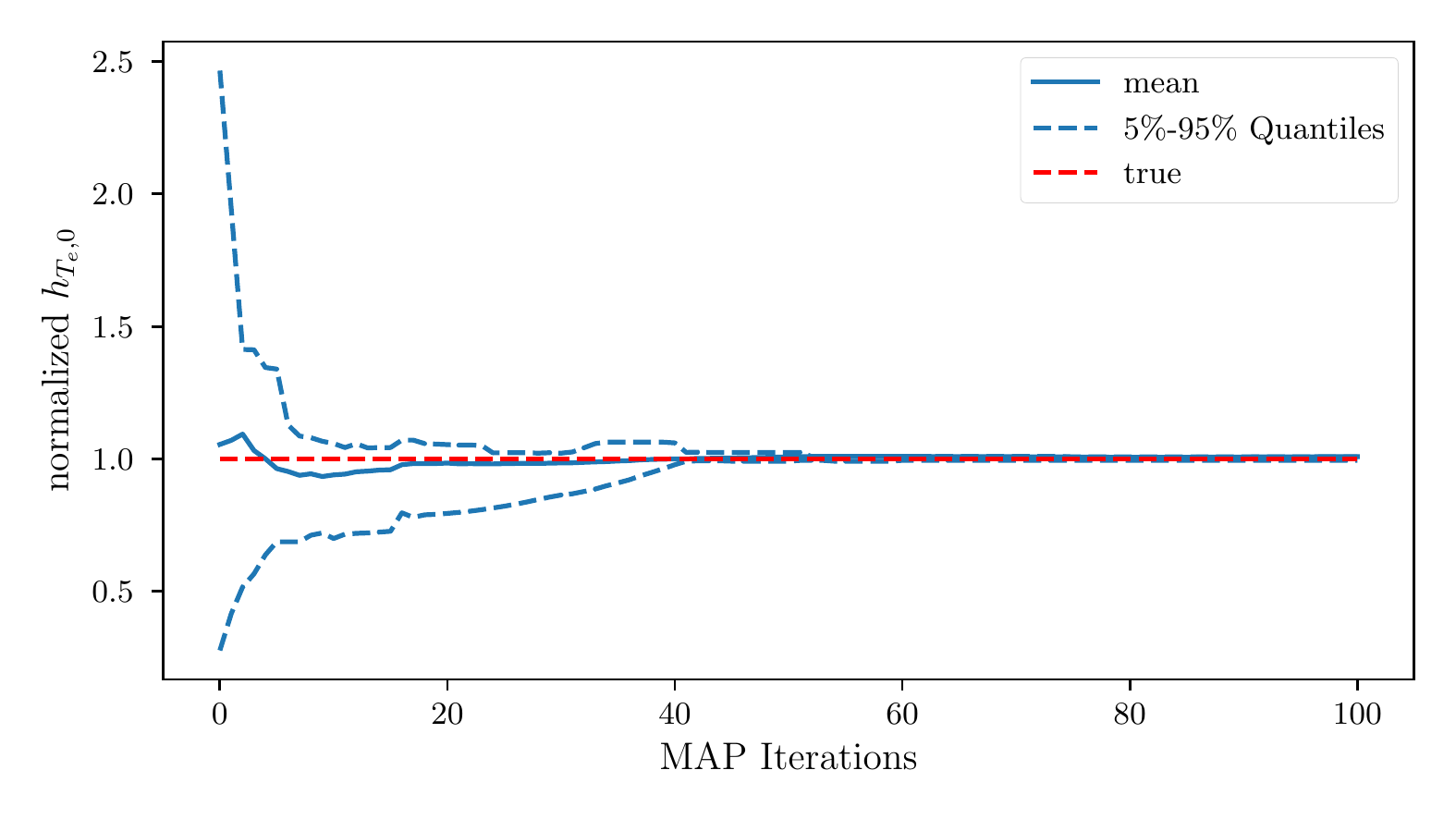;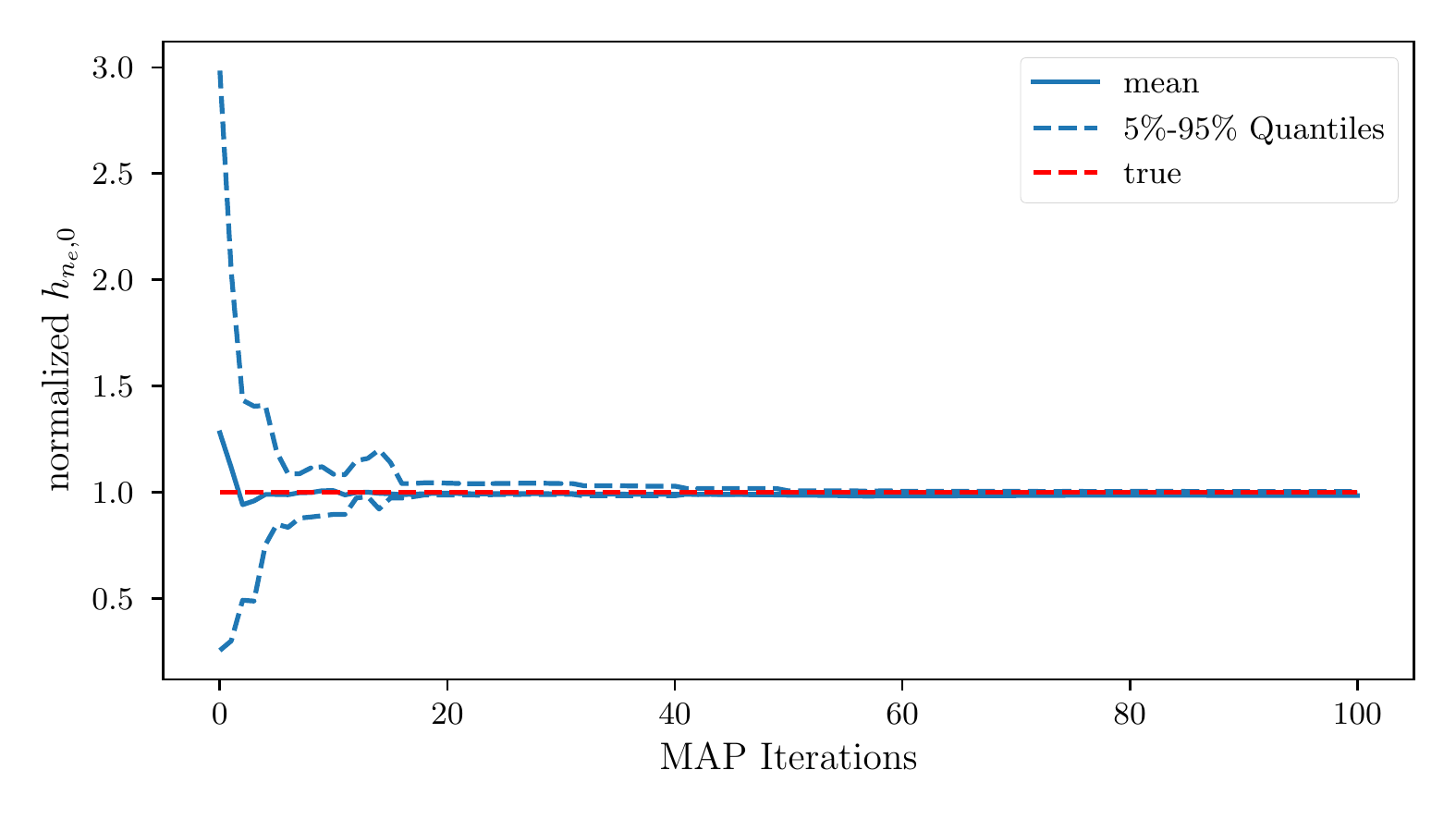}%
    {%
        {}{}%
    }%
    \multigraph[labels={fig:full-te-am1}{fig:full-ne-am1}]{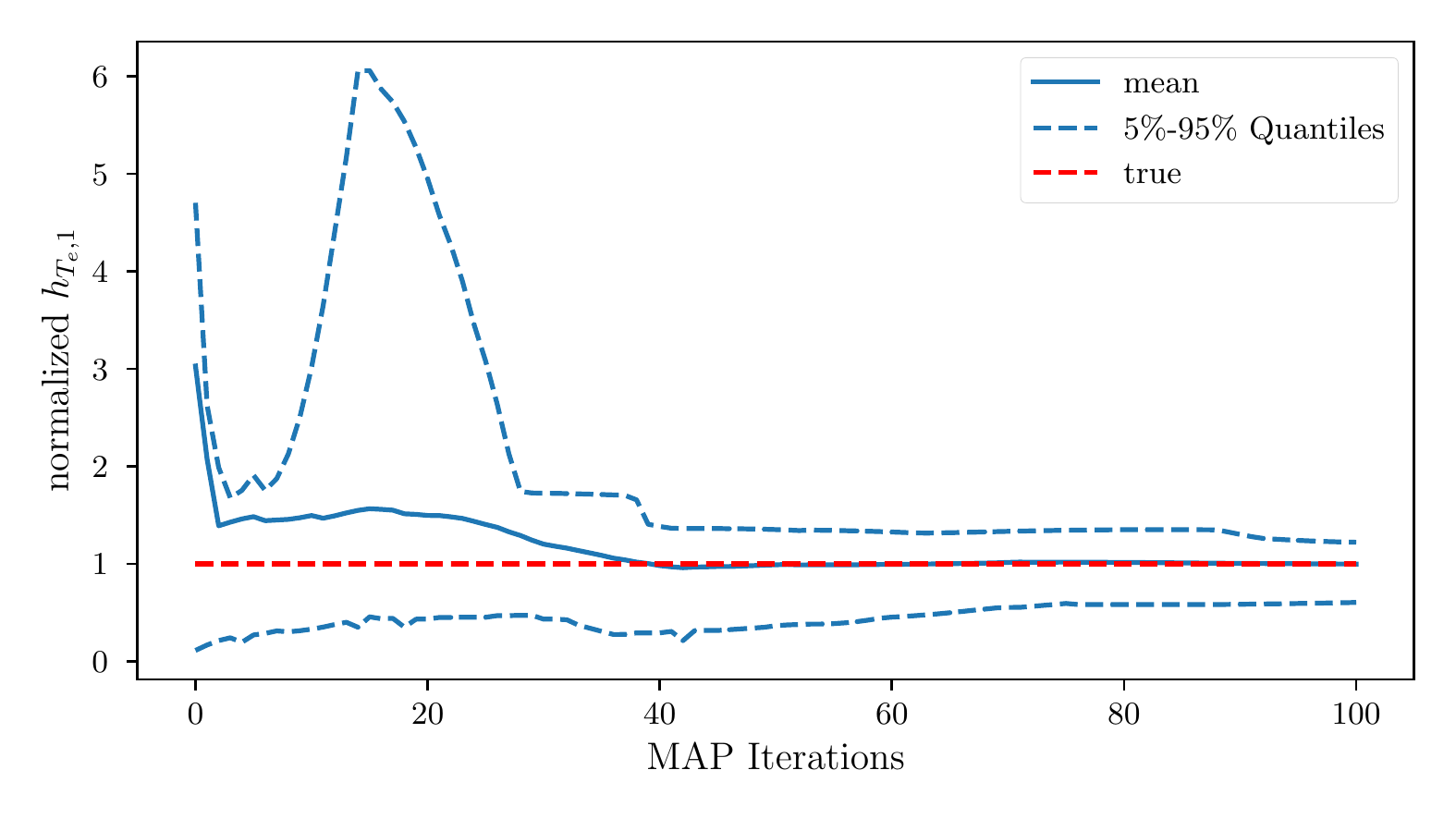;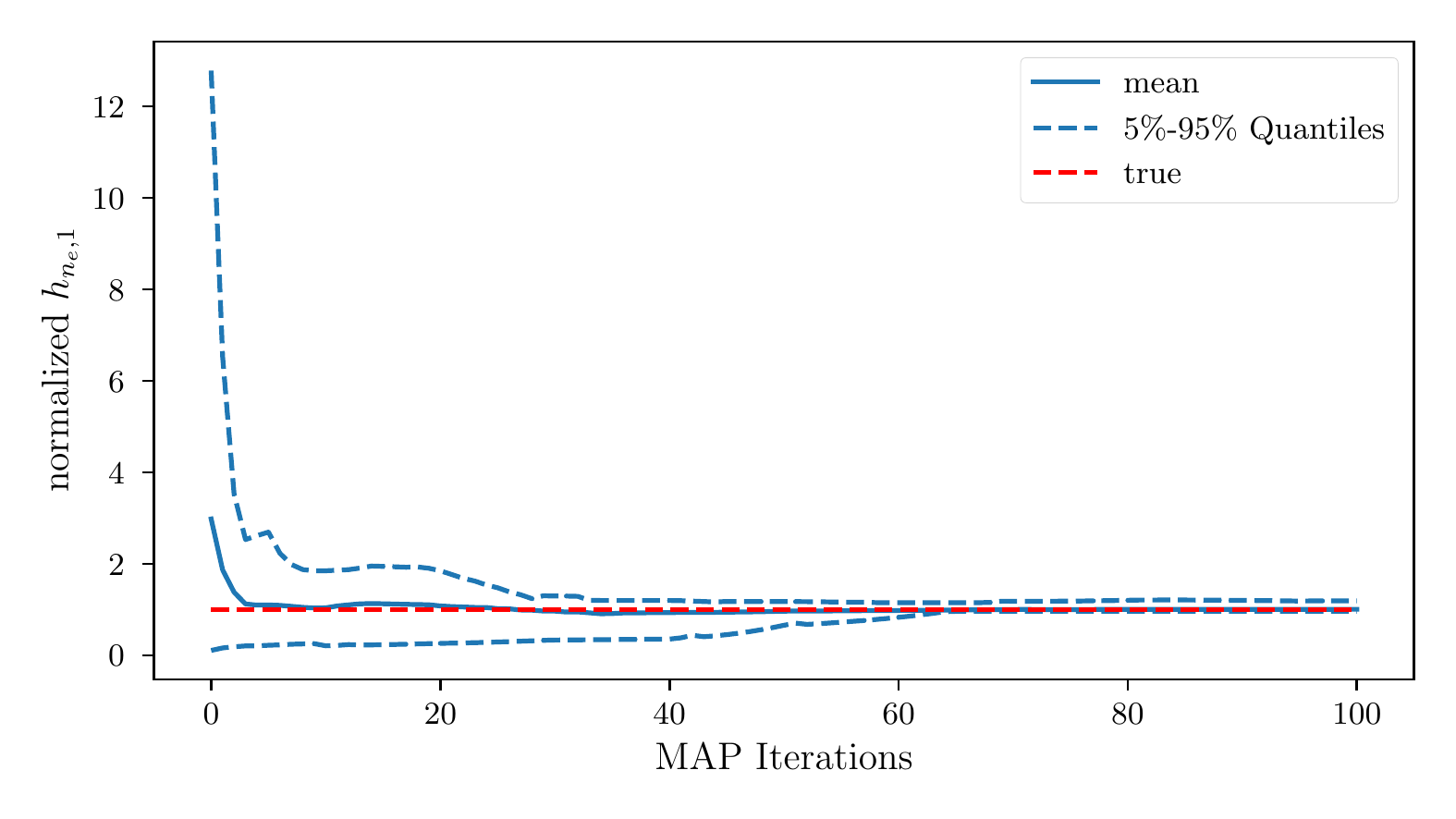}%
    {%
        {}{}%
    }%
    \multigraph[labels={fig:full-te-am2}{fig:full-ne-am2}]{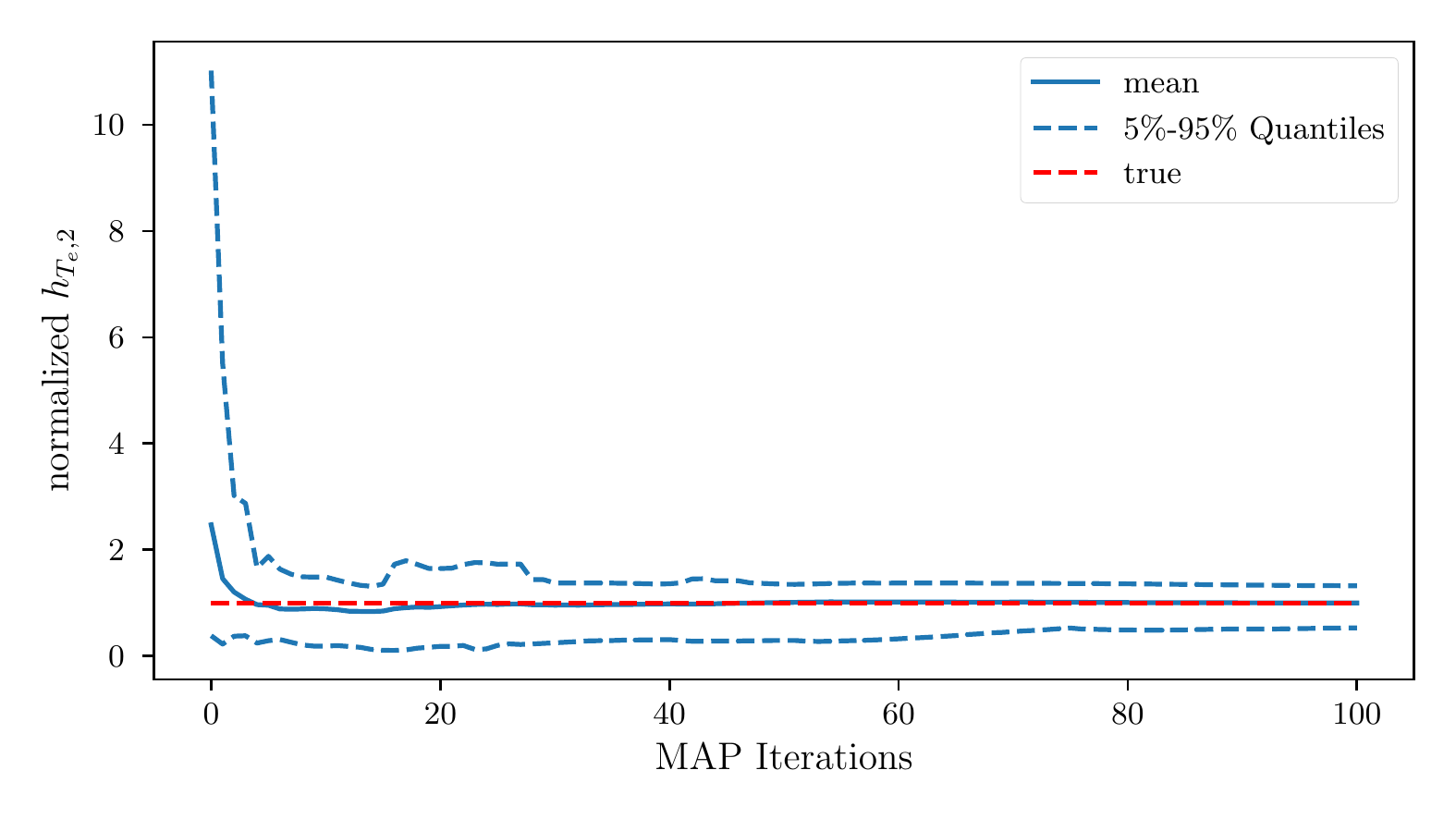;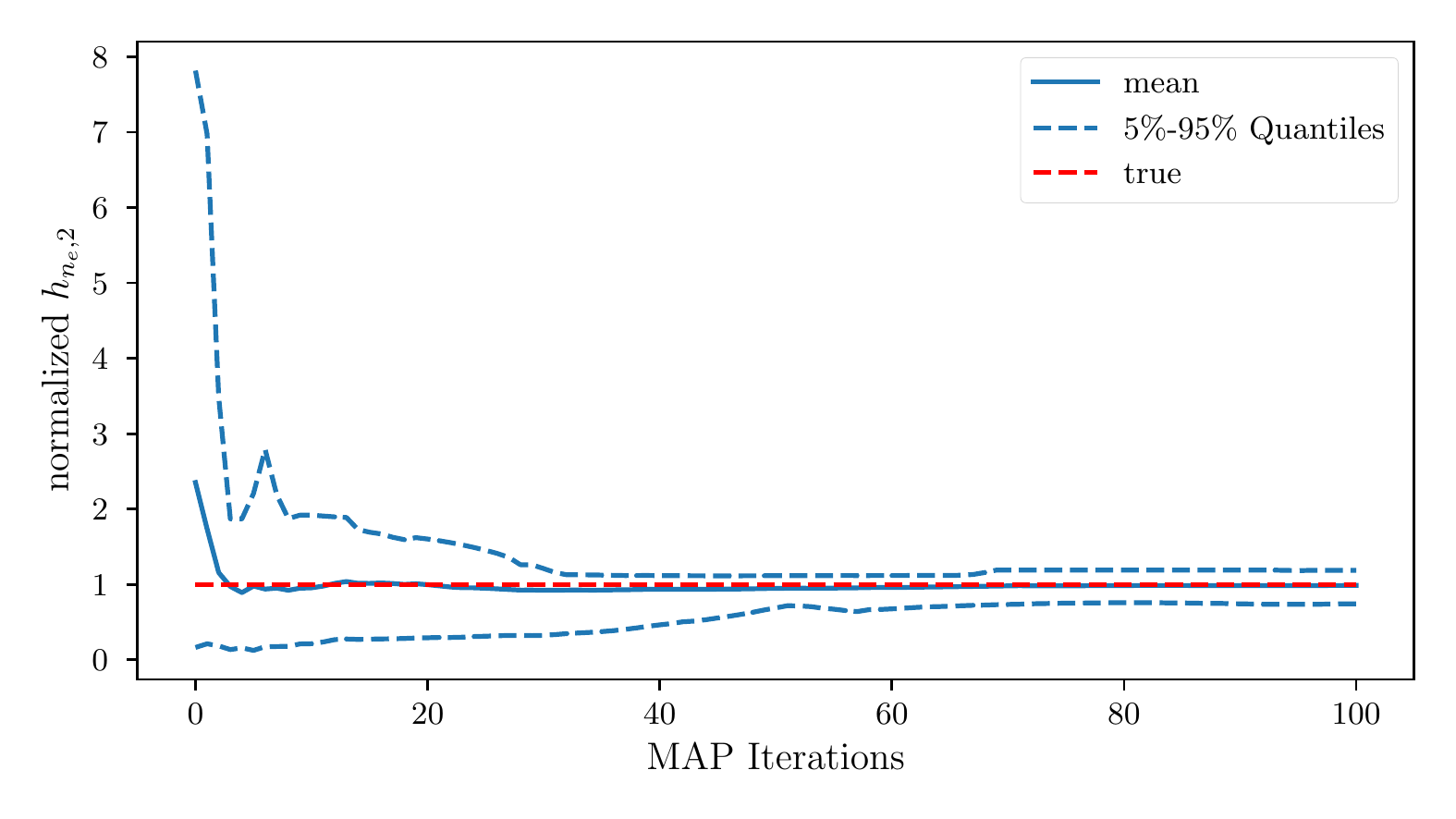}%
    {%
        {}{}%
    }%
    \caption{
        Convergence of the model parameters during inference for \num{25} independent trials.
        Each trial yields a randomly sampled set of true and initial guess values.
        Each figure shows the mean (solid line) and $\SI{5}{\percent}-\SI{95}{\percent}$ quantiles (dashed lines) of the normalized (to their true value) free parameters as a function of the \gls{MAP} optimization iterations.
        A red dashed line guides the eye to indicate the normalized target value (\ie, \num{1}).
        See~\cref{sec:forward} for the description of each parameter.
    }
\end{figure}

\subsection{Inference of improved confinement scenarios at \gls{W7X}}\label{sec:experimental}

In this section,
we replace the synthetic diagnostic data with experimental data.
We address two questions:
how does the approximation of the equilibria by the \gls{NN} model affect the inferred plasma parameters?
What are the implications of a self-consistent \gls{MHD} equilibrium for plasma parameter estimation?

To address both questions,
two plasma scenarios with $\Ti \simeq \Te$ at different \averagePlasmaBeta and magnetic configurations are considered:

\begin{description}
    \item[\expId{20181016.37}, $t=\SI{1.7}{\second}$:]{
          a $\averagePlasmaBeta>\SI{1}{\percent}$,
          improved confinement scenario following pellet injections in the \textit{standard} configuration~\cite{Bozhenkov2020}.
          }
    \item[\expId{20180808.5}, $t=\SI{20.0}{\second}$:]{
          a low-power,
          improved confinement scenario in the \textit{high-iota} configuration~\cite{Wurden2022}.
          }
\end{description}



In this section,
we compare three different ideal-\gls{MHD} equilibrium models:
a reference, fixed, finite-beta \gls{VMEC} equilibrium (see description below);
a self-consistent \gls{VMEC} equilibrium;
and a self-consistent \gls{NN} equilibrium.
Comparing the two self-consistent equilibrium models allows the investigation of the approximations provided by the \gls{NN} model,
while comparing the fixed and self-consistent equilibrium models enables the analysis of the implications of a self-consistent equilibrium on the inferred plasma profiles.

\TODO{Ask Sergey a better citation for the VMEC look-up table.}

The reference \gls{VMEC} equilibrium is taken from the set of precalculated finite-beta equilibria commonly used at \gls{W7X}~\cite{Andreeva2019}.
The pressure profile is assumed to be $p_0(1 - s)^{a_2}$,
where $p_0$ is the pressure scale,
and $a_2$ determines the pressure peaking factor.
Estimating $a_2$ from the experimental profiles,
and assuming $p_0$ so that \Wkin matches the measured \gls{Wdia} value (with a plasma volume of $V_p = \SI{30}{\meter\tothe{3}}$),
the closest precalculated equilibrium in terms of Euclidean distance in the $(p_0, a_2)$ space is selected.
The net toroidal current is less than \SI{2}{\kilo\ampere} in these experiments and can be neglected.
\Phiedge is set to the \replaced[id=AM]{toroidal flux enclosed by the \gls{LCFS} of the vacuum field }{vacuum }obtained from field line tracing.

In case of the self-consistent \gls{MHD} equilibria,
the equilibrium is not fixed,
but evaluated consistently with the proposed density and temperature profiles.
Nine free parameters represent the plasma state:
four profile parameters each for $n_e$ and \Te,
and a global calibration factor for the \gls{TS} (see~\cref{sec:forward}).
\Phiedge is held fixed\deleted[id=AM]{ to its vacuum value}.
As in~\cref{sec:full},
the toroidal current profile shape is fixed,
but the net toroidal current is set to the value measured by the Rogowski coil.

To improve the inference's convergence,
diagnostics raw data inform the initial guess:
\replaced[id=AM]{%
    \neZero is set to be consistent with the line average integrated density from the \gls{DI} (assuming a flat density profile and a \gls{LOS} length of \SI{1.1}{\meter}),
}{%
    the line average integrated density from the \gls{DI} sets \neZero,
}
and \teZero is chosen so that the plasma's kinetic energy matches the measured diamagnetic energy from the diamagnetic loop
(assuming a plasma volume of $V_p = \SI{30}{\meter\tothe{3}}$,
and a linear pressure profile in normalized toroidal flux).

For both considered experiments,
the inferred profile parameters with the three \gls{MHD} equilibrium models are compatible (\cref{tab:real-20181016-free-parameters,tab:real-20180808-free-parameters}):
after \num{100} \gls{MAP} iterations,
the parameter values are within one standard deviation away from each other.
The standard deviations are estimated using \num{30000} \gls{MCMC} samples,
of which the first \num{10000} are discarded,
and only \num{1} in every \num{50} are kept to reduce the sample autocorrelation.

It is worth highlighting the time needed to perform each inference
(\num{100} \gls{MAP} iterations on a single core,
given here in case of the \expId{20181016.37} shot):
using the self-consistent \gls{VMEC} equilibrium took \SI{1108}{\minute} and \SI{45}{\second},
using the fixed \gls{VMEC} equilibrium took \SI{9}{\minute} and \SI{35}{\second},
and using the \gls{NN} model took only \SI{3}{\minute} and \SI{31}{\second}.
The inference time is reduced by more than two orders of magnitude.

\begin{table}[h!]
    \caption{
        Comparison of the inferred free parameters in case of the \expId{20181016.37} shot at $t=\SI{1.7}{\second}$ with the three ideal-\gls{MHD} models:
        a fixed, finite-beta precalculated \gls{VMEC} equilibrium;
        a self-consistent \gls{VMEC} equilibrium;
        and the \gls{NN} model described in~\cref{sec:forward}.
        The free parameter values are obtained via a \gls{MAP} optimization,
        and the standard deviation (in brackets) is estimated with \gls{MCMC}.
        Because of the computational time required to run \gls{VMEC} for each \gls{MCMC} sample,
        the \gls{MCMC} procedure has not been performed in case of the self-consistent \gls{VMEC} equilibrium.
    }%
    \label{tab:real-20181016-free-parameters}
    \centering
    \begin{tabular}{lccc}
\toprule
  Parameter &    Fixed VMEC &  Self-consistent VMEC & Self-consistent NN \\
\midrule
$h_{T_e,0}$ &   3.39 (0.16) &                 3.403 &        3.40 (0.17) \\
$h_{T_e,1}$ &   1.04 (0.36) &                 1.038 &        1.03 (0.38) \\
$h_{T_e,2}$ &     4.7 (1.9) &                 4.834 &          4.7 (1.9) \\
$h_{T_e,3}$ & 1.129 (0.081) &                 1.138 &      1.130 (0.080) \\
$h_{n_e,0}$ &   8.27 (0.51) &                 8.341 &        8.35 (0.50) \\
$h_{n_e,1}$ &   0.16 (0.32) &                 0.140 &        0.14 (0.33) \\
$h_{n_e,2}$ &   2.77 (0.74) &                 2.726 &        2.71 (0.89) \\
$h_{n_e,3}$ & 1.330 (0.072) &                 1.330 &      1.328 (0.079) \\
\bottomrule
\end{tabular}

\end{table}

\begin{table}[h!]
    \caption{
        Comparison of the inferred free parameters in case of the \expId{20180808.5} shot at $t=\SI{20.0}{\second}$ with the three ideal-\gls{MHD} models.
        See~\cref{tab:real-20181016-free-parameters} for a description of each row.
    }%
    \label{tab:real-20180808-free-parameters}
    \centering
    \begin{tabular}{lccc}
\toprule
  Parameter &    Fixed VMEC &  Self-consistent VMEC & Self-consistent NN \\
\midrule
$h_{T_e,0}$ & 1.930 (0.089) &                 1.928 &      1.915 (0.088) \\
$h_{T_e,1}$ & 0.001 (0.323) &                 0.001 &      0.002 (0.344) \\
$h_{T_e,2}$ &     4.1 (1.4) &                 4.126 &          4.4 (1.4) \\
$h_{T_e,3}$ & 1.330 (0.087) &                 1.330 &      1.330 (0.086) \\
$h_{n_e,0}$ &   4.32 (0.49) &                 4.258 &        4.21 (0.50) \\
$h_{n_e,1}$ &   2.52 (0.65) &                 2.524 &        2.35 (0.68) \\
$h_{n_e,2}$ &     8.2 (1.9) &                 8.535 &          6.4 (1.9) \\
$h_{n_e,3}$ & 1.079 (0.075) &                 1.090 &      1.051 (0.072) \\
\bottomrule
\end{tabular}

\end{table}

\begin{figure}
    \centering
    \begin{subfigure}{.45\linewidth}
        \centering
        \includegraphics[width=\linewidth]{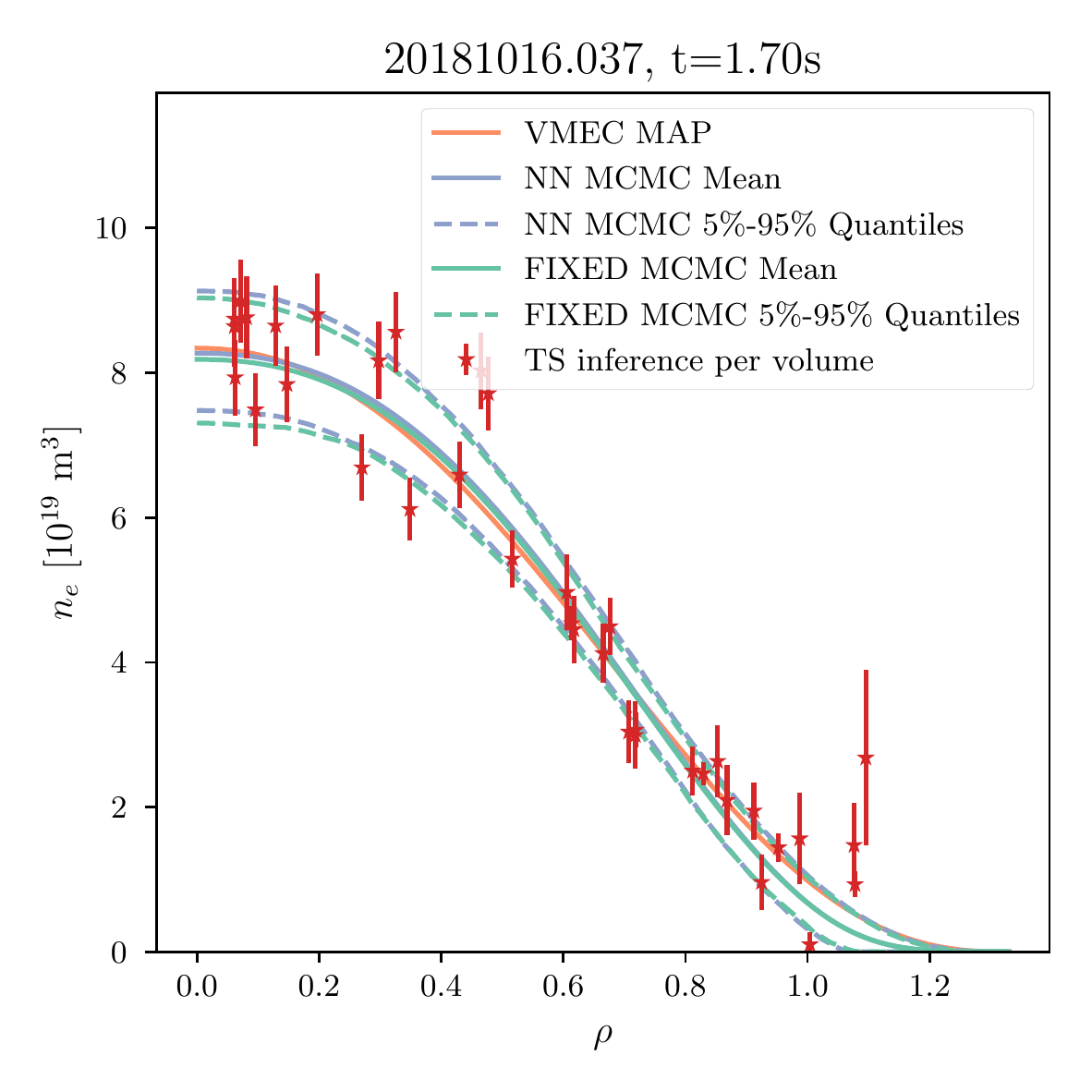}
        \caption{}\label{fig:experimental-20181016-ne}
    \end{subfigure}
    \begin{subfigure}{.45\linewidth}
        \centering
        \includegraphics[width=\linewidth]{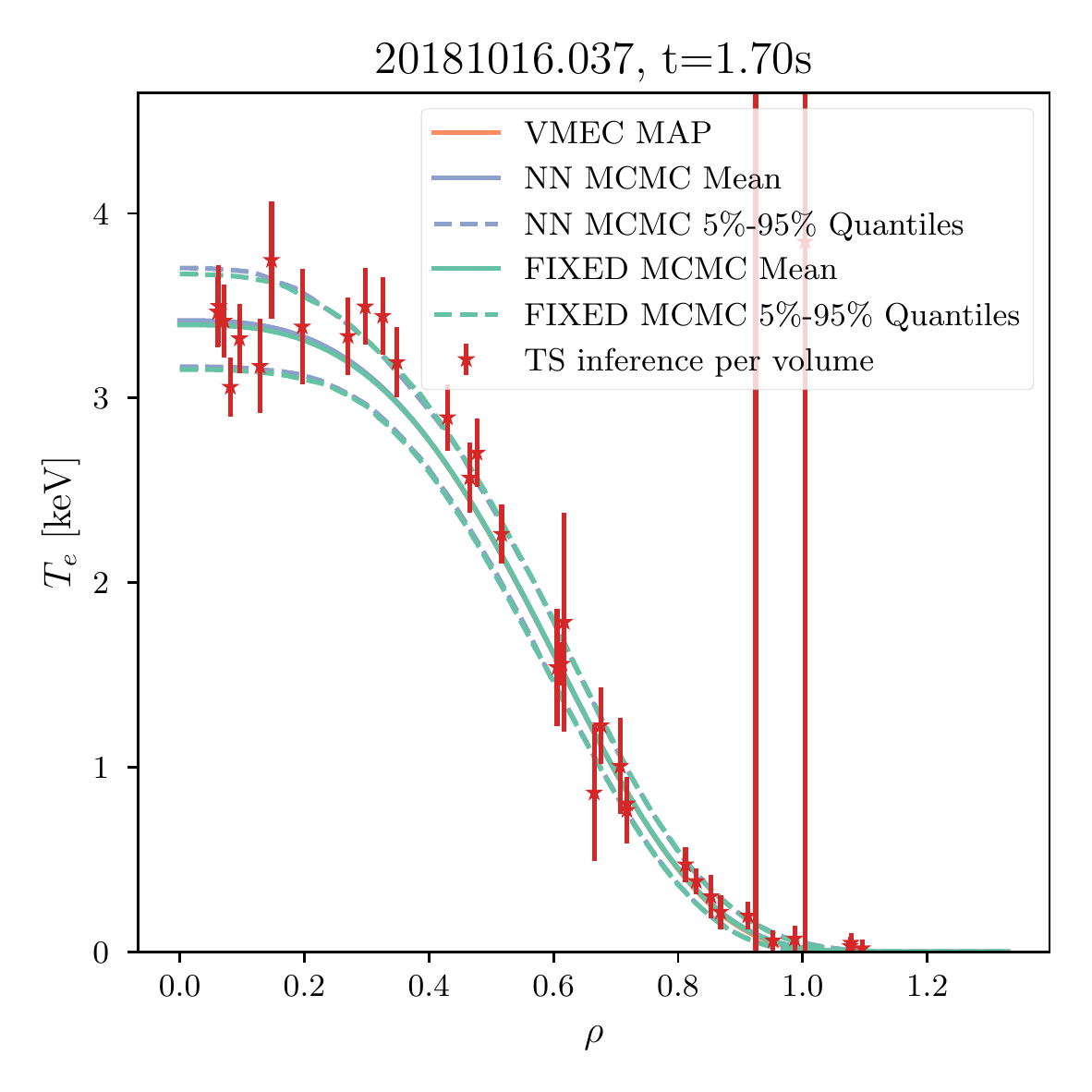}
        \caption{}\label{fig:experimental-20181016-te}
    \end{subfigure}

    \begin{subfigure}{.45\linewidth}
        \centering
        \includegraphics[width=\linewidth]{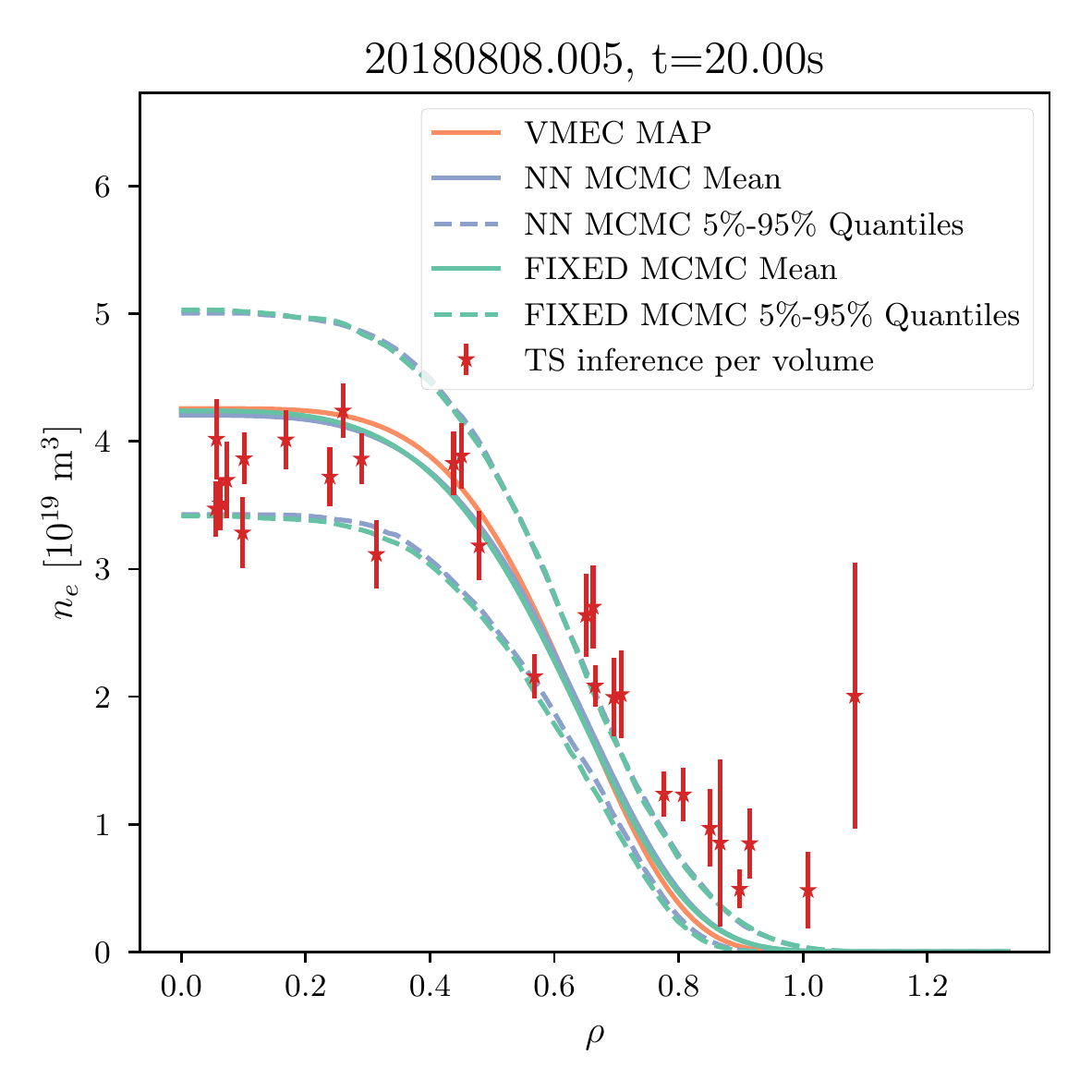}
        \caption{}\label{fig:experimental-20180808-ne}
    \end{subfigure}
    \begin{subfigure}{.45\linewidth}
        \centering
        \includegraphics[width=\linewidth]{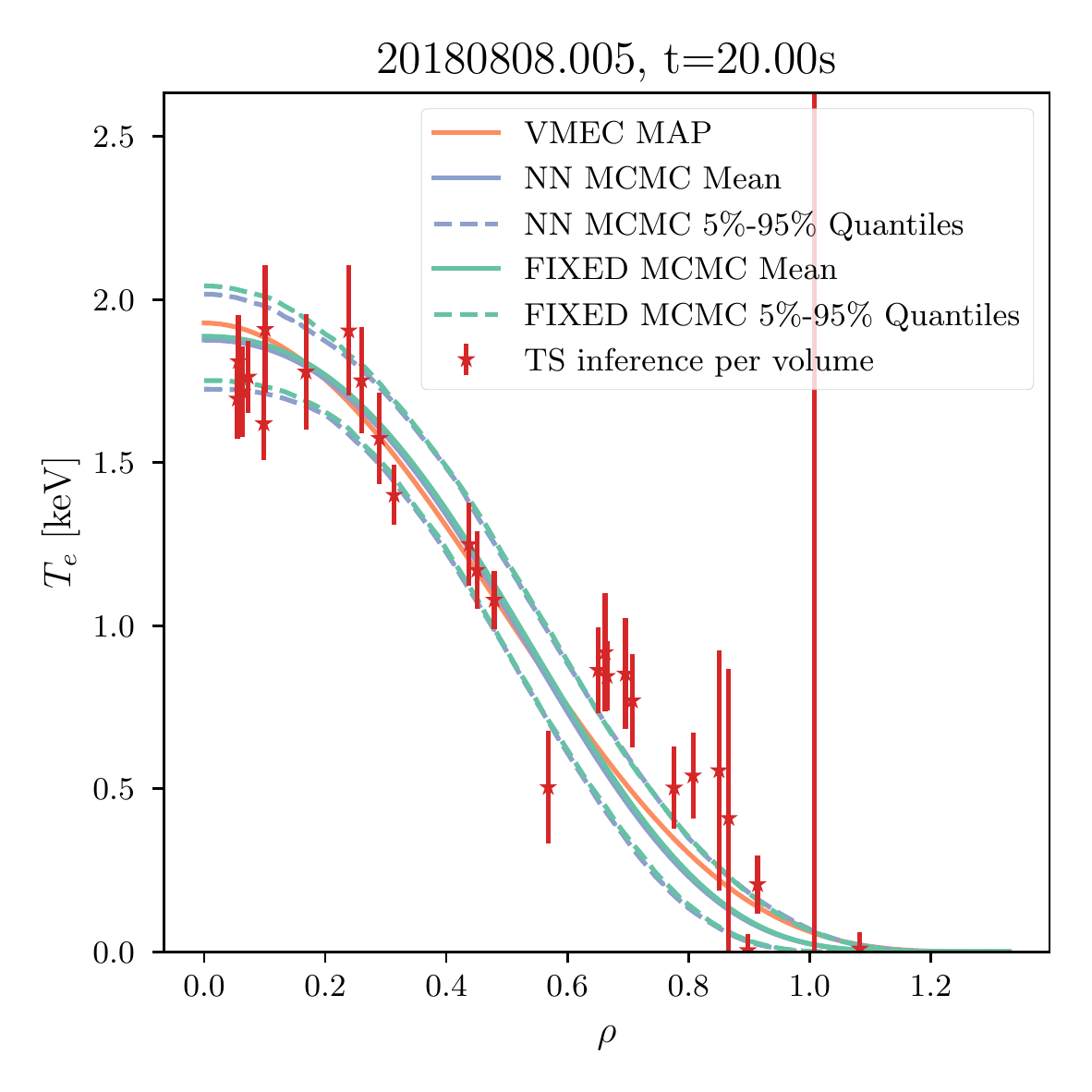}
        \caption{}\label{fig:experimental-20180808-te}
    \end{subfigure}
    \caption{
        Comparison of the inferred electron density (\cref{fig:experimental-20181016-ne,fig:experimental-20180808-ne}) and temperature (\cref{fig:experimental-20181016-te,fig:experimental-20180808-te}) profiles with the following \gls{MHD} models:
        a fixed, finite-beta precalculated \gls{VMEC} equilibrium;
        a self-consistent \gls{VMEC} equilibrium;
        and the self-consistent \gls{NN} model described in~\cref{sec:forward}.
        The solid lines represent the mean profile across all \gls{MCMC} samples
        (in case of the self-consistent \gls{VMEC} equilibrium,
        the solid line is the \gls{MAP} profile),
        and dashed lines represent the $\SI{5}{\percent}-\SI{95}{\percent}$ quantiles.
        In addition,
        the values (as well as their standard deviation) of the electron density and temperature at each \gls{TS} volume location,
        from a reference Minerva inference,
        are also shown (red stars).
        The shot id and time of each reconstructed slice are shown in each figure title.
    }
\end{figure}

As depicted in~\cref{tab:real-20181016-free-parameters,tab:real-20180808-free-parameters},
the three ideal-\gls{MHD} models yield compatible plasma profiles.
This is true not only in terms of \gls{MAP} profiles,
but also for their posterior distributions estimated with \gls{MCMC} (\cref{fig:experimental-20181016-ne,fig:experimental-20181016-te,fig:experimental-20180808-ne,fig:experimental-20180808-te}).

The inferred profiles are also consistent with a independently inferred electron temperatures and densities per \gls{TS} volume (\cref{fig:experimental-20181016-ne,fig:experimental-20181016-te,fig:experimental-20180808-ne,fig:experimental-20180808-te}).
For the reference Minerva inference,
the density and temperature values at each \gls{TS} volume are independently reconstructed using the fixed, finite-beta \gls{VMEC} equilibrium as \gls{MHD} model,
and the density is further scaled to match the line-integrated value from the \gls{DI}.
See~\cite[Section~5]{Bozhenkov2017} for a description of the inference procedure.

\Cref{tab:real-20181016-equilibrium-properties,tab:real-20180808-equilibrium-properties} show selected quantities of interest of the resulting equilibria.
In case of the inference performed with the \gls{NN} model,
\replaced[id=AM]{%
    the mean and standard deviation (in brackets) of the \gls{MCMC} samples are reported.
}{%
    the reported values are obtained from a \gls{VMEC} \textit{validation} run:
    an ideal-\gls{MHD} equilibrium computed by \gls{VMEC},
    but with the plasma profiles obtained through the inference procedure using the \gls{NN} model.
}

The inferred \gls{MHD} equilibria obtained using the self-consistent \gls{VMEC} and \gls{NN} models feature compatible properties (\cref{tab:real-20181016-equilibrium-properties,tab:real-20180808-equilibrium-properties})\replaced[id=AM]{: the inferred values with the self-consistent \gls{NN} model are within one standard deviation with respect to the values inferred with the self-consistent \gls{VMEC} equilibrium.}{.
    In particular,
    the inferred volume averaged beta,
    kinetic energy,
    pressure on axis,
    and pressure peaking factor are close to each other.
}
\deleted[id=AM]{%
    However,
    they do differ from the values obtained using the precalculated fixed equilibrium.
}
\added[id=AM]{%
    It is worth noting that,
    for the two experimental scenarios investigated,
    the assumed pressure profile in the precalculated fixed equilibrium (first column in the tables) differs from the inferred pressure profiles (second and third columns in the tables).
}


\begin{table}[h!]
    \caption{
        Comparison of selected quantities of the inferred equilibria of experiment\expId{20181016.37} at $t=\SI{1.7}{\second}$ with three \gls{MHD} models:
        a fixed, finite-beta precalculated \gls{VMEC} equilibrium;
        a self-consistent \gls{VMEC} equilibrium;
        and the self-consistent \gls{NN} model as described in~\cref{sec:forward}.
        \replaced[id=AM]{%
            In case of the inference performed with the self-consistent \gls{NN} model,
            the mean and standard deviation (in brackets) of the \gls{MCMC} samples are reported.
        }{%
            In case of the self-consistent \gls{MHD} equilibria,
            the \gls{MAP} plasma profiles set the equilibrium pressure profile.
        }
        \replaced[id=AM]{%
            In these experimental conditions,
            the measured diamagnetic energy is $\gls{Wdia} = \SI{1137.661}{\kilo\joule}$.
        }{%
            The relative difference between the inferred plasma kinetic energy and the measured diamagnetic energy,
            $\gls{Wdia} = \SI{1137.661}{\kilo\joule}$,
            is stated in brackets.
        }
        \DONE{DB: Uncertainties of quantities of interest? AM: When I perform the MCMC sampling I do no save all quantities of interest but only the free parameter values} 
        \DONE{DB: I do not understand why p0 and the pressure peaking factor is so differnt. Wouldnt this be strongly visible in the difference of the profiles? AM: As stated in the caption, the pressure profile used in the fixed VMEC case here is the one assumed a priori and not the one inferred in the reconstruction. I can better explain live when we discuss this.} 
    }%
    \label{tab:real-20181016-equilibrium-properties}
    \centering
    \begin{tabular}{lccc}
  \toprule
  Quantity                          & Fixed VMEC & Self-consistent VMEC & Self-consistent NN \\
  \midrule
  $\langle \beta \rangle$ [\%]      & 1.176      & 1.067                & 1.083 (0.064)      \\
  $\gls{ibar}_{\text{axis}}$        & 0.862      & 0.857                & 0.8597 (0.0012)    \\
  $\gls{ibar}_{\text{LCFS}}$        & 0.971      & 0.974                & 0.97414 (0.00014)  \\
  $W_{\text{kin}}$ [kJ]             & 1198       & 1089                 & 1090 (64)          \\ 
  $R_{\text{axis}} (\varphi=0)$ [m] & 5.983      & 5.987                & 5.9876 (0.0028)    \\
  $p_0$ [kPa]                       & 80.0       & 91.1                 & 90.6 (6.7)         \\ 
  pressure peaking factor           & 2.98       & 3.74                 & 3.65 (0.19)        \\ 
  \bottomrule
\end{tabular}

\end{table}

\begin{table}[h!]
    \caption{
      Comparison of selected quantities of the inferred equilibria of experiment\expId{20180808.5} at $t=\SI{20.0}{\second}$ with the three \gls{MHD} models, analogous to~\cref{tab:real-20181016-equilibrium-properties}.
      $\gls{Wdia} = \SI{232.188}{\kilo\joule}$.
    }%
    \label{tab:real-20180808-equilibrium-properties}
    \centering
    \begin{tabular}{lccc}
  \toprule
  Quantity                          & Fixed VMEC & Self-consistent VMEC & Self-consistent NN \\
  \midrule
  $\langle \beta \rangle$ [\%]      & 0.253      & 0.273                & 0.268 (0.034)      \\
  $\gls{ibar}_{\text{axis}}$        & 1.012      & 1.003                & 1.00253 (0.00026)  \\
  $\gls{ibar}_{\text{LCFS}}$        & 1.193      & 1.172                & 1.17385 (0.00012)  \\
  $W_{\text{kin}}$ [kJ]             & 218        & 236                  & 228 (29)           \\ 
  $R_{\text{axis}} (\varphi=0)$ [m] & 5.976      & 5.977                & 5.97641 (0.00092)  \\
  $p_0$ [kPa]                       & 20.0       & 26.3                 & 25.3 (3.2)         \\ 
  pressure peaking factor           & 3.47       & 4.22                 & 4.13 (0.31)        \\ 
  \bottomrule
\end{tabular}

\end{table}

\TODO{Consider to show reconstructed flux surfaces and iota profile.}

\section{Conclusions}\label{sec:conclusions}

In this work,
we found that employing the \gls{NN} model from~\cite{Merlo2023} as ideal-\gls{MHD} equilibrium model in the Minerva Bayesian framework yields inferred plasma parameters compatible with the parameters obtained using \gls{VMEC}.
We have also investigated the robustness of the \gls{NN} accelerated inference using synthetic data,
and we observed that it is robust across diverse \gls{W7X} plasma scenarios.

The inference time can be reduced by more than two orders of magnitude when using the \gls{NN} model.
Moreover,
given the fast computational time,
plasma parameter posterior distributions can be approximated using \gls{MCMC} samples with self-consistent ideal-\gls{MHD} equilibria.

This work has two major limitations:
no magnetic diagnostics (except for the Rogowski coil) are included in the Minerva graph,
owing to the lack of accuracy of the \gls{NN} model in faithfully \replaced[id=AM]{computing}{reproducing} the plasma current density\footnote{the plasma current density depends up to the secon-order derivatives of the equilibrium solution, and the equilibrium solution provided by the \gls{NN} model does not have smooth second-order radial derivatives~\cite{Merlo2023}},
which is required to predict the magnetic diagnostics observations (the plasma contribution to the magnetic field at a location outside the plasma volume depends on a volume integral of the plasma current density~\cite{Hirshman2004});
$T_i \sim \Te$ and $Z_{\text{eff}}\simeq\num{1}$ have been assumed,
as a result,
the approach can be tested only on a limited set of experimental conditions where these assumptions are close to being satisfied.
To infer additional plasma parameters,
more diagnostics can be included in the Minerva graph.
A \gls{XICS} can be used to constrain the ion temperature~\cite{Langenberg2019},
and a spectrometer can be used to infer the effective charge $Z_{\text{eff}}$~\cite{Pavone2019a}.

Using \gls{NN} ideal-\gls{MHD} models,
it is possible to quickly infer plasma parameters while taking into account finite-beta effects in the equilibrium.
If diagnostic data are available,
the \gls{MHD} self-consistent inference of plasma parameters can theoretically be performed between shots,
providing valuable insights for the conduction of operational campaigns.

\section{Author Statement}\label{sec:author-statement}


The contributions to this paper are described using the CRediT taxonomy~\cite{Brand2015}:

\begin{description}
    \item[Andrea Merlo] Conceptualization, Data Curation, Formal Analysis, Investigation, Methodology, Software, Visualization, Writing - original draft, Writing - review \& editing.
    \item[Andrea Pavone] Conceptualization, Data Curation, Formal Analysis, Investigation, Methodology, Software, Writing - review \& editing.
    \item[Daniel Böckenhoff] Methodology, Software, Supervision, Validation, Writing - review \& editing.
    \item[Ekkehard Pasch] Data Curation.
    \item[Kai Jakob Brunner] Data Curation.
    \item[Kian Rahbarnia] Data Curation.
    \item[Jonathan Schilling] Software.
    \item[Udo Höfel] Software.
    \item[Sehyun Kwak] Software.
    \item[Jakob Svensson] Software.
    \item[Thomas Sunn Pedersen] Funding acquisition, Supervision.
\end{description}

\section{Acknowledgement}\label{sec:acknowledgement}



We are indebted to the communities behind the multiple open-source software packages on which this work depends:
hydra~\cite{Yadan2019},
matplotlib~\cite{Hunter2007},
numpy~\cite{Harris2020},
pymc3~\cite{Salvatier2016},
pytorch~\cite{Paszke2019},
pytorch lightning~\cite{Falcon2019},
scipy~\cite{Virtanen2020}.

Financial support by the European Social Fund (ID: ESF/14-BM-A55-0007/19) and the Ministry of Education, Science and Culture of Mecklenburg-Vorpommern,
Germany via the project ``NEISS'' is gratefully acknowledged.
This work has been carried out within the framework of the EUROfusion Consortium,
funded by the European Union via the Euratom Research and Training Programme (Grant Agreement No 101052200 — EUROfusion).
Views and opinions expressed are however those of the author(s) only and do not necessarily reflect those of the European Union or the European Commission.
Neither the European Union nor the European Commission can be held responsible for them.


\ifthenelseproperty{compilation}{backmatter}{%
    \backmatter
}{}

\ifthenelseproperty{compilation}{bibliography}{%
    \printbibliography
}{}

\ifthenelseproperty{compilation}{glossaries}{%
    \ifthenelseproperty{compilation}{acronyms}{%
	\printglossary[type=\acronymtype,style=mcoltree]%
}{}%
\ifthenelseproperty{compilation}{los}{%
	\setlength\extrarowheight{5pt}%
	\printglossary
			[
				title=List of Symbols,
				type=symbols,
				style=customListOfSymbols,
			]
	\setlength\extrarowheight{0pt}%
}{}%

}{}

\ifthenelseproperty{compilation}{lof}{%
    \disabledprotrusion{\listoffigures}
}{}

\ifthenelseproperty{compilation}{lot}{%
    \disabledprotrusion{\listoftables}
}{}

\ifthenelseproperty{compilation}{lol}{%
    \disabledprotrusion{\lstlistoflistings}
}{}

\ifthenelseproperty{compilation}{appendix}{%
    \section{Appendix}\label{sec:appendix}

\subsection{Prior distributions of model parameters}\label{sec:prior-distributions}

\Cref{tab:prior-distributions} lists the prior distribution of the model free parameters.

\begin{table}[h!]
    \caption{
        The prior distributions of the model parameters.
        $U(a, b)$ means a uniform distribution defined on $(a, b)$.
        The bounds of uniform distribution of \Phiedge are set by the distribution of the training data on which the \gls{NN} model has been trained.
    }%
    \label{tab:prior-distributions}
    \centering
    \begin{tabular}{lcc}
        \toprule
        Parameter & Distribution                  & Unit                            \\
        \multicolumn{3}{l}{Electron temperature}                                    \\
        \midrule
        \teZero   & $U(\num{1.0}, \num{6.0})$     & \si{\kilo\eV}                   \\
        \teOne    & $U(\num{e-3}, \num{8.0})$     & -                               \\
        \teTwo    & $U(\num{e-1}, \num{e1})$      & -                               \\
        \teThree  & $U(\num{1.0}, \num{1.33})$    & -                               \\
        \midrule
        \multicolumn{3}{l}{Electron density}                                        \\
        \midrule
        \neZero   & $U(\num{1.0}, \num{15.0})$    & $10^{19}$ \si{\meter\tothe{-3}} \\
        \neOne    & $U(\num{e-3}, \num{8.0})$     & -                               \\
        \neTwo    & $U(\num{e-1}, \num{e1})$      & -                               \\
        \neThree  & $U(\num{1.0}, \num{1.33})$    & -                               \\
        \midrule
        \multicolumn{3}{l}{Net toroidal flux}                                       \\
        \midrule
        \Phiedge  & $U(\num{-2.31}, \num{-1.35})$ & \si{\weber}                     \\
        \bottomrule
    \end{tabular}
\end{table}

\subsection{Modeling the uncertainty due to the \gls{TS} laser misalignment}\label{sec:misalignment}

\TODO{Consider to also plot the line integrated density in the first figure, or something like the overview plot.}

This section briefly introduces how the uncertainty due to the \gls{TS} laser misalignment is modeled in the Minerva graph.

To estimate the uncertainties due to the \gls{TS} laser misalignment,
we make use of the fact that the inferred electron density linearly depends on the \gls{TS} observations (see~\cref{eq:ts}).
In this work,
we assume that only two factors affect electron density:
a change in the macroscopic plasma state (\eg, pellet injection, \gls{NBI} heating) or a shift in the \gls{TS} laser.
To rule out the former,
we consider time windows from multiple shots where all the other plasma parameters apart from the inferred electron density and temperature from the \gls{TS} are constant.
Previous Minerva reconstruction are used for this estimation.
\Cref{tab:shots} lists the considered shots and time windows.

\begin{table}[h!]
    \caption{
        The \gls{W7X} shots and time windows used to inform the modeling of the uncertainty due to the \gls{TS} laser misalignment.
        The magnetic configuration and line integrated density
        (as measured by the \gls{DI})
        are also listed.
    }%
    \label{tab:shots}
    \centering
    \begin{tabular}{lccc}
        \toprule
        Shot                 & Time Window [s]     & Configuration & $\int n_e dl$ [$10^{19}$m$^{-2}$] \\%
        \midrule
        \expId{20180808.005} & \num{45} - \num{54} & FTM+\num{252} & \num{3}                           \\%
        \expId{20180809.039} & \num{3} - \num{4}   & KKM+\num{252} & \num{5}                           \\%
        \expId{20181016.013} & \num{4} - \num{15}  & EJM+\num{262} & \num{12}                          \\%
        \expId{20181016.014} & \num{4} - \num{19}  & EJM+\num{262} & \num{12}                          \\%
        \expId{20181016.026} & \num{16} - \num{36} & EJM+\num{262} & \num{6}                           \\%
        \expId{20181017.016} & \num{12} - \num{32} & FTM+\num{251} & \num{5}                           \\%
        \expId{20181017.025} & \num{8} - \num{28}  & EJM+\num{262} & \num{12}                          \\%
        \bottomrule
    \end{tabular}
\end{table}

\Cref{fig:ne-time} shows an example of these shot,
in which the inferred electron density of three \gls{TS} volumes is plotted over time.
Given that all the other plasma parameters are constant in this time window,
we assume that even the actual electron density is constant as well,
and that the variations in the inferred values are due to the \gls{TS} laser shifts.
Therefore,
we assume the time average (over the whole time window) of the electron density to be the \textit{actual} density,
and we consider the standard variation (in time) of the inferred electron density as a proxy to model the uncertainty due to the \gls{TS} misalignment.

\begin{figure}[h!]
    \centering
    \igraph[]{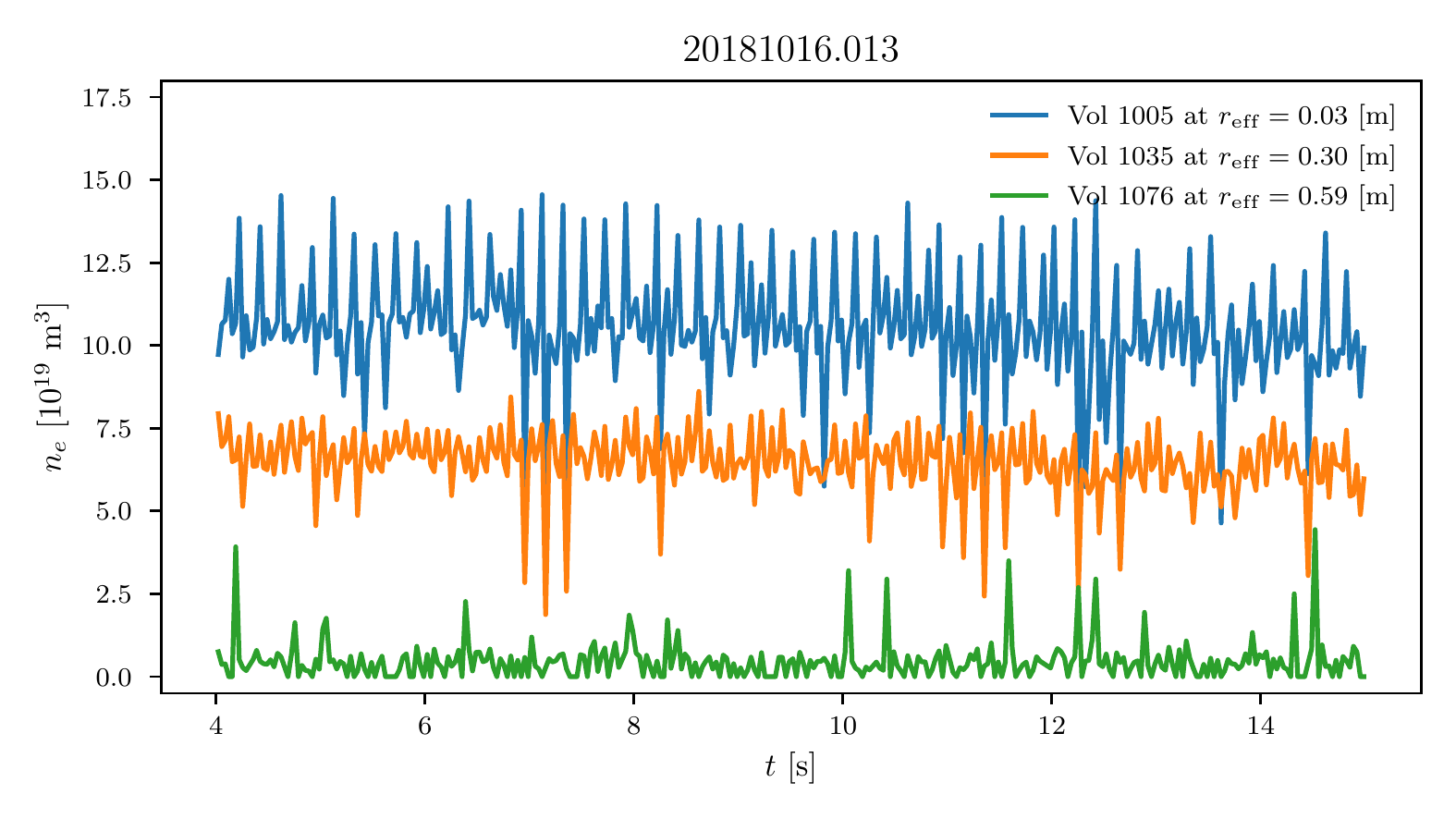}%
    \caption{%
        The inferred electron density for three \gls{TS} volumes in stationary plasma conditions.
        In the selected time window,
        all other macroscopic plasma parameters are nearly constant.
        Volumes in the core (\num{1005}),
        mid-radius (\num{1035}),
        and plasma edge (\num{1076}) are depicted.
        Data from previous Minerva reconstructions are used.
    }%
    \label{fig:ne-time}
\end{figure}

How does the standard deviation depend on the inferred value?
\Cref{fig:ne-std-mean} shows the relationship between the standard deviation and average electron density for all the evaluated shots.
When considering only the volumes inside the \gls{LCFS} (stars),
a simple linear model reasonably describes the data:
$\sigma_{n_e} \sim \mathcal{TN}(\alpha \mu_{n_e}, \sigma_{\sigma_{n_e}})$,
where $\mathcal{TN}$ stands for a truncated normal distribution.
However,
the volumes outside the \gls{LCFS} (dots) do not seem to follow the same trend.
The posterior distributions of $\alpha$ and $\sigma_{\sigma_{n_e}}$ are numerically estimated with \num{50000} \gls{MCMC} samples.
The mean of the posterior distribution yields $\alpha = \num{0.18}$.

\begin{figure}[h!]
    \centering
    \igraph[]{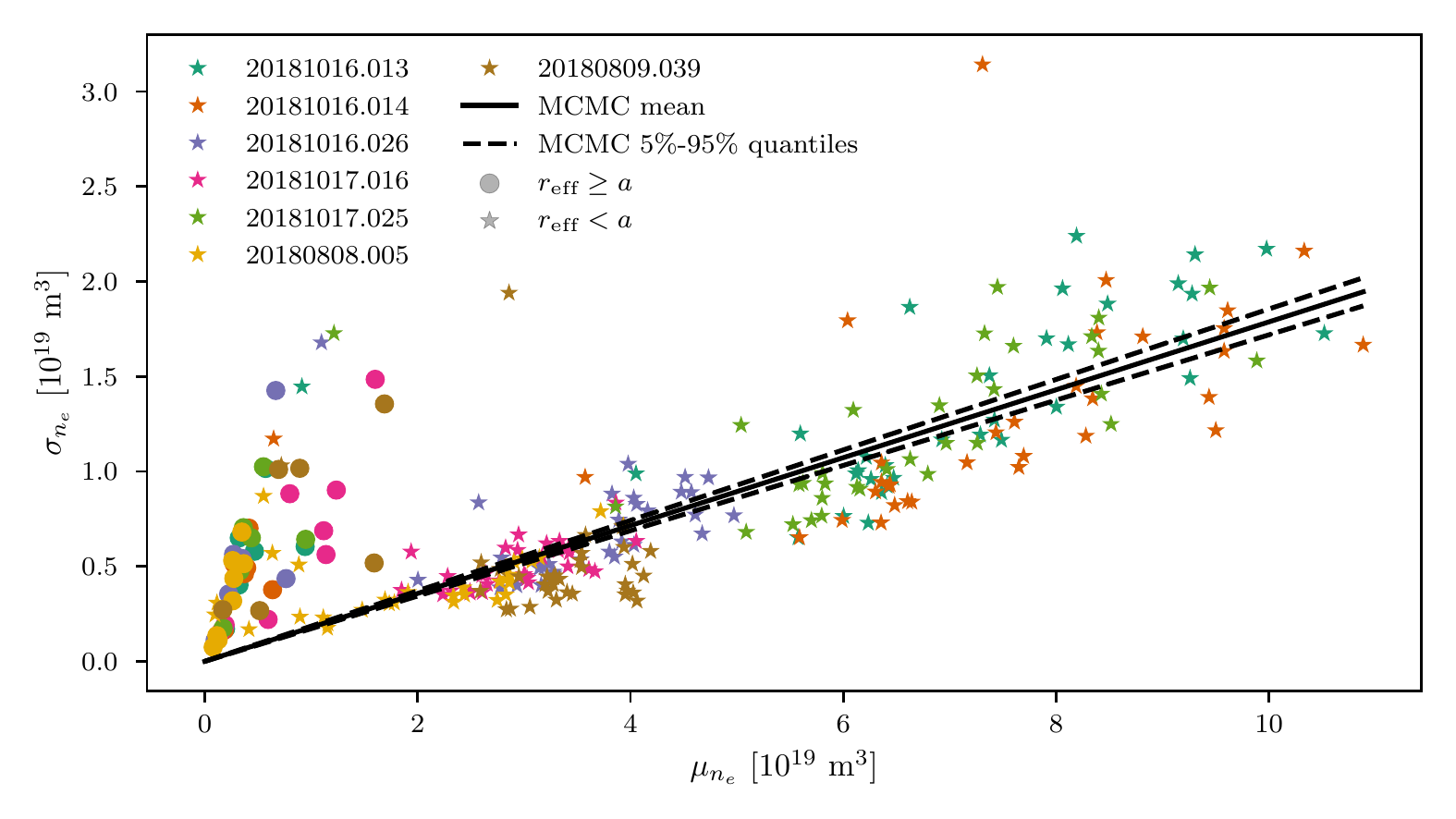}%
    \caption{%
        The inferred electron density standard deviation as a function of the average value in each time window.
        Colors indicate different shots,
        stars indicate \gls{TS} volumes inside the \gls{LCFS},
        and dots indicate \gls{TS} volumes outside the \gls{LCFS}.
        Given a simple linear model of the form $\sigma_{n_e} \sim \mathcal{TN}(\alpha \mu_{n_e}, \sigma_{\sigma_{n_e}})$,
        where $\mathcal{TN}$ stands for a truncated normal distribution,
        the solid and dashed lines represent the mean and \SI{5}{\percent}-\SI{95}{\percent} quantiles of $\alpha$,
        respectively.
    }%
    \label{fig:ne-std-mean}
\end{figure}

Finally,
we model the uncertainty of a generic \gls{TS} volume due to \gls{TS} laser misalignment using the linear relationship between the electron density and the \gls{TS} observed data as follows:

\begin{align}
    \sigma_{\text{laser}}^i = \max(\sigma_0, \alpha D_{\text{TS}}^i) ,
\end{align}

where $\sigma_0$ accounts for the standard deviation of the \gls{TS} volumes outside the \gls{LCFS} not well modeled by the linear relationship.
$\sigma_0$ is usually set to a small value compared to the observed \gls{TS} data:
$\sigma_0 \simeq \frac{2.5}{100} \max_i (D_{\text{TS}}^i)$.
}{}
\ifthenelseproperty{compilation}{listofpublications}{%
    \begin{refsection}
	\newrefcontext[sorting=ndymdt]
	\nocite{*}
    \ifthenelseproperty{compilation}{clsdefineschapter}{%
		\ifKOMA
			\addchap[Publications as first author]{Publications as first author}\label{sec:publications_as_first_author}
		\else
	    	\chapter[Publications as first author]{Publications as first author}\label{sec:publications_as_first_author}
	    \fi
    }{%
		\ifKOMA
			\addsec[Publications as first author]{Publications as first author}\label{sec:publications_as_first_author}
		\else
	    	\section[Publications as first author]{Publications as first author}\label{sec:publications_as_first_author}
	    \fi
    }
	\printbibliography[
						keyword=firstAuthor,
						keyword=refereed,
						heading=subbibliography,
						title={Peer-reviewed articles},
						resetnumbers=true
					]
\end{refsection}
\begin{refsection}
	\newrefcontext[sorting=ndymdt]
	\nocite{*}
    \ifthenelseproperty{compilation}{clsdefineschapter}{%
		\ifKOMA
			\addchap[Publications as coauthor]{Publications as coauthor}\label{sec:publications_as_coauthor}
		\else
	    	\chapter[Publications as coauthor]{Publications as coauthor}\label{sec:publications_as_coauthor}
	    \fi
    }{%
		\ifKOMA
			\addsec[Publications as coauthor]{Publications as coauthor}\label{sec:publications_as_coauthor}
		\else
	    	\section[Publications as coauthor]{Publications as coauthor}\label{sec:publications_as_coauthor}
	    \fi
    }
	\setboolean{isCoauthorList}{true}
	\printbibliography[
						keyword=coAuthor,
						keyword=refereed,
						heading=subbibliography,
						title={Peer-reviewed articles},
						resetnumbers=true
					]

@inproceedings{Aaronson2011,
  title = {The {{Computational Complexity}} of {{Linear Optics}}},
  booktitle = {Proceedings of the {{Forty-third Annual ACM Symposium}} on {{Theory}} of {{Computing}}},
  author = {Aaronson, Scott and Arkhipov, Alex},
  year = {2011},
  series = {{{STOC}} '11},
  pages = {333--342},
  publisher = {{ACM}},
  address = {{New York, NY, USA}},
  doi = {10.1145/1993636.1993682},
  urldate = {2019-09-01},
  isbn = {978-1-4503-0691-1}
}

@article{Abbate2021,
  title = {Data-Driven Profile Prediction for {{DIII-D}}},
  author = {Abbate, J. and Conlin, R. and Kolemen, E.},
  year = {2021},
  month = mar,
  journal = {Nuclear Fusion},
  volume = {61},
  number = {4},
  pages = {046027},
  publisher = {{IOP Publishing}},
  issn = {0029-5515},
  doi = {10.1088/1741-4326/abe08d},
  urldate = {2021-10-06},
  langid = {english}
}

@article{Abbate2023,
  title = {A General Infrastructure for Data-Driven Control Design and Implementation in Tokamaks},
  author = {Abbate, Joseph and Conlin, Rory and Shousha, Ricardo and Erickson, Keith and Kolemen, Egemen},
  year = {2023},
  month = feb,
  journal = {Journal of Plasma Physics},
  volume = {89},
  number = {1},
  pages = {895890102},
  publisher = {{Cambridge University Press}},
  issn = {0022-3778, 1469-7807},
  doi = {10.1017/S0022377822001040},
  urldate = {2023-01-23},
  langid = {english}
}

@article{Abdou2020,
  title = {Physics and Technology Considerations for the Deuterium\textendash Tritium Fuel Cycle and Conditions for Tritium Fuel Self Sufficiency},
  author = {Abdou, Mohamed and Riva, Marco and Ying, Alice and Day, Christian and Loarte, Alberto and Baylor, L. R. and Humrickhouse, Paul and Fuerst, Thomas F. and Cho, Seungyon},
  year = {2020},
  month = nov,
  journal = {Nuclear Fusion},
  volume = {61},
  number = {1},
  pages = {013001},
  publisher = {{IOP Publishing}},
  issn = {0029-5515},
  doi = {10.1088/1741-4326/abbf35},
  urldate = {2022-06-02},
  langid = {english}
}

@article{Abrams2021,
  title = {Evaluation of Silicon Carbide as a Divertor Armor Material in {{DIII-D H-mode}} Discharges},
  author = {Abrams, T. and Bringuier, S. and Thomas, D. M. and Sinclair, G. and Gonderman, S. and Holland, L. and Rudakov, D. L. and Wilcox, R. S. and Unterberg, E. A. and Scotti, F.},
  year = {2021},
  month = apr,
  journal = {Nuclear Fusion},
  volume = {61},
  number = {6},
  pages = {066005},
  publisher = {{IOP Publishing}},
  issn = {0029-5515},
  doi = {10.1088/1741-4326/abecee},
  urldate = {2021-06-24},
  langid = {english}
}

@article{Achiam2017,
  title = {Constrained {{Policy Optimization}}},
  author = {Achiam, Joshua and Held, David and Tamar, Aviv and Abbeel, Pieter},
  year = {2017},
  month = may,
  journal = {arXiv:1705.10528 [cs]},
  eprint = {1705.10528},
  primaryclass = {cs},
  urldate = {2020-10-14},
  archiveprefix = {arxiv},
  langid = {english}
}

@article{Achilles2004,
  title = {Photon-Number-Resolving Detection Using Time-Multiplexing},
  author = {Achilles, Daryl and Silberhorn, Christine and Sliwa, Cezary and Banaszek, Konrad and Walmsley, Ian A. and Fitch, Michael J. and Jacobs, Bryan C. and Pittman, Todd B. and Franson, James D.},
  year = {2004},
  month = jun,
  journal = {Journal of Modern Optics},
  volume = {51},
  number = {9-10},
  pages = {1499--1515},
  issn = {0950-0340},
  doi = {10.1080/09500340408235288},
  urldate = {2019-08-31}
}

@techreport{Agency2012,
  type = {Text},
  title = {Fusion {{Physics}}},
  author = {Agency, International Atomic Energy},
  year = {2012},
  journal = {Fusion Physics},
  pages = {1--1129},
  institution = {{International Atomic Energy Agency}},
  urldate = {2023-03-13},
  isbn = {9789201304100},
  langid = {english}
}

@article{Aharonov2008,
  title = {Fault-{{Tolerant Quantum Computation}} with {{Constant Error Rate}}},
  author = {Aharonov, D. and {Ben-Or}, M.},
  year = {2008},
  month = jan,
  journal = {SIAM Journal on Computing},
  volume = {38},
  number = {4},
  pages = {1207--1282},
  issn = {0097-5397},
  doi = {10.1137/S0097539799359385},
  urldate = {2019-09-01}
}

@article{Albergo,
  title = {Exploring Adversarial and Variational Learning of Particle Physics Data},
  author = {Albergo, Michael S and College, Downing},
  pages = {118},
  langid = {english}
}

@article{Albert2020,
  title = {Accelerated Methods for Direct Computation of Fusion Alpha Particle Losses within, Stellarator Optimization},
  author = {Albert, Christopher G. and Kasilov, Sergei V. and Kernbichler, Winfried},
  year = {2020},
  journal = {Journal of Plasma Physics},
  volume = {86},
  number = {2},
  publisher = {{Cambridge University Press}},
  doi = {10.1017/S0022377820000203}
}

@article{Albert2020a,
  title = {Symplectic Integration with Non-Canonical Quadrature for Guiding-Center Orbits in Magnetic Confinement Devices},
  author = {Albert, Christopher G. and Kasilov, Sergei V. and Kernbichler, Winfried},
  year = {2020},
  month = feb,
  journal = {Journal of Computational Physics},
  volume = {403},
  pages = {109065},
  issn = {0021-9991},
  doi = {10.1016/j.jcp.2019.109065},
  urldate = {2022-07-24},
  langid = {english}
}

@article{Albertsson2018,
  title = {Machine {{Learning}} in {{High Energy Physics Community White Paper}}},
  author = {Albertsson, Kim and Altoe, Piero and Anderson, Dustin and Andrews, Michael and Araque Espinosa, Juan Pedro and Aurisano, Adam and Basara, Laurent and Bevan, Adrian and Bhimji, Wahid and Bonacorsi, Daniele and Calafiura, Paolo and Campanelli, Mario and Capps, Louis and Carminati, Federico and Carrazza, Stefano and Childers, Taylor and Coniavitis, Elias and Cranmer, Kyle and David, Claire and Davis, Douglas and Duarte, Javier and Erdmann, Martin and Eschle, Jonas and Farbin, Amir and Feickert, Matthew and Castro, Nuno Filipe and Fitzpatrick, Conor and Floris, Michele and Forti, Alessandra and {Garra-Tico}, Jordi and Gemmler, Jochen and Girone, Maria and Glaysher, Paul and Gleyzer, Sergei and Gligorov, Vladimir and Golling, Tobias and Graw, Jonas and Gray, Lindsey and Greenwood, Dick and Hacker, Thomas and Harvey, John and Hegner, Benedikt and Heinrich, Lukas and Hooberman, Ben and Junggeburth, Johannes and Kagan, Michael and Kane, Meghan and Kanishchev, Konstantin and Karpi{\'n}ski, Przemys{\l}aw and Kassabov, Zahari and Kaul, Gaurav and Kcira, Dorian and Keck, Thomas and Klimentov, Alexei and Kowalkowski, Jim and Kreczko, Luke and Kurepin, Alexander and Kutschke, Rob and Kuznetsov, Valentin and K{\"o}hler, Nicolas and Lakomov, Igor and Lannon, Kevin and Lassnig, Mario and Limosani, Antonio and Louppe, Gilles and Mangu, Aashrita and Mato, Pere and Meinhard, Helge and Menasce, Dario and Moneta, Lorenzo and Moortgat, Seth and Narain, Meenakshi and Neubauer, Mark and Newman, Harvey and Pabst, Hans and Paganini, Michela and Paulini, Manfred and Perdue, Gabriel and Perez, Uzziel and Picazio, Attilio and Pivarski, Jim and Prosper, Harrison and Psihas, Fernanda and Radovic, Alexander and Reece, Ryan and Rinkevicius, Aurelius and Rodrigues, Eduardo and Rorie, Jamal and Rousseau, David and Sauers, Aaron and Schramm, Steven and Schwartzman, Ariel and Severini, Horst and Seyfert, Paul and Siroky, Filip and Skazytkin, Konstantin and Sokoloff, Mike and Stewart, Graeme and Stienen, Bob and Stockdale, Ian and Strong, Giles and Thais, Savannah and Tomko, Karen and Upfal, Eli and Usai, Emanuele and Ustyuzhanin, Andrey and Vala, Martin and Vallecorsa, Sofia and Vasel, Justin and Verzetti, Mauro and {Vilas{\'i}s-Cardona}, Xavier and Vlimant, Jean-Roch and Vukotic, Ilija and Wang, Sean-Jiun and Watts, Gordon and Williams, Michael and Wu, Wenjing and Wunsch, Stefan and Zapata, Omar},
  year = {2018},
  month = sep,
  journal = {Journal of Physics: Conference Series},
  volume = {1085},
  pages = {022008},
  issn = {1742-6588, 1742-6596},
  doi = {10.1088/1742-6596/1085/2/022008},
  urldate = {2020-10-14}
}

@article{Alcuson2020,
  title = {Suppression of Electrostatic Micro-Instabilities in Maximum-{{J}} Stellarators},
  author = {Alcus{\'o}n, J. A. and Xanthopoulos, P. and Plunk, G. G. and Helander, P. and Wilms, F. and Turkin, Y. and von Stechow, A. and Grulke, O.},
  year = {2020},
  month = jan,
  journal = {Plasma Physics and Controlled Fusion},
  volume = {62},
  number = {3},
  pages = {035005},
  publisher = {{IOP Publishing}},
  issn = {0741-3335},
  doi = {10.1088/1361-6587/ab630e},
  urldate = {2021-04-27},
  langid = {english}
}

@article{Aleynikova2018,
  title = {Kinetic Ballooning Modes in Tokamaks and Stellarators},
  author = {Aleynikova, K. and Zocco, A. and Xanthopoulos, P. and Helander, P. and N{\"u}hrenberg, C.},
  year = {2018},
  journal = {Journal of Plasma Physics},
  volume = {84},
  number = {6},
  publisher = {{Cambridge University Press}},
  doi = {10.1017/S0022377818001186}
}

@article{Aleynikova2021,
  title = {Model for Current Drive Induced Crash Cycles in {{W7-X}}},
  author = {Aleynikova, K. and Hudson, S. R. and Helander, P. and Kumar, A. and Geiger, J. and Hirsch, M. and Loizu, J. and N{\"u}hrenberg, C. and Rahbarnia, K. and Qu, Z. and Gao, Y. and Thomsen, H. and Turkin, Y. and Zanini, M. and Team, the W7-X.},
  year = {2021},
  month = dec,
  journal = {Nuclear Fusion},
  volume = {61},
  number = {12},
  pages = {126040},
  publisher = {{IOP Publishing}},
  issn = {0029-5515},
  doi = {10.1088/1741-4326/ac2ab9},
  urldate = {2022-01-10},
  langid = {english}
}

@article{Aleynikova2022,
  title = {Influence of Magnetic Configuration Properties on Kinetic Ballooning Modes in {{W7-X}}},
  author = {Aleynikova, K. and Zocco, A. and Geiger, J.},
  year = {2022},
  month = aug,
  journal = {Journal of Plasma Physics},
  volume = {88},
  number = {4},
  pages = {905880411},
  publisher = {{Cambridge University Press}},
  issn = {0022-3778, 1469-7807},
  doi = {10.1017/S0022377822000745},
  urldate = {2022-09-27},
  langid = {english}
}

@article{Ali2019,
  title = {Initial Results from the Hotspot Detection Scheme for Protection of Plasma Facing Components in {{Wendelstein}} 7-{{X}}},
  author = {Ali, A. and Niemann, H. and Jakubowski, M. and Pedersen, T. Sunn and Neu, R. and Corre, Y. and Drewelow, P. and Sitjes, A. Puig and Wurden, G. and Pisano, F. and Cannas, B. and Gao, Y. and {\'S}l{\k{e}}czka, M.},
  year = {2019},
  month = may,
  journal = {Nuclear Materials and Energy},
  volume = {19},
  pages = {335--339},
  issn = {23521791},
  doi = {10.1016/j.nme.2019.03.006},
  urldate = {2020-10-14},
  langid = {english}
}

@misc{Alladio2017,
  title = {The {{PROTO-SPHERA}} Experiment, an Innovative Confinement Scheme for {{Fusion}}},
  author = {Alladio, Franco and Micozzi, P. and Apruzzese, G.M. and Boncagni, L. and D'Arcangelo, Ocleto and Giovannozzi, Edmondo and Grosso, L.A. and Iafrati, Matteo and Lampasi, Alessandro and Maffia, G. and Mancuso, A. and Piergotti, V. and Rocchi, G. and Sibio, A. and Tilia, B. and Tudisco, Onofrio and Zanza, V.},
  year = {2017},
  month = sep,
  doi = {10.13140/RG.2.2.26419.32801}
}

@article{Allen1992,
  title = {Neural Network Approach to Energy Confinement Scaling in {{Tokamaks}}},
  author = {Allen, L. and Bishop, C. M.},
  year = {1992},
  month = jul,
  journal = {Plasma Physics and Controlled Fusion},
  volume = {34},
  number = {7},
  pages = {1291},
  issn = {0741-3335},
  doi = {10.1088/0741-3335/34/7/008},
  urldate = {2023-03-22},
  langid = {english}
}

@book{Alley2003,
  title = {The Craft of Scientific Presentations},
  author = {Alley, Michael},
  year = {2003},
  publisher = {{Springer}}
}

@article{Alonso2021,
  title = {Physics Design Point of High-Field Stellarator Reactors},
  author = {Alonso, J. A. and Calvo, I. and Carralero, D. and Velasco, J. L. and {Garc{\'i}a-Rega{\~n}a}, J. M. and Palermo, I. and Rapisarda, D.},
  year = {2021},
  month = oct,
  journal = {arXiv:2109.15189 [physics]},
  eprint = {2109.15189},
  primaryclass = {physics},
  urldate = {2021-12-05},
  archiveprefix = {arxiv}
}

@article{Alvarez1957,
  title = {Catalysis of Nuclear Reactions by {$\mu$} Mesons},
  author = {Alvarez, Luis W. and Bradner, H. and Crawford Jr, F. S. and Crawford, J. A. and {Falk-Vairant}, P. and Good, M. L. and Gow, J. D. and Rosenfeld, A. H. and Solmitz, F. and Stevenson, M. L.},
  year = {1957},
  journal = {Physical Review},
  volume = {105},
  number = {3},
  pages = {1127},
  publisher = {{APS}},
  doi = {10.1103/PhysRev.105.1127}
}

@book{Ames1992,
  title = {Numerical Methods for Partial Differential Equations},
  author = {Ames, William F.},
  year = {1992},
  series = {Computer Science and Scientific Computing},
  edition = {3rd ed},
  publisher = {{Academic Press}},
  address = {{Boston}},
  isbn = {978-0-12-056761-4},
  langid = {english},
  lccn = {QA374 .A46 1992}
}

@article{Amini2019,
  title = {Application of {{Machine Learning}} and {{Artificial Intelligence}} in {{Proxy Modeling}} for {{Fluid Flow}} in {{Porous Media}}},
  author = {Amini, Shohreh and Mohaghegh, Shahab},
  year = {2019},
  month = jul,
  journal = {Fluids},
  volume = {4},
  number = {3},
  pages = {126},
  issn = {2311-5521},
  doi = {10.3390/fluids4030126},
  urldate = {2020-10-14},
  langid = {english}
}

@article{Amini2020,
  title = {Deep {{Evidential Regression}}},
  author = {Amini, Alexander and Schwarting, Wilko and Soleimany, Ava and Rus, Daniela},
  year = {2020},
  month = nov,
  journal = {arXiv:1910.02600 [cs, stat]},
  eprint = {1910.02600},
  primaryclass = {cs, stat},
  urldate = {2022-02-28},
  archiveprefix = {arxiv}
}

@incollection{Amrouche2020,
  title = {The {{Tracking Machine Learning}} Challenge : {{Accuracy}} Phase},
  shorttitle = {The {{Tracking Machine Learning}} Challenge},
  author = {Amrouche, Sabrina and Basara, Laurent and Calafiura, Paolo and Estrade, Victor and Farrell, Steven and Ferreira, Diogo R. and Finnie, Liam and Finnie, Nicole and Germain, C{\'e}cile and Gligorov, Vladimir Vava and Golling, Tobias and Gorbunov, Sergey and Gray, Heather and Guyon, Isabelle and Hushchyn, Mikhail and Innocente, Vincenzo and Kiehn, Moritz and Moyse, Edward and Puget, Jean-Francois and Reina, Yuval and Rousseau, David and Salzburger, Andreas and Ustyuzhanin, Andrey and Vlimant, Jean-Roch and Wind, Johan Sokrates and Xylouris, Trian and Yilmaz, Yetkin},
  year = {2020},
  eprint = {1904.06778},
  primaryclass = {hep-ex, physics:physics},
  pages = {231--264},
  doi = {10.1007/978-3-030-29135-8_9},
  urldate = {2022-10-07},
  archiveprefix = {arxiv}
}

@article{Anand,
  title = {{{IMPLEMENTATION OF}} 3-{{D EFFECTS OF THE PLASMA-FACING COMPONENTS IN A}} 2-{{D REAL-TIME MODEL-BASED APPROACH FOR WALL HEAT FLUX CONTROL ON ITER}}},
  author = {Anand, H and Pitts, R A and Nespoli, F and Galperti, C and Coda, S and Labit, B and Zabeo, L and Nunes, I and Brank, M},
  pages = {8},
  langid = {english}
}

@misc{Anderson2020,
  title = {Meaningful Uncertainties from Deep Neural Network Surrogates of Large-Scale Numerical Simulations},
  author = {Anderson, Gemma J. and Gaffney, Jim A. and Spears, Brian K. and Bremer, Peer-Timo and Anirudh, Rushil and Thiagarajan, Jayaraman J.},
  year = {2020},
  month = oct,
  number = {arXiv:2010.13749},
  eprint = {arXiv:2010.13749},
  publisher = {{arXiv}},
  doi = {10.48550/arXiv.2010.13749},
  urldate = {2022-10-06},
  archiveprefix = {arxiv}
}

@article{Andreeva2002,
  title = {Vacuum Magnetic Configurations of {{Wendelstein}} 7-{{X}}},
  author = {Andreeva, Tamara},
  year = {2002},
  publisher = {{Max-Planck-Institut f\"ur Plasmaphysik}}
}

@article{Andreeva2019,
  title = {Equilibrium Evaluation for {{Wendelstein}} 7-{{X}} Experiment Programs in the First Divertor Phase},
  author = {Andreeva, T. and Alonso, J. A. and Bozhenkov, S. and Brandt, C. and Endler, M. and Fuchert, G. and Geiger, J. and Grahl, M. and Klinger, T. and Krychowiak, M. and Langenberg, A. and Lazerson, S. and Neuner, U. and Rahbarnia, K. and Pablant, N. and Pavone, A. and Schilling, J. and Schmitt, J. and Thomsen, H. and Turkin, Y.},
  year = {2019},
  month = sep,
  journal = {Fusion Engineering and Design},
  series = {{{SI}}:{{SOFT-30}}},
  volume = {146},
  pages = {299--302},
  issn = {0920-3796},
  doi = {10.1016/j.fusengdes.2018.12.050},
  urldate = {2023-02-28},
  langid = {english}
}

@article{Angioni2003,
  title = {Effects of Localized Electron Heating and Current Drive on the Sawtooth Period},
  author = {Angioni, C. and Goodman, T. P. and Henderson, M. A. and Sauter, O.},
  year = {2003},
  month = may,
  journal = {Nuclear Fusion},
  volume = {43},
  number = {6},
  pages = {455--468},
  publisher = {{IOP Publishing}},
  issn = {0029-5515},
  doi = {10.1088/0029-5515/43/6/308},
  urldate = {2022-01-10},
  langid = {english}
}

@article{Anitescu2019a,
  title = {Artificial {{Neural Network Methods}} for the {{Solution}} of {{Second Order Boundary Value Problems}}},
  author = {Anitescu, Cosmin and Atroshchenko, Elena and Alajlan, Naif and Rabczuk, Timon},
  year = {2019},
  journal = {Computers, Materials \& Continua},
  volume = {59},
  number = {1},
  pages = {345--359},
  issn = {1546-2226},
  doi = {10.32604/cmc.2019.06641},
  urldate = {2021-04-07},
  langid = {english}
}

@article{Antonsen2019,
  title = {Adjoint Approach to Calculating Shape Gradients for Three-Dimensional Magnetic Confinement Equilibria},
  author = {Antonsen, Thomas and Paul, Elizabeth J. and Landreman, Matt},
  year = {2019},
  month = apr,
  journal = {Journal of Plasma Physics},
  volume = {85},
  number = {2},
  publisher = {{Cambridge University Press}},
  issn = {0022-3778, 1469-7807},
  doi = {10.1017/S0022377819000254},
  urldate = {2021-05-12},
  langid = {english}
}

@article{Ardente2005,
  title = {Life Cycle Assessment of a Solar Thermal Collector: Sensitivity Analysis, Energy and Environmental Balances},
  shorttitle = {Life Cycle Assessment of a Solar Thermal Collector},
  author = {Ardente, Fulvio and Beccali, Giorgio and Cellura, Maurizio and Brano, Valerio Lo},
  year = {2005},
  journal = {Renewable Energy},
  volume = {30},
  number = {2},
  pages = {109--130},
  publisher = {{Elsevier}},
  doi = {10.1016/j.renene.2004.05.006}
}

@book{Ariola2016,
  title = {Magnetic {{Control}} of {{Tokamak Plasmas}}},
  author = {Ariola, Marco and Pironti, Alfredo},
  year = {2016},
  series = {Advances in {{Industrial Control}}},
  publisher = {{Springer International Publishing}},
  address = {{Cham}},
  doi = {10.1007/978-3-319-29890-0},
  urldate = {2020-10-14},
  isbn = {978-3-319-29888-7 978-3-319-29890-0},
  langid = {english}
}

@article{ArmstrongMcKay2022,
  title = {Exceeding 1.5\textdegree{{C}} Global Warming Could Trigger Multiple Climate Tipping Points},
  author = {Armstrong McKay, David I. and Staal, Arie and Abrams, Jesse F. and Winkelmann, Ricarda and Sakschewski, Boris and Loriani, Sina and Fetzer, Ingo and Cornell, Sarah E. and Rockstr{\"o}m, Johan and Lenton, Timothy M.},
  year = {2022},
  month = sep,
  journal = {Science},
  volume = {377},
  number = {6611},
  pages = {eabn7950},
  publisher = {{American Association for the Advancement of Science}},
  doi = {10.1126/science.abn7950},
  urldate = {2022-09-20}
}

@article{Artaud2018,
  title = {Metis: A Fast Integrated Tokamak Modelling Tool for Scenario Design},
  shorttitle = {Metis},
  author = {Artaud, Jean-Fran{\c c}ois and Imbeaux, Fr{\'e}d{\'e}ric and Garcia, J. and Giruzzi, G. and Aniel, T. and Basiuk, V. and B{\'e}coulet, A. and Bourdelle, C. and Buravand, Y. and Decker, J.},
  year = {2018},
  journal = {Nuclear Fusion},
  volume = {58},
  number = {10},
  pages = {105001},
  publisher = {{IOP Publishing}},
  doi = {10.1088/1741-4326/aad5b1}
}

@article{Artola2021,
  title = {{{3D}} Simulations of Vertical Displacement Events in Tokamaks: {{A}} Benchmark of {{M3D-C1}}, {{NIMROD}}, and {{JOREK}}},
  shorttitle = {{{3D}} Simulations of Vertical Displacement Events in Tokamaks},
  author = {Artola, F. J. and Sovinec, C. R. and Jardin, S. C. and Hoelzl, M. and Krebs, I. and Clauser, C.},
  year = {2021},
  month = may,
  journal = {Physics of Plasmas},
  volume = {28},
  number = {5},
  pages = {052511},
  publisher = {{American Institute of Physics}},
  issn = {1070-664X},
  doi = {10.1063/5.0037115},
  urldate = {2021-12-14}
}

@article{Asunta2019,
  title = {{{ST40}} Data and Control},
  author = {Asunta, O. and Buxton, P.F. and Lister, J.B. and Pinkney, E. and Scott, L.},
  year = {2019},
  month = sep,
  journal = {Fusion Engineering and Design},
  volume = {146},
  pages = {2194--2198},
  issn = {09203796},
  doi = {10.1016/j.fusengdes.2019.03.151},
  urldate = {2020-10-14},
  langid = {english}
}

@article{Attenberger1987,
  title = {Some Practical Considerations Involving Spectral Representations of {{3D}} Plasma Equilibria},
  author = {Attenberger, S. E and Houlberg, W. A and Hirshman, S. P},
  year = {1987},
  month = oct,
  journal = {Journal of Computational Physics},
  volume = {72},
  number = {2},
  pages = {435--448},
  issn = {0021-9991},
  doi = {10.1016/0021-9991(87)90092-1},
  urldate = {2021-10-13},
  langid = {english}
}

@article{Avaria2019,
  title = {Hard {{X-Ray Emission Detection Using Deep Learning Analysis}} of the {{Radiated UHF Electromagnetic Signal From}} a {{Plasma Focus Discharge}}},
  author = {Avaria, Gonzalo and {Ardila-Rey}, Jorge and Davis, Sergio and Orellana, Luis and Cevallos, Benjamin and Pavez, Cristian and Soto, Leopoldo},
  year = {2019},
  journal = {IEEE Access},
  volume = {7},
  pages = {74899--74908},
  issn = {2169-3536},
  doi = {10.1109/ACCESS.2019.2921288},
  urldate = {2020-10-14},
  langid = {english}
}

@article{Avrutskiy2020,
  title = {Neural Networks Catching up with Finite Differences in Solving Partial Differential Equations in Higher Dimensions},
  author = {Avrutskiy, V. I.},
  year = {2020},
  month = sep,
  journal = {Neural Computing and Applications},
  volume = {32},
  number = {17},
  eprint = {1712.05067},
  pages = {13425--13440},
  issn = {0941-0643, 1433-3058},
  doi = {10/gjnkn6},
  urldate = {2021-04-06},
  archiveprefix = {arxiv}
}

@article{Ayed2019,
  title = {Learning {{Dynamical Systems}} from {{Partial Observations}}},
  author = {Ayed, Ibrahim and {de B{\'e}zenac}, Emmanuel and Pajot, Arthur and Brajard, Julien and Gallinari, Patrick},
  year = {2019},
  month = feb,
  journal = {arXiv:1902.11136 [physics]},
  eprint = {1902.11136},
  primaryclass = {physics},
  urldate = {2020-10-14},
  archiveprefix = {arxiv},
  langid = {english}
}

@article{Bader2017,
  ids = {Bader2017a},
  title = {{{HSX}} as an Example of a Resilient Non-Resonant Divertor},
  author = {Bader, A. and Boozer, A. H. and Hegna, C. C. and Lazerson, S. A. and Schmitt, J. C.},
  year = {2017},
  month = mar,
  journal = {Physics of Plasmas},
  volume = {24},
  number = {3},
  pages = {032506},
  publisher = {{American Institute of Physics}},
  issn = {1070-664X},
  doi = {10.1063/1.4978494},
  urldate = {2021-12-14}
}

@article{Bader2019,
  title = {Stellarator Equilibria with Reactor Relevant Energetic Particle Losses},
  author = {Bader, Aaron and Drevlak, M. and Anderson, D. T. and Faber, B. J. and Hegna, C. C. and Likin, K. M. and Schmitt, J. C. and Talmadge, J. N.},
  year = {2019},
  month = oct,
  journal = {Journal of Plasma Physics},
  volume = {85},
  number = {5},
  publisher = {{Cambridge University Press}},
  issn = {0022-3778, 1469-7807},
  doi = {10.1017/S0022377819000680},
  urldate = {2020-11-20},
  langid = {english}
}

@article{Bader2020,
  title = {A {{New Optimized Quasihelically SymmetricStellarator}}},
  author = {Bader, A. and Faber, B. J. and Schmitt, J. C. and Anderson, D. T. and Drevlak, M. and Duff, J. M. and Frerichs, H. and Hegna, C. C. and Kruger, T. G. and Landreman, M. and McKinney, I. J. and Singh, L. and Schroeder, J. M. and Terry, P. W. and Ware, A. S.},
  year = {2020},
  month = apr,
  journal = {arXiv:2004.11426 [physics]},
  eprint = {2004.11426},
  primaryclass = {physics},
  urldate = {2020-11-18},
  archiveprefix = {arxiv}
}

@article{Bader2020a,
  title = {Advancing the Physics Basis for Quasi-Helically Symmetric Stellarators},
  author = {Bader, A. and Faber, B. J. and Schmitt, J. C. and Anderson, D. T. and Drevlak, M. and Duff, J. M. and Frerichs, H. and Hegna, C. C. and Kruger, T. G. and Landreman, M. and McKinney, I. J. and Singh, L. and Schroeder, J. M. and Terry, P. W. and Ware, A. S.},
  year = {2020},
  month = oct,
  journal = {Journal of Plasma Physics},
  volume = {86},
  number = {5},
  publisher = {{Cambridge University Press}},
  issn = {0022-3778, 1469-7807},
  doi = {10.1017/S0022377820000963},
  urldate = {2022-03-17},
  langid = {english}
}

@article{Bader2021,
  title = {Modeling of Energetic Particle Transport in Optimized Stellarators},
  author = {Bader, A. and Anderson, D. T. and Drevlak, M. and Faber, B. J. and Hegna, C. C. and Henneberg, S. and Landreman, M. and Schmitt, J. C. and Suzuki, Y. and Ware, A.},
  year = {2021},
  month = oct,
  journal = {Nuclear Fusion},
  volume = {61},
  number = {11},
  pages = {116060},
  publisher = {{IOP Publishing}},
  issn = {0029-5515},
  doi = {10.1088/1741-4326/ac2991},
  urldate = {2022-09-24},
  langid = {english}
}

@article{Badrinarayanan2016,
  title = {{{SegNet}}: {{A Deep Convolutional Encoder-Decoder Architecture}} for {{Image Segmentation}}},
  shorttitle = {{{SegNet}}},
  author = {Badrinarayanan, Vijay and Kendall, Alex and Cipolla, Roberto},
  year = {2016},
  month = oct,
  journal = {arXiv:1511.00561 [cs]},
  eprint = {1511.00561},
  primaryclass = {cs},
  urldate = {2020-10-15},
  archiveprefix = {arxiv}
}

@article{Bahri2021,
  title = {Explaining Neural Scaling Laws},
  author = {Bahri, Yasaman and Dyer, Ethan and Kaplan, Jared and Lee, Jaehoon and Sharma, Utkarsh},
  year = {2021},
  journal = {arXiv preprint arXiv:2102.06701},
  eprint = {2102.06701},
  archiveprefix = {arxiv}
}

@article{Baillod2021,
  title = {Stellarator Optimization for Nested Magnetic Surfaces at Finite \$\textbackslash beta\$ and Toroidal Current},
  author = {Baillod, A. and Loizu, J. and Graves, J. P. and Landreman, M.},
  year = {2021},
  month = nov,
  journal = {arXiv:2111.15564 [physics]},
  eprint = {2111.15564},
  primaryclass = {physics},
  urldate = {2021-12-07},
  archiveprefix = {arxiv}
}

@article{Baker*2018,
  title = {Accelerating {{Neural Architecture Search}} Using {{Performance Prediction}}},
  author = {Baker*, Bowen and Gupta*, Otkrist and Raskar, Ramesh and Naik, Nikhil},
  year = {2018},
  month = feb,
  urldate = {2021-10-28},
  langid = {english}
}

@article{Baltz2017,
  title = {Achievement of {{Sustained Net Plasma Heating}} in a {{Fusion Experiment}} with the {{Optometrist Algorithm}}},
  author = {Baltz, E. A. and Trask, E. and Binderbauer, M. and Dikovsky, M. and Gota, H. and Mendoza, R. and Platt, J. C. and Riley, P. F.},
  year = {2017},
  month = dec,
  journal = {Scientific Reports},
  volume = {7},
  number = {1},
  pages = {6425},
  issn = {2045-2322},
  doi = {10.1038/s41598-017-06645-7},
  urldate = {2020-10-14},
  langid = {english}
}

@article{Banacloche2020,
  title = {Socioeconomic and Environmental Impacts of Bringing the Sun to Earth: {{A}} Sustainability Analysis of a Fusion Power Plant Deployment},
  shorttitle = {Socioeconomic and Environmental Impacts of Bringing the Sun to Earth},
  author = {Banacloche, Santacruz and Gamarra, Ana R. and Lechon, Yolanda and Bustreo, Chiara},
  year = {2020},
  month = oct,
  journal = {Energy},
  volume = {209},
  pages = {118460},
  issn = {0360-5442},
  doi = {10.1016/j.energy.2020.118460},
  urldate = {2021-05-15},
  langid = {english}
}

@article{Banino2021,
  title = {{{PonderNet}}: {{Learning}} to {{Ponder}}},
  shorttitle = {{{PonderNet}}},
  author = {Banino, Andrea and Balaguer, Jan and Blundell, Charles},
  year = {2021},
  month = sep,
  journal = {arXiv:2107.05407 [cs]},
  eprint = {2107.05407},
  primaryclass = {cs},
  urldate = {2021-11-15},
  archiveprefix = {arxiv}
}

@article{Bar2019,
  title = {Mesh-{{Free Unsupervised Learning-Based PDE Solver}} of {{Forward}} and {{Inverse}} Problems},
  author = {Bar, Leah and Sochen, Nir},
  year = {2019}
}

@article{Bar2021,
  title = {Strong {{Solutions}} for {{PDE-Based Tomography}} by {{Unsupervised Learning}}},
  author = {Bar, Leah and Sochen, Nir},
  year = {2021},
  month = jan,
  journal = {SIAM Journal on Imaging Sciences},
  volume = {14},
  number = {1},
  pages = {128--155},
  publisher = {{Society for Industrial and Applied Mathematics}},
  doi = {10.1137/20M1332827},
  urldate = {2021-05-13}
}

@article{Barenco1995,
  title = {Elementary Gates for Quantum Computation},
  author = {Barenco, Adriano and Bennett, Charles H. and Cleve, Richard and DiVincenzo, David P. and Margolus, Norman and Shor, Peter and Sleator, Tycho and Smolin, John A. and Weinfurter, Harald},
  year = {1995},
  month = nov,
  journal = {Physical Review A},
  volume = {52},
  number = {5},
  pages = {3457--3467},
  issn = {1050-2947, 1094-1622},
  doi = {10.1103/PhysRevA.52.3457},
  urldate = {2019-08-25},
  langid = {english}
}

@article{Barillas2014,
  title = {{{SCR-1}}: {{Design}} and {{Construction}} of a {{Small Modular Stellarator}} for {{Magnetic Confinement}} of {{Plasma}}},
  shorttitle = {{{SCR-1}}},
  author = {Barillas, L and Vargas, V I and Alpizar, A and Asenjo, J and Carranza, J M and Cerdas, F and Guti{\'e}rrez, R and Monge, J I and Mora, J and Morera, J and Peraza, H and Queral, V and Rojas, C and Rozen, D and Saenz, F and S{\'a}nchez, G and Sandoval, M and Trimi{\~n}o, H and Uma{\~n}a, J and Villegas, L F},
  year = {2014},
  month = may,
  journal = {Journal of Physics: Conference Series},
  volume = {511},
  pages = {012037},
  issn = {1742-6596},
  doi = {10.1088/1742-6596/511/1/012037},
  urldate = {2020-10-14},
  langid = {english}
}

@article{Barnes2019,
  title = {\$\textbackslash texttt\{stella\}\$: A Mixed Implicit-Explicit, Delta-f Gyrokinetic Code for General Magnetic Field Configurations},
  shorttitle = {\$\textbackslash texttt\{stella\}\$},
  author = {Barnes, Michael and {Parra-Diaz}, Felix and Landreman, Matt},
  year = {2019},
  month = aug,
  journal = {Journal of Computational Physics},
  volume = {391},
  eprint = {1806.02162},
  pages = {365--380},
  issn = {00219991},
  doi = {10.1016/j.jcp.2019.01.025},
  urldate = {2020-10-14},
  archiveprefix = {arxiv},
  langid = {english}
}

@inproceedings{Barton2015,
  title = {Simultaneous Closed-Loop Control of the Current Profile and the Electron Temperature Profile in the {{TCV}} Tokamak},
  booktitle = {2015 {{American Control Conference}} ({{ACC}})},
  author = {Barton, Justin E. and Wehner, William P. and Schuster, Eugenio and Felici, Federico and Sauter, Olivier},
  year = {2015},
  month = jul,
  pages = {3316--3321},
  issn = {2378-5861},
  doi = {10.1109/ACC.2015.7171844}
}

@misc{Basri2019,
  title = {The {{Convergence Rate}} of {{Neural Networks}} for {{Learned Functions}} of {{Different Frequencies}}},
  author = {Basri, Ronen and Jacobs, David and Kasten, Yoni and Kritchman, Shira},
  year = {2019},
  month = dec,
  number = {arXiv:1906.00425},
  eprint = {arXiv:1906.00425},
  publisher = {{arXiv}},
  urldate = {2022-10-07},
  archiveprefix = {arxiv}
}

@article{Battaglia2018,
  title = {Relational Inductive Biases, Deep Learning, and Graph Networks},
  author = {Battaglia, Peter W. and Hamrick, Jessica B. and Bapst, Victor and {Sanchez-Gonzalez}, Alvaro and Zambaldi, Vinicius and Malinowski, Mateusz and Tacchetti, Andrea and Raposo, David and Santoro, Adam and Faulkner, Ryan and Gulcehre, Caglar and Song, Francis and Ballard, Andrew and Gilmer, Justin and Dahl, George and Vaswani, Ashish and Allen, Kelsey and Nash, Charles and Langston, Victoria and Dyer, Chris and Heess, Nicolas and Wierstra, Daan and Kohli, Pushmeet and Botvinick, Matt and Vinyals, Oriol and Li, Yujia and Pascanu, Razvan},
  year = {2018},
  month = jun,
  doi = {10.48550/arXiv.1806.01261},
  urldate = {2022-10-22},
  langid = {english}
}

@book{Bauer1978,
  title = {A {{Computational Method}} in {{Plasma Physics}}},
  author = {Bauer, Frances and Betancourt, Octavio and Garabedian, Paul},
  year = {1978},
  publisher = {{Springer Berlin Heidelberg}},
  address = {{Berlin, Heidelberg}},
  doi = {10.1007/978-3-642-85470-5},
  urldate = {2020-10-14},
  isbn = {978-3-642-85472-9 978-3-642-85470-5},
  langid = {english}
}

@book{Bauer1984,
  title = {Magnetohydrodynamic {{Equilibrium}} and {{Stability}} of {{Stellarators}}},
  author = {Bauer, Frances and Betancourt, Octavio and Garabedian, Paul},
  year = {1984},
  publisher = {{Springer New York}},
  address = {{New York, NY}},
  doi = {10.1007/978-1-4612-5240-5},
  urldate = {2020-10-14},
  isbn = {978-1-4612-9753-6 978-1-4612-5240-5},
  langid = {english}
}

@article{Beck2020,
  title = {An Overview on Deep Learning-Based Approximation Methods for Partial Differential Equations},
  author = {Beck, Christian and Hutzenthaler, Martin and Jentzen, Arnulf and Kuckuck, Benno},
  year = {2020},
  journal = {arXiv preprint arXiv:2012.12348},
  eprint = {2012.12348},
  archiveprefix = {arxiv}
}

@article{Beidler1990,
  title = {Physics and {{Engineering Design}} for {{Wendelstein VII-X}}},
  author = {Beidler, Craig and Grieger, G{\"u}nter and Herrnegger, Franz and Harmeyer, Ewald and Kisslinger, Johann and Lotz, Wolf and Maassberg, Henning and Merkel, Peter and N{\"u}hrenberg, J{\"u}rgen and Rau, Fritz and Sapper, J{\"o}rg and Sardei, Francesco and Scardovelli, Ruben and Schl{\"u}ter, Arnulf and Wobig, Horst},
  year = {1990},
  month = jan,
  journal = {Fusion Technology},
  volume = {17},
  number = {1},
  pages = {148--168},
  publisher = {{Taylor \& Francis}},
  issn = {0748-1896},
  doi = {10.13182/FST90-A29178},
  urldate = {2022-09-15}
}

@article{Beidler2011,
  title = {Benchmarking of the Mono-Energetic Transport Coefficients\textemdash Results from the {{International Collaboration}} on {{Neoclassical Transport}} in {{Stellarators}} ({{ICNTS}})},
  author = {Beidler, C. D. and Allmaier, K. and Isaev, M. Yu and Kasilov, S. V. and Kernbichler, W. and Leitold, G. O. and Maa{\ss}berg, H. and Mikkelsen, D. R. and Murakami, S. and Schmidt, M. and Spong, D. A. and Tribaldos, V. and Wakasa, A.},
  year = {2011},
  month = jun,
  journal = {Nuclear Fusion},
  volume = {51},
  number = {7},
  pages = {076001},
  publisher = {{IOP Publishing}},
  issn = {0029-5515},
  doi = {10.1088/0029-5515/51/7/076001},
  urldate = {2021-08-16},
  langid = {english}
}

@article{Beidler2021,
  title = {Demonstration of Reduced Neoclassical Energy Transport in {{Wendelstein}} 7-{{X}}},
  author = {Beidler, C. D. and Smith, H. M. and Alonso, A. and Andreeva, T. and Baldzuhn, J. and Beurskens, M. N. A. and Borchardt, M. and Bozhenkov, S. A. and Brunner, K. J. and Damm, H. and Drevlak, M. and Ford, O. P. and Fuchert, G. and Geiger, J. and Helander, P. and Hergenhahn, U. and Hirsch, M. and H{\"o}fel, U. and Kazakov, Ye O. and Kleiber, R. and Krychowiak, M. and Kwak, S. and Langenberg, A. and Laqua, H. P. and Neuner, U. and Pablant, N. A. and Pasch, E. and Pavone, A. and Pedersen, T. S. and Rahbarnia, K. and Schilling, J. and Scott, E. R. and Stange, T. and Svensson, J. and Thomsen, H. and Turkin, Y. and Warmer, F. and Wolf, R. C. and Zhang, D.},
  year = {2021},
  month = aug,
  journal = {Nature},
  volume = {596},
  number = {7871},
  pages = {221--226},
  publisher = {{Nature Publishing Group}},
  issn = {1476-4687},
  doi = {10.1038/s41586-021-03687-w},
  urldate = {2021-12-22},
  copyright = {2021 The Author(s)},
  langid = {english}
}

@article{Belkin2019,
  title = {Reconciling Modern Machine Learning Practice and the Bias-Variance Trade-Off},
  author = {Belkin, Mikhail and Hsu, Daniel and Ma, Siyuan and Mandal, Soumik},
  year = {2019},
  month = sep,
  journal = {arXiv:1812.11118 [cs, stat]},
  eprint = {1812.11118},
  primaryclass = {cs, stat},
  urldate = {2022-05-12},
  archiveprefix = {arxiv}
}

@book{Bellan2008,
  title = {Fundamentals of Plasma Physics},
  author = {Bellan, Paul M.},
  year = {2008},
  publisher = {{Cambridge University Press}}
}

@article{Bellemare2017,
  title = {A {{Distributional Perspective}} on {{Reinforcement Learning}}},
  author = {Bellemare, Marc G. and Dabney, Will and Munos, R{\'e}mi},
  year = {2017},
  month = jul,
  journal = {arXiv:1707.06887 [cs, stat]},
  eprint = {1707.06887},
  primaryclass = {cs, stat},
  urldate = {2020-10-14},
  archiveprefix = {arxiv},
  langid = {english}
}

@article{Belli2008,
  title = {Kinetic Calculation of Neoclassical Transport Including Self-Consistent Electron and Impurity Dynamics},
  author = {Belli, E. A. and Candy, J.},
  year = {2008},
  month = jul,
  journal = {Plasma Physics and Controlled Fusion},
  volume = {50},
  number = {9},
  pages = {095010},
  publisher = {{IOP Publishing}},
  issn = {0741-3335},
  doi = {10.1088/0741-3335/50/9/095010},
  urldate = {2022-07-29},
  langid = {english}
}

@article{Belli2011,
  title = {Full Linearized {{Fokker}}\textendash{{Planck}} Collisions in Neoclassical Transport Simulations},
  author = {Belli, E. A. and Candy, J.},
  year = {2011},
  month = dec,
  journal = {Plasma Physics and Controlled Fusion},
  volume = {54},
  number = {1},
  pages = {015015},
  publisher = {{IOP Publishing}},
  issn = {0741-3335},
  doi = {10.1088/0741-3335/54/1/015015},
  urldate = {2022-07-20},
  langid = {english}
}

@inproceedings{Bender2021,
  title = {On the {{Dangers}} of {{Stochastic Parrots}}: {{Can Language Models Be Too Big}}? \&\#x1f99c;},
  shorttitle = {On the {{Dangers}} of {{Stochastic Parrots}}},
  booktitle = {Proceedings of the 2021 {{ACM Conference}} on {{Fairness}}, {{Accountability}}, and {{Transparency}}},
  author = {Bender, Emily M. and Gebru, Timnit and {McMillan-Major}, Angelina and Shmitchell, Shmargaret},
  year = {2021},
  month = mar,
  series = {{{FAccT}} '21},
  pages = {610--623},
  publisher = {{Association for Computing Machinery}},
  address = {{New York, NY, USA}},
  doi = {10.1145/3442188.3445922},
  urldate = {2021-12-10},
  isbn = {978-1-4503-8309-7}
}

@inproceedings{Bengio2009,
  title = {Curriculum Learning},
  booktitle = {Proceedings of the 26th Annual International Conference on Machine Learning},
  author = {Bengio, Yoshua and Louradour, J{\'e}r{\^o}me and Collobert, Ronan and Weston, Jason},
  year = {2009},
  pages = {41--48},
  doi = {10.1145/1553374.1553380}
}

@article{Bengio2012,
  title = {Practical Recommendations for Gradient-Based Training of Deep Architectures},
  author = {Bengio, Yoshua},
  year = {2012},
  month = sep,
  journal = {arXiv:1206.5533 [cs]},
  eprint = {1206.5533},
  primaryclass = {cs},
  urldate = {2022-01-01},
  archiveprefix = {arxiv}
}

@article{Berg2018,
  title = {A Unified Deep Artificial Neural Network Approach to Partial Differential Equations in Complex Geometries},
  author = {Berg, Jens and Nystr{\"o}m, Kaj},
  year = {2018},
  month = nov,
  journal = {Neurocomputing},
  volume = {317},
  eprint = {1711.06464},
  pages = {28--41},
  issn = {09252312},
  doi = {10/gfvrt8},
  urldate = {2021-04-07},
  archiveprefix = {arxiv}
}

@article{Bergstra2011,
  title = {Algorithms for {{Hyper-Parameter Optimization}}},
  author = {Bergstra, James and Bardenet, R{\'e}mi and Bengio, Yoshua and K{\'e}gl, Bal{\'a}zs},
  year = {2011},
  journal = {Advances in Neural Information Processing Systems},
  volume = {24},
  pages = {2546--2554},
  urldate = {2020-12-22},
  langid = {english}
}

@inproceedings{Bergstra2013,
  title = {Making a {{Science}} of {{Model Search}}: {{Hyperparameter Optimization}} in {{Hundreds}} of {{Dimensions}} for {{Vision Architectures}}},
  shorttitle = {Making a {{Science}} of {{Model Search}}},
  booktitle = {International {{Conference}} on {{Machine Learning}}},
  author = {Bergstra, James and Yamins, Daniel and Cox, David},
  year = {2013},
  month = feb,
  volume = {28},
  pages = {115--123},
  publisher = {{PMLR}},
  issn = {1938-7228},
  urldate = {2020-12-22},
  langid = {english}
}

@article{Berlinguette2019,
  title = {Revisiting the Cold Case of Cold Fusion},
  author = {Berlinguette, Curtis P. and Chiang, Yet-Ming and Munday, Jeremy N. and Schenkel, Thomas and Fork, David K. and Koningstein, Ross and Trevithick, Matthew D.},
  year = {2019},
  month = jun,
  journal = {Nature},
  volume = {570},
  number = {7759},
  pages = {45--51},
  publisher = {{Nature Publishing Group}},
  issn = {1476-4687},
  doi = {10.1038/s41586-019-1256-6},
  urldate = {2021-07-02},
  copyright = {2019 Springer Nature Limited},
  langid = {english}
}

@article{Berry1984,
  title = {Quantal Phase Factors Accompanying Adiabatic Changes},
  author = {Berry, Michael Victor},
  year = {1984},
  month = mar,
  journal = {Proceedings of the Royal Society of London. A. Mathematical and Physical Sciences},
  volume = {392},
  number = {1802},
  pages = {45--57},
  issn = {2053-9169},
  doi = {10.1098/rspa.1984.0023},
  urldate = {2020-10-14},
  langid = {english}
}

@article{Bertschinger2022,
  title = {Training {{Fully Connected Neural Networks}} Is \$\textbackslash exists\textbackslash mathbb\{\vphantom\}{{R}}\vphantom\{\}\$-{{Complete}}},
  author = {Bertschinger, Daniel and Hertrich, Christoph and Jungeblut, Paul and Miltzow, Tillmann and Weber, Simon},
  year = {2022},
  month = apr,
  journal = {arXiv:2204.01368 [cs]},
  eprint = {2204.01368},
  primaryclass = {cs},
  urldate = {2022-04-22},
  archiveprefix = {arxiv}
}

@book{Bertsekas1995,
  title = {Dynamic Programming and Optimal Control},
  author = {Bertsekas, Dimitri P. and Bertsekas, Dimitri P. and Bertsekas, Dimitri P. and Bertsekas, Dimitri P.},
  year = {1995},
  volume = {1},
  publisher = {{Athena scientific Belmont, MA}}
}

@book{Bertsekas2019,
  title = {Reinforcement Learning and Optimal Control},
  author = {Bertsekas, Dimitri P.},
  year = {2019},
  publisher = {{Athena Scientific Belmont, MA}}
}

@article{Besshou1986,
  title = {Diamagnetism and Beta in Beam Heated Currentless Plasmas of Heliotron {{E}}},
  author = {Besshou, S. and Thomas, C. E. and Ohba, T. and Iiyoshi, A. and Uo, K.},
  year = {1986},
  month = oct,
  journal = {Nuclear Fusion},
  volume = {26},
  number = {10},
  pages = {1339},
  issn = {0029-5515},
  doi = {2011012504290300},
  urldate = {2023-01-24},
  langid = {english}
}

@article{Betancourt1988,
  title = {{{BETAS}}, a Spectral Code for Three-Dimensional Magnetohydrodynamic Equilibrium and Nonlinear Stability Calculations},
  author = {Betancourt, Octavio},
  year = {1988},
  journal = {Communications on Pure and Applied Mathematics},
  volume = {41},
  number = {5},
  pages = {551--568},
  publisher = {{Wiley Online Library}},
  doi = {10.1002/cpa.3160410504}
}

@article{Beucler2021,
  title = {Enforcing {{Analytic Constraints}} in {{Neural Networks Emulating Physical Systems}}},
  author = {Beucler, Tom and Pritchard, Michael and Rasp, Stephan and Ott, Jordan and Baldi, Pierre and Gentine, Pierre},
  year = {2021},
  month = mar,
  journal = {Physical Review Letters},
  volume = {126},
  number = {9},
  pages = {098302},
  publisher = {{American Physical Society}},
  doi = {10.1103/PhysRevLett.126.098302},
  urldate = {2023-03-28}
}

@misc{Bindel2023,
  title = {Direct {{Optimization}} of {{Fast-Ion Confinement}} in {{Stellarators}}},
  author = {Bindel, David and Landreman, Matt and Padidar, Misha},
  year = {2023},
  month = feb,
  number = {arXiv:2302.11369},
  eprint = {arXiv:2302.11369},
  publisher = {{arXiv}},
  doi = {10.48550/arXiv.2302.11369},
  urldate = {2023-02-26},
  archiveprefix = {arxiv}
}

@misc{BindingEnergyCurve,
  title = {Binding {{Energy Curve}}},
  journal = {Wikipedia},
  urldate = {2023-03-29},
  howpublished = {https://it.m.wikipedia.org/wiki/File:Binding\_energy\_curve\_-\_common\_isotopes.svg}
}

@book{Bishop1980,
  title = {Tensor {{Analysis}} on {{Manifolds}}},
  author = {Bishop, Richard L},
  year = {1980},
  langid = {english}
}

@book{Bishop1995,
  title = {Neural Networks for Pattern Recognition},
  author = {Bishop, Christopher M.},
  year = {1995},
  publisher = {{Oxford university press}}
}

@article{Bishop1995a,
  title = {Real-{{Time Control}} of a {{Tokamak Plasma Using Neural Networks}}},
  author = {Bishop, Chris M. and Haynes, Paul S. and Smith, Mike E. U. and Todd, Tom N. and Trotman, David L.},
  year = {1995},
  month = jan,
  journal = {Neural Computation},
  volume = {7},
  number = {1},
  pages = {206--217},
  issn = {0899-7667},
  doi = {10.1162/neco.1995.7.1.206},
  urldate = {2022-03-04}
}

@book{Bittencourt2013,
  title = {Fundamentals of Plasma Physics},
  author = {Bittencourt, Jos{\'e} A.},
  year = {2013},
  publisher = {{Springer Science \& Business Media}}
}

@article{Blalock2020,
  title = {What Is the {{State}} of {{Neural Network Pruning}}?},
  author = {Blalock, Davis and Ortiz, Jose Javier Gonzalez and Frankle, Jonathan and Guttag, John},
  year = {2020},
  month = mar,
  journal = {arXiv:2003.03033 [cs, stat]},
  eprint = {2003.03033},
  primaryclass = {cs, stat},
  urldate = {2021-02-25},
  archiveprefix = {arxiv}
}

@article{Blatzheim2019,
  title = {Neural Network Regression Approaches to Reconstruct Properties of Magnetic Configuration from {{Wendelstein}} 7-{{X}} Modeled Heat Load Patterns},
  author = {Blatzheim, Marko and B{\"o}ckenhoff, Daniel and {the Wendelstein 7-X Team}},
  year = {2019},
  month = dec,
  journal = {Nuclear Fusion},
  volume = {59},
  number = {12},
  pages = {126029},
  issn = {0029-5515, 1741-4326},
  doi = {10.1088/1741-4326/ab4123},
  urldate = {2020-10-14},
  langid = {english}
}

@article{Blatzheim2019a,
  title = {Neural Network Performance Enhancement for Limited Nuclear Fusion Experiment Observations Supported by Simulations},
  author = {Blatzheim, Marko and B{\"o}ckenhoff, Daniel and H{\"o}lbe, Hauke and Pedersen, Thomas Sunn and Labahn, Roger and {The W7-X Team}},
  year = {2019},
  month = jan,
  journal = {Nuclear Fusion},
  volume = {59},
  number = {1},
  pages = {016012},
  issn = {0029-5515, 1741-4326},
  doi = {10.1088/1741-4326/aaefaf},
  urldate = {2020-10-14},
  langid = {english}
}

@article{Blechschmidt2021,
  title = {Three {{Ways}} to {{Solve Partial Differential Equations}} with {{Neural Networks}} -- {{A Review}}},
  author = {Blechschmidt, Jan and Ernst, Oliver G.},
  year = {2021},
  month = feb,
  journal = {arXiv:2102.11802 [cs, math]},
  eprint = {2102.11802},
  primaryclass = {cs, math},
  urldate = {2021-04-06},
  archiveprefix = {arxiv}
}

@book{Board2015,
  title = {Climate Intervention: {{Reflecting}} Sunlight to Cool Earth},
  shorttitle = {Climate Intervention},
  author = {Board, Ocean Studies and Council, National Research},
  year = {2015},
  publisher = {{National Academies Press}}
}

@incollection{Boarin2021,
  title = {Economics and Financing of Small Modular Reactors ({{SMRs}})},
  booktitle = {Handbook of {{Small Modular Nuclear Reactors}}},
  author = {Boarin, Sara and Mancini, Mauro and Ricotti, Marco and Locatelli, Giorgio},
  year = {2021},
  pages = {241--278},
  publisher = {{Elsevier}}
}

@article{Bockenhoff2018,
  title = {Reconstruction of Magnetic Configurations in {{W7-X}} Using Artificial Neural Networks},
  author = {B{\"o}ckenhoff, Daniel and Blatzheim, Marko and H{\"o}lbe, Hauke and Niemann, Holger and Pisano, Fabio and Labahn, Roger and Pedersen, Thomas Sunn and {The W7-X Team}},
  year = {2018},
  month = may,
  journal = {Nuclear Fusion},
  volume = {58},
  number = {5},
  pages = {056009},
  issn = {0029-5515, 1741-4326},
  doi = {10.1088/1741-4326/aab22d},
  urldate = {2020-10-14},
  langid = {english}
}

@article{Bockenhoff2019,
  title = {Application of Improved Analysis of Convective Heat Loads on Plasma Facing Components to {{Wendelstein}} 7-{{X}}},
  author = {B{\"o}ckenhoff, Daniel and Blatzheim, Marko and {the W7-X Team}},
  year = {2019},
  month = aug,
  journal = {Nuclear Fusion},
  volume = {59},
  number = {8},
  pages = {086031},
  issn = {0029-5515, 1741-4326},
  doi = {10.1088/1741-4326/ab201e},
  urldate = {2020-10-14},
  langid = {english}
}

@article{Bockenhoff2020,
  title = {En {{Route Towards Heat Load Control}} for {{Wendelstein}} 7-{{X}} with {{Machine Learning Approaches}}},
  author = {B{\"o}ckenhoff, Daniel},
  year = {2020}
}

@article{Boixo2018,
  title = {Characterizing Quantum Supremacy in Near-Term Devices},
  author = {Boixo, Sergio and Isakov, Sergei V. and Smelyanskiy, Vadim N. and Babbush, Ryan and Ding, Nan and Jiang, Zhang and Bremner, Michael J. and Martinis, John M. and Neven, Hartmut},
  year = {2018},
  month = jun,
  journal = {Nature Physics},
  volume = {14},
  number = {6},
  pages = {595--600},
  issn = {1745-2481},
  doi = {10.1038/s41567-018-0124-x},
  urldate = {2019-09-01},
  copyright = {2018 The Author(s)},
  langid = {english}
}

@article{Bonfiglio2018,
  title = {Multi-Fidelity Optimization of Super-Cavitating Hydrofoils},
  author = {Bonfiglio, L. and Perdikaris, P. and Brizzolara, S. and Karniadakis, G. E.},
  year = {2018},
  journal = {Computer Methods in Applied Mechanics and Engineering},
  volume = {332},
  pages = {63--85},
  publisher = {{Elsevier}},
  doi = {10.1016/j.cma.2017.12.009}
}

@article{Bonou2016,
  title = {Life Cycle Assessment of Onshore and Offshore Wind Energy-from Theory to Application},
  author = {Bonou, Alexandra and Laurent, Alexis and Olsen, Stig I.},
  year = {2016},
  month = oct,
  journal = {Applied Energy},
  volume = {180},
  pages = {327--337},
  issn = {0306-2619},
  doi = {10.1016/j.apenergy.2016.07.058},
  urldate = {2021-05-12},
  langid = {english}
}

@book{Booth2003,
  title = {The Craft of Research},
  author = {Booth, Wayne C. and Booth, William C. and Colomb, Gregory G. and Williams, Joseph M. and Colomb, Gregory G. and Williams, Joseph M.},
  year = {2003},
  publisher = {{University of Chicago press}}
}

@article{Boozer1983,
  title = {Evaluation of the Structure of Ergodic Fields},
  author = {Boozer, Allen H.},
  year = {1983},
  journal = {The Physics of Fluids},
  volume = {26},
  number = {5},
  pages = {1288--1291},
  publisher = {{American Institute of Physics}},
  doi = {10.1063/1.864289}
}

@article{Boozer2005,
  ids = {Boozer2005a},
  title = {Physics of Magnetically Confined Plasmas},
  author = {Boozer, Allen H.},
  year = {2005},
  month = jan,
  journal = {Reviews of Modern Physics},
  volume = {76},
  number = {4},
  pages = {1071--1141},
  publisher = {{American Physical Society}},
  issn = {0034-6861, 1539-0756},
  doi = {10.1103/RevModPhys.76.1071},
  urldate = {2020-10-14},
  langid = {english}
}

@article{Boozer2008,
  title = {Stellarators and the Path from {{ITER}} to {{DEMO}}},
  author = {Boozer, Allen H.},
  year = {2008},
  month = nov,
  journal = {Plasma Physics and Controlled Fusion},
  volume = {50},
  number = {12},
  pages = {124005},
  publisher = {{IOP Publishing}},
  issn = {0741-3335},
  doi = {10.1088/0741-3335/50/12/124005},
  urldate = {2021-06-29},
  langid = {english}
}

@article{Boozer2015,
  title = {Non-Axisymmetric Magnetic Fields and Toroidal Plasma Confinement},
  author = {Boozer, Allen H.},
  year = {2015},
  month = jan,
  journal = {Nuclear Fusion},
  volume = {55},
  number = {2},
  pages = {025001},
  publisher = {{IOP Publishing}},
  issn = {0029-5515},
  doi = {10.1088/0029-5515/55/2/025001},
  urldate = {2021-12-14},
  langid = {english}
}

@article{Boozer2017,
  title = {Runaway Electrons and {{ITER}}},
  author = {Boozer, Allen H.},
  year = {2017},
  month = mar,
  journal = {Nuclear Fusion},
  volume = {57},
  number = {5},
  pages = {056018},
  publisher = {{IOP Publishing}},
  issn = {0029-5515},
  doi = {10.1088/1741-4326/aa6355},
  urldate = {2021-01-02},
  langid = {english}
}

@article{Boozer2020,
  title = {Why Carbon Dioxide Makes Stellarators so Important},
  author = {Boozer, Allen H.},
  year = {2020},
  month = jun,
  journal = {Nuclear Fusion},
  volume = {60},
  number = {6},
  eprint = {1912.06289},
  pages = {065001},
  issn = {0029-5515, 1741-4326},
  doi = {10.1088/1741-4326/ab87af},
  urldate = {2020-11-06},
  archiveprefix = {arxiv}
}

@article{Boozer2022,
  title = {Constraints on Stellarator Divertors from {{Hamiltonian}} Mechanics},
  author = {Boozer, Allen H.},
  year = {2022},
  journal = {arXiv preprint arXiv:2206.06368},
  eprint = {2206.06368},
  archiveprefix = {arxiv}
}

@misc{Boozer2022a,
  title = {Omnigenous Toroidal Plasma Equilibria},
  author = {Boozer, Allen H.},
  year = {2022},
  month = aug,
  number = {arXiv:2208.02391},
  eprint = {arXiv:2208.02391},
  publisher = {{arXiv}},
  doi = {10.48550/arXiv.2208.02391},
  urldate = {2022-09-14},
  archiveprefix = {arxiv}
}

@book{Borutzky2010,
  title = {Bond {{Graph Methodology}}},
  author = {Borutzky, Wolfgang},
  year = {2010},
  publisher = {{Springer London}},
  address = {{London}},
  doi = {10.1007/978-1-84882-882-7},
  urldate = {2020-10-14},
  isbn = {978-1-84882-881-0 978-1-84882-882-7},
  langid = {english}
}

@article{Bosch2013,
  title = {Technical Challenges in the Construction of the Steady-State Stellarator {{Wendelstein}} 7-{{X}}},
  author = {Bosch, H.-S. and Wolf, R. C. and Andreeva, T. and Baldzuhn, J. and Birus, D. and Bluhm, T. and Br{\"a}uer, T. and Braune, H. and Bykov, V. and Cardella, A. and Durodi{\'e}, F. and Endler, M. and Erckmann, V. and Gantenbein, G. and Hartmann, D. and Hathiramani, D. and Heimann, P. and Heinemann, B. and Hennig, C. and Hirsch, M. and Holtum, D. and Jagielski, J. and Jelonnek, J. and Kasparek, W. and Klinger, T. and K{\"o}nig, R. and Kornejew, P. and Kroiss, H. and Krom, J. G. and K{\"u}hner, G. and Laqua, H. and Laqua, H. P. and Lechte, C. and Lewerentz, M. and Maier, J. and McNeely, P. and Messiaen, A. and Michel, G. and Ongena, J. and Peacock, A. and Pedersen, T. S. and Riedl, R. and Riemann, H. and Rong, P. and Rust, N. and Schacht, J. and Schauer, F. and Schroeder, R. and Schweer, B. and Spring, A. and St{\"a}bler, A. and Thumm, M. and Turkin, Y. and Wegener, L. and Werner, A. and Zhang, D. and Zilker, M. and Akijama, T. and Alzbutas, R. and Ascasibar, E. and Balden, M. and Banduch, M. and Baylard, Ch and Behr, W. and Beidler, C. and Benndorf, A. and Bergmann, T. and Biedermann, C. and Bieg, B. and Biel, W. and Borchardt, M. and Borowitz, G. and Borsuk, V. and Bozhenkov, S. and Brakel, R. and Brand, H. and Brown, T. and Brucker, B. and Burhenn, R. and Buscher, K.-P. and {Caldwell-Nichols}, C. and Cappa, A. and Cardella, A. and Carls, A. and Carvalho, P. and Ciupi{\'n}ski, {\L} and Cole, M. and Collienne, J. and Czarnecka, A. and Czymek, G. and Dammertz, G. and Dhard, C. P. and Davydenko, V. I. and Dinklage, A. and Drevlak, M. and Drotziger, S. and Dudek, A. and Dumortier, P. and Dundulis, G. and v Eeten, P. and Egorov, K. and Estrada, T. and Faugel, H. and Fellinger, J. and Feng, Y. and Fernandes, H. and Fietz, W. H. and Figacz, W. and Fischer, F. and Fontdecaba, J. and Freund, A. and Funaba, T. and F{\"u}nfgelder, H. and Galkowski, A. and Gates, D. and Giannone, L. and Rega{\~n}a, J. M. Garc{\'i}a and Geiger, J. and Gei{\ss}ler, S. and Greuner, H. and Grahl, M. and Gro{\ss}, S. and Grosman, A. and Grote, H. and Grulke, O. and Haas, M. and Haiduk, L. and Hartfu{\ss}, H.-J. and Harris, J. H. and Haus, D. and Hein, B. and Heitzenroeder, P. and Helander, P. and Heller, R. and Hidalgo, C. and Hildebrandt, D. and H{\"o}hnle, H. and Holtz, A. and Holzhauer, E. and Holzth{\"u}m, R. and Huber, A. and Hunger, H. and Hurd, F. and Ihrke, M. and Illy, S. and Ivanov, A. and Jablonski, S. and Jaksic, N. and Jakubowski, M. and Jaspers, R. and Jensen, H. and Jenzsch, H. and Kacmarczyk, J. and Kaliatk, T. and Kallmeyer, J. and Kamionka, U. and Karaleviciu, R. and Kern, S. and Keunecke, M. and Kleiber, R. and Knauer, J. and Koch, R. and Kocsis, G. and K{\"o}nies, A. and K{\"o}ppen, M. and Koslowski, R. and Koshurinov, J. and {Kr{\"a}mer-Flecken}, A. and Krampitz, R. and Kravtsov, Y. and Krychowiak, M. and Krzesinski, G. and Ksiazek, I. and Kubkowska, M. and Kus, A. and Langish, S. and Laube, R. and Laux, M. and Lazerson, S. and Lennartz, M. and Li, C. and Lietzow, R. and Lohs, A. and Lorenz, A. and Louche, F. and Lubyako, L. and Lumsdaine, A. and Lyssoivan, A. and Maa{\ss}berg, H. and Marek, P. and Martens, C. and Marushchenko, N. and Mayer, M. and Mendelevitch, B. and Mertens, Ph and Mikkelsen, D. and Mishchenko, A. and Missal, B. and Mizuuchi, T. and Modrow, H. and M{\"o}nnich, T. and Morizaki, T. and Murakami, S. and Musielok, F. and Nagel, M. and Naujoks, D. and Neilson, H. and Neubauer, O. and Neuner, U. and Nocentini, R. and Noterdaeme, J.-M. and N{\"u}hrenberg, C. and Obermayer, S. and Offermanns, G. and Oosterbeek, H. and Otte, M. and Panin, A. and Pap, M. and Paquay, S. and Pasch, E. and Peng, X. and Petrov, S. and Pilopp, D. and Pirsch, H. and Plaum, B. and Pompon, F. and Povilaitis, M. and Preinhaelter, J. and Prinz, O. and Purps, F. and Rajna, T. and R{\'e}csei, S. and Reiman, A. and Reiter, D. and Remmel, J. and Renard, S. and Rhode, V. and Riemann, J. and Rimkevicius, S. and Ri{\ss}e, K. and Rodatos, A. and Rodin, I. and Rom{\'e}, M. and Roscher, H.-J. and Rummel, K. and Rummel, Th and Runov, A. and Ryc, L. and Sachtleben, J. and Samartsev, A. and Sanchez, M. and Sano, F. and Scarabosio, A. and Schmid, M. and Schmitz, H. and Schmitz, O. and Schneider, M. and Schneider, W. and Scheibl, L. and Scholz, M. and Schr{\"o}der, G. and Schr{\"o}der, M. and Schruff, J. and Schumacher, H. and Shikhovtsev, I. V. and Shoji, M. and Siegl, G. and Skodzik, J. and Smirnow, M. and Speth, E. and Spong, D. A. and Stadler, R. and Sulek, Z. and Szab{\'o}, V. and Szabolics, T. and Szetefi, T. and {Sz{\"o}kefalvi-Nagy}, Z. and Tereshchenko, A. and Thomsen, H. and Thumm, M. and Timmermann, D. and Tittes, H. and Toi, K. and Tournianski, M. and v Toussaint, U. and Tretter, J. and Tulip{\'a}n, S. and Turba, P. and Uhlemann, R. and Urban, J. and Urbonavicius, E. and Urlings, P. and Valet, S. and Eester, D. Van and Schoor, M. Van and Vervier, M. and Viebke, H. and Vilbrandt, R. and Vrancken, M. and Wauters, T. and Weissgerber, M. and Wei{\ss}, E. and Weller, A. and Wendorf, J. and Wenzel, U. and Windisch, T. and Winkler, E. and Winkler, M. and Wolowski, J. and Wolters, J. and Wrochna, G. and Xanthopoulos, P. and Yamada, H. and Yokoyama, M. and Zacharias, D. and Zajac, J. and Zangl, G. and Zarnstorff, M. and Zeplien, H. and Zoletnik, S. and Zuin, M.},
  year = {2013},
  month = nov,
  journal = {Nuclear Fusion},
  volume = {53},
  number = {12},
  pages = {126001},
  publisher = {{IOP Publishing and International Atomic Energy Agency}},
  issn = {0029-5515},
  doi = {10.1088/0029-5515/53/12/126001},
  urldate = {2023-03-29},
  langid = {english}
}

@article{Bosnjakovic2020,
  title = {The {{Perspective}} of {{Large-Scale Production}} of {{Algae Biodiesel}}},
  author = {Bo{\v s}njakovi{\'c}, Mladen and Sinaga, Nazaruddin},
  year = {2020},
  journal = {Applied Sciences},
  volume = {10},
  number = {22},
  pages = {8181},
  publisher = {{Multidisciplinary Digital Publishing Institute}},
  doi = {10.3390/app10228181}
}

@article{Boyer2020,
  title = {Toward Fusion Plasma Scenario Planning for {{NSTX-U}} Using Machine-Learning-Accelerated Models},
  author = {Boyer, Mark},
  year = {2020},
  month = jun,
  urldate = {2021-02-25},
  langid = {english}
}

@article{Bozhenkov2017,
  title = {The {{Thomson}} Scattering Diagnostic at {{Wendelstein}} 7-{{X}} and Its Performance in the First Operation Phase},
  author = {Bozhenkov, S. A. and Beurskens, M. and Molin, A. Dal and Fuchert, G. and Pasch, E. and Stoneking, M. R. and Hirsch, M. and H{\"o}fel, U. and Knauer, J. and Svensson, J. and Mora, H. Trimino and Wolf, R. C.},
  year = {2017},
  month = oct,
  journal = {Journal of Instrumentation},
  volume = {12},
  number = {10},
  pages = {P10004},
  issn = {1748-0221},
  doi = {2017101015130600},
  urldate = {2023-02-12},
  langid = {english}
}

@inproceedings{Bozhenkov2018,
  title = {High Density and High Performance Operation with Pellet Injection in {{W7-X}}},
  booktitle = {27th {{IAEA Fusion Energy Conference}}, {{Gandhinagar}}},
  author = {Bozhenkov, S. and Kazakov, Y. and Baldzuhn, J. and Laqua, H. P. and Alonso, J. A. and Beurskens, M. N. A. and Brandt, C. and Brunner, K. J. and Damm, H. and Fuchert, G.},
  year = {2018}
}

@article{Bozhenkov2020,
  title = {High-Performance Plasmas after Pellet Injections in {{Wendelstein}} 7-{{X}}},
  author = {Bozhenkov, S. A. and Kazakov, Y. and Ford, O. P. and Beurskens, M. N. A. and Alcus{\'o}n, J. and Alonso, J. A. and Baldzuhn, J. and Brandt, C. and Brunner, K. J. and Damm, H. and Fuchert, G. and Geiger, J. and Grulke, O. and Hirsch, M. and H{\"o}fel, U. and Huang, Z. and Knauer, J. and Krychowiak, M. and Langenberg, A. and Laqua, H. P. and Lazerson, S. and Marushchenko, N. B. and Moseev, D. and Otte, M. and Pablant, N. and Pasch, E. and Pavone, A. and Proll, J. H. E. and Rahbarnia, K. and Scott, E. R. and Smith, H. M. and Stange, T. and von Stechow, A. and Thomsen, H. and Turkin, Yu and Wurden, G. and Xanthopoulos, P. and Zhang, D. and {and}, R. C. Wolf},
  year = {2020},
  month = may,
  journal = {Nuclear Fusion},
  volume = {60},
  number = {6},
  pages = {066011},
  publisher = {{IOP Publishing}},
  issn = {0029-5515},
  doi = {10.1088/1741-4326/ab7867},
  urldate = {2022-09-15},
  langid = {english}
}

@article{Braams1991,
  title = {The Interpretation of Tokamak Magnetic Diagnostics},
  author = {Braams, B. J.},
  year = {1991},
  month = jul,
  journal = {Plasma Physics and Controlled Fusion},
  volume = {33},
  number = {7},
  pages = {715},
  issn = {0741-3335},
  doi = {10.1088/0741-3335/33/7/001},
  urldate = {2023-04-12},
  langid = {english}
}

@article{Bradshaw2011,
  title = {Is Nuclear Fusion a Sustainable Energy Form?},
  author = {Bradshaw, A. M. and Hamacher, T. and Fischer, U.},
  year = {2011},
  month = oct,
  journal = {Fusion Engineering and Design},
  series = {Proceedings of the 26th {{Symposium}} of {{Fusion Technology}} ({{SOFT-26}})},
  volume = {86},
  number = {9},
  pages = {2770--2773},
  issn = {0920-3796},
  doi = {10.1016/j.fusengdes.2010.11.040},
  urldate = {2022-06-02},
  langid = {english}
}

@techreport{Brakel2018,
  title = {Magnetic {{Configurations}} for {{W7-X}} Operation},
  author = {Brakel, Rudolf},
  year = {2018},
  number = {1-JDB00-T0001.1},
  pages = {3},
  langid = {english}
}

@article{Brand2015,
  title = {Beyond Authorship: Attribution, Contribution, Collaboration, and Credit},
  shorttitle = {Beyond Authorship},
  author = {Brand, Amy and Allen, Liz and Altman, Micah and Hlava, Marjorie and Scott, Jo},
  year = {2015},
  journal = {Learned Publishing},
  volume = {28},
  number = {2},
  pages = {151--155},
  issn = {1741-4857},
  doi = {10.1087/20150211},
  urldate = {2021-08-13},
  langid = {english}
}

@misc{Brandstetter2023,
  title = {Clifford {{Neural Layers}} for {{PDE Modeling}}},
  author = {Brandstetter, Johannes and van den Berg, Rianne and Welling, Max and Gupta, Jayesh K.},
  year = {2023},
  month = mar,
  number = {arXiv:2209.04934},
  eprint = {arXiv:2209.04934},
  publisher = {{arXiv}},
  doi = {10.48550/arXiv.2209.04934},
  urldate = {2023-03-30},
  archiveprefix = {arxiv}
}

@article{Brandt2020,
  title = {Soft X-Ray Tomography Measurements in the {{Wendelstein}} 7-{{X}} Stellarator},
  author = {Brandt, C. and Schilling, J. and Thomsen, H. and Broszat, T. and Laube, R. and Schr{\"o}der, T. and Andreeva, T. and Beurskens, M. N. A. and Bozhenkov, S. A. and Brunner, K. J. and Card, A. and Cordes, C. and Damm, H. and Fuchert, G. and Gallowski, K. and Gutzmann, R. and Knauer, J. and Laqua, H. P. and Marquardt, M. and Nelde, Ph and Neuner, U. and Pasch, E. and Rahbarnia, K. and Recknagel, J. and Sch{\"u}lke, M. and Scott, E. R. and {and}, T. Sieber},
  year = {2020},
  month = jan,
  journal = {Plasma Physics and Controlled Fusion},
  volume = {62},
  number = {3},
  pages = {035010},
  publisher = {{IOP Publishing}},
  issn = {0741-3335},
  doi = {10.1088/1361-6587/ab630d},
  urldate = {2021-05-04},
  langid = {english}
}

@book{Bransden2003,
  title = {Physics of Atoms and Molecules},
  author = {Bransden, Brian Harold and Joachain, Charles Jean},
  year = {2003},
  publisher = {{Pearson Education India}}
}

@article{Brezinsek2005,
  title = {Characterization of the Deuterium Recycling Flux in Front of a Graphite Surface in the {{TEXTOR}} Tokamak},
  author = {Brezinsek, S. and Sergienko, G. and Pospieszczyk, A. and Mertens, Ph and Samm, U. and Greenland, P. T.},
  year = {2005},
  month = mar,
  journal = {Plasma Physics and Controlled Fusion},
  volume = {47},
  number = {4},
  pages = {615--634},
  publisher = {{IOP Publishing}},
  issn = {0741-3335},
  doi = {10.1088/0741-3335/47/4/003},
  urldate = {2021-08-19},
  langid = {english}
}

@article{Brezis2008,
  title = {Big Pharma and Health Care: Unsolvable Conflict of Interests between Private Enterprise and Public Health},
  shorttitle = {Big Pharma and Health Care},
  author = {Brezis, Mayer},
  year = {2008},
  journal = {Israel Journal of Psychiatry and Related Sciences},
  volume = {45},
  number = {2},
  pages = {83}
}

@misc{Britain2019,
  title = {Rising to the Climate Emergency, {{Centre}} for {{Alternative Technology}}},
  author = {Britain, Zero Carbon}
}

@article{Brod2019,
  title = {Photonic Implementation of Boson Sampling: A Review},
  shorttitle = {Photonic Implementation of Boson Sampling},
  author = {Brod, Daniel J. and Galv{\~a}o, Ernesto F. and Crespi, Andrea and Osellame, Roberto and Spagnolo, Nicol{\`o} and Sciarrino, Fabio},
  year = {2019},
  month = may,
  journal = {Advanced Photonics},
  volume = {1},
  number = {3},
  pages = {034001},
  issn = {2577-5421, 2577-5421},
  doi = {10.1117/1.AP.1.3.034001},
  urldate = {2019-09-04}
}

@inproceedings{Brown2015,
  title = {Engineering Optimization of Stellarator Coils Lead to Improvements in Device Maintenance},
  booktitle = {2015 {{IEEE}} 26th {{Symposium}} on {{Fusion Engineering}} ({{SOFE}})},
  author = {Brown, T. and Breslau, J. and Gates, D. and Pomphrey, N. and Zolfaghari, A.},
  year = {2015},
  month = may,
  pages = {1--6},
  issn = {2155-9953},
  doi = {10.1109/SOFE.2015.7482426}
}

@article{Brown2020,
  title = {Language {{Models}} Are {{Few-Shot Learners}}},
  author = {Brown, Tom B. and Mann, Benjamin and Ryder, Nick and Subbiah, Melanie and Kaplan, Jared and Dhariwal, Prafulla and Neelakantan, Arvind and Shyam, Pranav and Sastry, Girish and Askell, Amanda and Agarwal, Sandhini and {Herbert-Voss}, Ariel and Krueger, Gretchen and Henighan, Tom and Child, Rewon and Ramesh, Aditya and Ziegler, Daniel M. and Wu, Jeffrey and Winter, Clemens and Hesse, Christopher and Chen, Mark and Sigler, Eric and Litwin, Mateusz and Gray, Scott and Chess, Benjamin and Clark, Jack and Berner, Christopher and McCandlish, Sam and Radford, Alec and Sutskever, Ilya and Amodei, Dario},
  year = {2020},
  month = jul,
  journal = {arXiv:2005.14165 [cs]},
  eprint = {2005.14165},
  primaryclass = {cs},
  urldate = {2020-12-03},
  archiveprefix = {arxiv}
}

@book{Bruss2007,
  title = {Lectures on Quantum Information},
  editor = {Bruss, Dagmar and Leuchs, Gerd},
  year = {2007},
  publisher = {{Wiley-VCH}},
  address = {{Weinheim}},
  isbn = {978-3-527-40527-5},
  langid = {english},
  lccn = {QA76.889 .L438 2007}
}

@book{Bruss2019,
  title = {Quantum {{Information}}: {{From Foundations}} to {{Quantum Technology Applications}}},
  shorttitle = {Quantum {{Information}}},
  author = {Bruss, Dagmar and Leuchs, Gerd},
  year = {2019},
  publisher = {{John Wiley \& Sons}}
}

@article{Bu2021,
  title = {Quadratic {{Residual Networks}}: {{A New Class}} of {{Neural Networks}} for {{Solving Forward}} and {{Inverse Problems}} in {{Physics Involving PDEs}}},
  shorttitle = {Quadratic {{Residual Networks}}},
  author = {Bu, Jie and Karpatne, Anuj},
  year = {2021},
  month = jan,
  journal = {arXiv:2101.08366 [cs]},
  eprint = {2101.08366},
  primaryclass = {cs},
  urldate = {2021-03-10},
  archiveprefix = {arxiv}
}

@article{Budker2016,
  title = {Rules for Collaborative Scientific Writing},
  author = {Budker, Dmitry and Kimball, Derek F. Jackson},
  year = {2016},
  journal = {arXiv preprint arXiv:1607.02942},
  eprint = {1607.02942},
  archiveprefix = {arxiv}
}

@book{Budynas2011,
  title = {Shigley's Mechanical Engineering Design},
  author = {Budynas, Richard Gordon and Nisbett, J. Keith},
  year = {2011},
  volume = {9},
  publisher = {{McGraw-Hill New York}}
}

@article{Buratti2020,
  title = {Analytical Studies of {{PROTO-SPHERA}} Equilibria},
  author = {Buratti, P. and Tirozzi, B. and Alladio, F. and Micozzi, P.},
  year = {2020},
  journal = {Journal of Plasma Physics},
  volume = {86},
  number = {6},
  publisher = {{CAMBRIDGE UNIV PRESS 32 AVENUE OF THE AMERICAS, NEW YORK, NY 10013-2473 USA}},
  doi = {10/gjqqvw}
}

@article{Burhenn2009,
  title = {On Impurity Handling in High Performance Stellarator/Heliotron Plasmas},
  author = {Burhenn, R. and Feng, Y. and Ida, K. and Maassberg, H. and McCarthy, K. J. and Kalinina, D. and Kobayashi, M. and Morita, S. and Nakamura, Y. and Nozato, H. and Okamura, S. and Sudo, S. and Suzuki, C. and Tamura, N. and Weller, A. and Yoshinuma, M. and Zurro, B.},
  year = {2009},
  month = apr,
  journal = {Nuclear Fusion},
  volume = {49},
  number = {6},
  pages = {065005},
  publisher = {{IOP Publishing}},
  issn = {0029-5515},
  doi = {10.1088/0029-5515/49/6/065005},
  urldate = {2021-09-30},
  langid = {english}
}

@book{Buscha2013,
  title = {Begegnungen},
  author = {Buscha, Anne and Szita, Szilvia},
  year = {2013},
  volume = {1}
}

@article{Bustreo2015,
  title = {The {{Monte Carlo}} Approach to the Economics of a {{DEMO-like}} Power Plant},
  author = {Bustreo, Chiara and Bolzonella, Tommaso and Zollino, Giuseppe},
  year = {2015},
  month = oct,
  journal = {Fusion Engineering and Design},
  series = {Proceedings of the 28th {{Symposium On Fusion Technology}} ({{SOFT-28}})},
  volume = {98--99},
  pages = {2108--2111},
  issn = {0920-3796},
  doi = {10.1016/j.fusengdes.2015.02.014},
  urldate = {2021-05-15},
  langid = {english}
}

@article{Bustreo2019,
  title = {How Fusion Power Can Contribute to a Fully Decarbonized {{European}} Power Mix after 2050},
  author = {Bustreo, C. and Giuliani, U. and Maggio, D. and Zollino, G.},
  year = {2019},
  month = sep,
  journal = {Fusion Engineering and Design},
  series = {{{SI}}:{{SOFT-30}}},
  volume = {146},
  pages = {2189--2193},
  issn = {0920-3796},
  doi = {10.1016/j.fusengdes.2019.03.150},
  urldate = {2022-06-01},
  langid = {english}
}

@article{Buttery2021,
  title = {The Advanced Tokamak Path to a Compact Net Electric Fusion Pilot Plant},
  author = {Buttery, R. J. and Park, J. M. and McClenaghan, J. T. and Weisberg, D. and Canik, J. and Ferron, J. and Garofalo, A. and Holcomb, C. T. and Leuer, J. and Snyder, P. B. and Team, The Atom Project},
  year = {2021},
  month = mar,
  journal = {Nuclear Fusion},
  volume = {61},
  number = {4},
  pages = {046028},
  publisher = {{IOP Publishing}},
  issn = {0029-5515},
  doi = {10.1088/1741-4326/abe4af},
  urldate = {2021-04-27},
  langid = {english}
}

@article{Cachola2020,
  title = {{{TLDR}}: {{Extreme Summarization}} of {{Scientific Documents}}},
  shorttitle = {{{TLDR}}},
  author = {Cachola, Isabel and Lo, Kyle and Cohan, Arman and Weld, Daniel S.},
  year = {2020},
  month = oct,
  journal = {arXiv:2004.15011 [cs]},
  eprint = {2004.15011},
  primaryclass = {cs},
  urldate = {2021-07-30},
  archiveprefix = {arxiv}
}

@article{Cai2021,
  title = {Physics-{{Informed Neural Networks}} ({{PINNs}}) for {{Heat Transfer Problems}}},
  author = {Cai, Shengze and Wang, Zhicheng and Wang, Sifan and Perdikaris, Paris and Karniadakis, George},
  year = {2021},
  month = mar,
  journal = {Journal of Heat Transfer},
  issn = {0022-1481},
  doi = {10/gjnkwv},
  urldate = {2021-04-06}
}

@article{Cai2021a,
  title = {{{DeepM}}\&{{Mnet}}: {{Inferring}} the Electroconvection Multiphysics Fields Based on Operator Approximation by Neural Networks},
  shorttitle = {{{DeepM}}\&{{Mnet}}},
  author = {Cai, Shengze and Wang, Zhicheng and Lu, Lu and Zaki, Tamer A. and Karniadakis, George Em},
  year = {2021},
  month = jul,
  journal = {Journal of Computational Physics},
  volume = {436},
  eprint = {2009.12935},
  pages = {110296},
  issn = {00219991},
  doi = {10.1016/j.jcp.2021.110296},
  urldate = {2021-04-30},
  archiveprefix = {arxiv}
}

@article{Callaghan1999,
  title = {Fast Equilibrium Interpretation on the {{W7-AS}} Stellarator Using Function Parameterization},
  author = {Callaghan, H. P. and McCarthy, P. J. and Geiger, J.},
  year = {1999},
  month = apr,
  journal = {Nuclear Fusion},
  volume = {39},
  number = {4},
  pages = {509--523},
  publisher = {{IOP Publishing}},
  issn = {0029-5515},
  doi = {10.1088/0029-5515/39/4/308},
  urldate = {2021-08-18},
  langid = {english}
}

@article{Callaghan2000,
  title = {Pressure Profile Recovery on {{W7-AS}} with Idealized External Magnetic Measurements Using Function Parametrization},
  author = {Callaghan, H. P. and McCarthy, P. J. and Geiger, J.},
  year = {2000},
  month = oct,
  journal = {Plasma Physics and Controlled Fusion},
  volume = {42},
  number = {10},
  pages = {1013--1022},
  publisher = {{IOP Publishing}},
  issn = {0741-3335},
  doi = {2011041911192500},
  urldate = {2021-08-18},
  langid = {english}
}

@book{Capinski2012,
  title = {Stochastic Calculus for Finance},
  author = {Capi{\'n}ski, Marek and Kopp, Ekkehard and Traple, Janusz},
  year = {2012},
  publisher = {{Cambridge University Press}}
}

@article{Carayannis2022,
  title = {{{REVIEWING FUSION ENERGY TO ADDRESS CLIMATE CHANGE BY}} 2050},
  author = {Carayannis, Elias G. and Draper, John and Crumpton, Charles David},
  year = {2022}
}

@article{Carlton-Jones2020,
  title = {Computing the Shape Gradient of Stellarator Coil Complexity with Respect to the Plasma Boundary},
  author = {{Carlton-Jones}, Arthur and Paul, Elizabeth J. and Dorland, William},
  year = {2020},
  month = nov,
  journal = {arXiv:2011.03702 [physics]},
  eprint = {2011.03702},
  primaryclass = {physics},
  urldate = {2020-12-26},
  archiveprefix = {arxiv}
}

@article{Carolan2015,
  title = {Universal Linear Optics},
  author = {Carolan, J. and Harrold, C. and Sparrow, C. and {Martin-Lopez}, E. and Russell, N. J. and Silverstone, J. W. and Shadbolt, P. J. and Matsuda, N. and Oguma, M. and Itoh, M. and Marshall, G. D. and Thompson, M. G. and Matthews, J. C. F. and Hashimoto, T. and O'Brien, J. L. and Laing, A.},
  year = {2015},
  month = aug,
  journal = {Science},
  volume = {349},
  number = {6249},
  pages = {711--716},
  issn = {0036-8075, 1095-9203},
  doi = {10.1126/science.aab3642},
  urldate = {2019-08-25},
  langid = {english}
}

@phdthesis{Carpanese2021,
  title = {Development of Free-Boundary Equilibrium and Transport Solvers for Simulation and Real-Time Interpretation of Tokamak Experiments},
  author = {Carpanese, Francesco},
  year = {2021},
  address = {{Lausanne}},
  doi = {10.5075/epfl-thesis-7914},
  langid = {english},
  school = {EPFL}
}

@article{Carreras1988,
  title = {Low-Aspect-Ratio Torsatron Configurations},
  author = {Carreras, B. A. and Dominguez, N. and Garcia, L. and Lynch, V. E. and Lyon, J. F. and Cary, J. R. and Hanson, J. D. and Navarro, A. P.},
  year = {1988},
  month = jul,
  journal = {Nuclear Fusion},
  volume = {28},
  number = {7},
  pages = {1195--1207},
  publisher = {{IOP Publishing}},
  issn = {0029-5515},
  doi = {10.1088/0029-5515/28/7/004},
  urldate = {2021-08-17},
  langid = {english}
}

@book{Cartea2015,
  title = {Algorithmic and High-Frequency Trading},
  author = {Cartea, {\'A}lvaro and Jaimungal, Sebastian and Penalva, Jos{\'e}},
  year = {2015},
  publisher = {{Cambridge University Press}}
}

@inproceedings{Caruana2004,
  title = {Ensemble Selection from Libraries of Models},
  booktitle = {Proceedings of the Twenty-First International Conference on {{Machine}} Learning},
  author = {Caruana, Rich and {Niculescu-Mizil}, Alexandru and Crew, Geoff and Ksikes, Alex},
  year = {2004},
  pages = {18}
}

@article{Cary1997,
  title = {Omnigenity and Quasihelicity in Helical Plasma Confinement Systems},
  author = {Cary, John R. and Shasharina, Svetlana G.},
  year = {1997},
  month = sep,
  journal = {Physics of Plasmas},
  volume = {4},
  number = {9},
  pages = {3323--3333},
  publisher = {{American Institute of Physics}},
  issn = {1070-664X},
  doi = {10.1063/1.872473},
  urldate = {2022-07-20}
}

@article{Cerf1998,
  title = {Optical Simulation of Quantum Logic},
  author = {Cerf, N. J. and Adami, C. and Kwiat, P. G.},
  year = {1998},
  month = mar,
  journal = {Physical Review A},
  volume = {57},
  number = {3},
  pages = {R1477-R1480},
  doi = {10.1103/PhysRevA.57.R1477},
  urldate = {2019-08-30}
}

@article{Cerfon2010,
  title = {``{{One}} Size Fits All'' Analytic Solutions to the {{Grad}}\textendash{{Shafranov}} Equation},
  author = {Cerfon, Antoine J. and Freidberg, Jeffrey P.},
  year = {2010},
  month = mar,
  journal = {Physics of Plasmas},
  volume = {17},
  number = {3},
  pages = {032502},
  publisher = {{American Institute of Physics}},
  issn = {1070-664X},
  doi = {10.1063/1.3328818},
  urldate = {2022-05-14}
}

@article{Cesaro2021,
  title = {Ammonia to Power: {{Forecasting}} the Levelized Cost of Electricity from Green Ammonia in Large-Scale Power Plants},
  shorttitle = {Ammonia to Power},
  author = {Cesaro, Zac and Ives, Matthew and {Nayak-Luke}, Richard and Mason, Mike and {Ba{\~n}ares-Alc{\'a}ntara}, Ren{\'e}},
  year = {2021},
  journal = {Applied Energy},
  volume = {282},
  pages = {116009},
  publisher = {{Elsevier}},
  doi = {10.1016/j.apenergy.2020.116009}
}

@article{Chandrasekhar1957,
  title = {On Force-Free Magnetic Fields.},
  author = {Chandrasekhar, Subramanyan and Kendall, Paul C.},
  year = {1957},
  journal = {The Astrophysical Journal},
  volume = {126},
  pages = {457},
  doi = {10.1086/146413}
}

@article{Chang2018,
  title = {Heat Transfer Prediction of Supercritical Water with Artificial Neural Networks},
  author = {Chang, Wanli and Chu, Xu and Binte Shaik Fareed, Anes Fatima and Pandey, Sandeep and Luo, Jiayu and Weigand, Bernhard and Laurien, Eckart},
  year = {2018},
  month = feb,
  journal = {Applied Thermal Engineering},
  volume = {131},
  pages = {815--824},
  issn = {13594311},
  doi = {10.1016/j.applthermaleng.2017.12.063},
  urldate = {2020-10-14},
  langid = {english}
}

@article{Chapman2010,
  title = {Controlling Sawtooth Oscillations in Tokamak Plasmas},
  author = {Chapman, I. T.},
  year = {2010},
  month = nov,
  journal = {Plasma Physics and Controlled Fusion},
  volume = {53},
  number = {1},
  pages = {013001},
  publisher = {{IOP Publishing}},
  issn = {0741-3335},
  doi = {10.1088/0741-3335/53/1/013001},
  urldate = {2022-01-10},
  langid = {english}
}

@inproceedings{Char2019,
  title = {Offline {{Contextual Bayesian Optimization}}},
  booktitle = {Advances in {{Neural Information Processing Systems}}},
  author = {Char, Ian and Chung, Youngseog and Neiswanger, Willie and Kandasamy, Kirthevasan and Nelson, Andrew Oakleigh and Boyer, Mark and Kolemen, Egemen and Schneider, Jeff},
  year = {2019},
  volume = {32},
  publisher = {{Curran Associates, Inc.}},
  urldate = {2022-02-08}
}

@misc{Charton2021,
  title = {Learning Advanced Mathematical Computations from Examples},
  author = {Charton, Fran{\c c}ois and Hayat, Amaury and Lample, Guillaume},
  year = {2021},
  month = mar,
  number = {arXiv:2006.06462},
  eprint = {arXiv:2006.06462},
  publisher = {{arXiv}},
  doi = {10.48550/arXiv.2006.06462},
  urldate = {2022-07-29},
  archiveprefix = {arxiv}
}

@article{Chen1995,
  title = {Universal Approximation to Nonlinear Operators by Neural Networks with Arbitrary Activation Functions and Its Application to Dynamical Systems},
  author = {Chen, Tianping and Chen, Hong},
  year = {1995},
  journal = {IEEE Transactions on Neural Networks},
  volume = {6},
  number = {4},
  pages = {911--917},
  publisher = {{IEEE}},
  doi = {10.1109/72.392253}
}

@article{Chen1995a,
  title = {Approximation Capability to Functions of Several Variables, Nonlinear Functionals, and Operators by Radial Basis Function Neural Networks},
  author = {Chen, Tianping and Chen, Hong},
  year = {1995},
  month = jul,
  journal = {IEEE Transactions on Neural Networks},
  volume = {6},
  number = {4},
  pages = {904--910},
  issn = {1941-0093},
  doi = {10.1109/72.392252}
}

@book{Chen2011,
  title = {An Indispensable Truth: How Fusion Power Can Save the Planet},
  shorttitle = {An Indispensable Truth},
  author = {Chen, Francis},
  year = {2011},
  publisher = {{Springer Science \& Business Media}}
}

@book{Chen2016,
  title = {Introduction to {{Plasma Physics}} and {{Controlled Fusion}}},
  author = {Chen, Francis F.},
  year = {2016},
  publisher = {{Springer International Publishing}},
  address = {{Cham}},
  doi = {10.1007/978-3-319-22309-4},
  urldate = {2020-10-14},
  isbn = {978-3-319-22308-7 978-3-319-22309-4},
  langid = {english}
}

@article{Chen2019,
  title = {Neural {{Ordinary Differential Equations}}},
  author = {Chen, Ricky T. Q. and Rubanova, Yulia and Bettencourt, Jesse and Duvenaud, David},
  year = {2019},
  month = dec,
  journal = {arXiv:1806.07366 [cs, stat]},
  eprint = {1806.07366},
  primaryclass = {cs, stat},
  urldate = {2021-05-12},
  archiveprefix = {arxiv}
}

@article{Chen2020,
  title = {A Comparison Study of Deep {{Galerkin}} Method and Deep {{Ritz}} Method for Elliptic Problems with Different Boundary Conditions},
  author = {Chen, Jingrun and Du, Rui and Wu, Keke},
  year = {2020},
  month = jun,
  journal = {Communications in Mathematical Research},
  volume = {36},
  number = {3},
  eprint = {2005.04554},
  pages = {354--376},
  issn = {1674-5647, 2707-8523},
  doi = {10/gjnmg9},
  urldate = {2021-04-06},
  archiveprefix = {arxiv}
}

@article{Chen2020a,
  title = {Solving {{Inverse Stochastic Problems}} from {{Discrete Particle Observations Using}} the {{Fokker-Planck Equation}} and {{Physics-informed Neural Networks}}},
  author = {Chen, Xiaoli and Yang, Liu and Duan, Jinqiao and Karniadakis, George Em},
  year = {2020},
  month = aug,
  journal = {arXiv:2008.10653 [physics, stat]},
  eprint = {2008.10653},
  primaryclass = {physics, stat},
  urldate = {2021-11-18},
  archiveprefix = {arxiv}
}

@article{Chen2021,
  title = {Physics-Informed Learning of Governing Equations from Scarce Data},
  author = {Chen, Zhao and Liu, Yang and Sun, Hao},
  year = {2021},
  month = oct,
  journal = {Nature Communications},
  volume = {12},
  number = {1},
  pages = {6136},
  publisher = {{Nature Publishing Group}},
  issn = {2041-1723},
  doi = {10.1038/s41467-021-26434-1},
  urldate = {2023-03-27},
  copyright = {2021 The Author(s)},
  langid = {english}
}

@article{Chen2022,
  title = {Automated Discovery of Fundamental Variables Hidden in Experimental Data},
  author = {Chen, Boyuan and Huang, Kuang and Raghupathi, Sunand and Chandratreya, Ishaan and Du, Qiang and Lipson, Hod},
  year = {2022},
  journal = {Nature Computational Science},
  volume = {2},
  number = {7},
  pages = {433--442},
  publisher = {{Nature Publishing Group}},
  doi = {10.1038/s43588-022-00281-6}
}

@article{Chilenski2015,
  title = {Improved Profile Fitting and Quantification of Uncertainty in Experimental Measurements of Impurity Transport Coefficients Using {{Gaussian}} Process Regression},
  author = {Chilenski, M. A. and Greenwald, M. and Marzouk, Y. and Howard, N. T. and White, A. E. and Rice, J. E. and Walk, J. R.},
  year = {2015},
  month = jan,
  journal = {Nuclear Fusion},
  volume = {55},
  number = {2},
  pages = {023012},
  publisher = {{IOP Publishing}},
  issn = {0029-5515},
  doi = {10.1088/0029-5515/55/2/023012},
  urldate = {2020-11-19},
  langid = {english}
}

@article{Chlechowitz2015,
  title = {Magnetic Diagnostic Optimization for Plasma Equilibrium Reconstruction in the {{HSX}} Stellarator},
  author = {Chlechowitz, E. and Talmadge, J. N. and Hanson, J. D. and Anderson, F. S. B. and Anderson, D. T.},
  year = {2015},
  month = sep,
  journal = {Nuclear Fusion},
  volume = {55},
  number = {11},
  pages = {113012},
  publisher = {{IOP Publishing}},
  issn = {0029-5515},
  doi = {10.1088/0029-5515/55/11/113012},
  urldate = {2023-01-24},
  langid = {english}
}

@article{Choudhary2020,
  title = {Physics-Enhanced Neural Networks Learn Order and Chaos},
  author = {Choudhary, Anshul and Lindner, John F. and Holliday, Elliott G. and Miller, Scott T. and Sinha, Sudeshna and Ditto, William L.},
  year = {2020},
  journal = {Physical Review E},
  volume = {101},
  number = {6},
  pages = {062207},
  publisher = {{APS}},
  doi = {10/gg86t5}
}

@article{Chow2019,
  title = {Lyapunov-Based {{Safe Policy Optimization}} for {{Continuous Control}}},
  author = {Chow, Yinlam and Nachum, Ofir and Faust, Aleksandra and {Duenez-Guzman}, Edgar and Ghavamzadeh, Mohammad},
  year = {2019},
  month = feb,
  journal = {arXiv:1901.10031 [cs, stat]},
  eprint = {1901.10031},
  primaryclass = {cs, stat},
  urldate = {2020-01-30},
  archiveprefix = {arxiv},
  langid = {english}
}

@misc{Chowdhury2019,
  title = {Efficient {{Parameter Sampling}} for {{Neural Network Construction}}},
  author = {Chowdhury, Drimik Roy and Kasim, Muhammad Firmansyah},
  year = {2019},
  month = dec,
  number = {arXiv:1912.10559},
  eprint = {arXiv:1912.10559},
  publisher = {{arXiv}},
  urldate = {2022-10-07},
  archiveprefix = {arxiv}
}

@article{Chu2021,
  title = {Slowed Canonical Progress in Large Fields of Science},
  author = {Chu, Johan S. G. and Evans, James A.},
  year = {2021},
  month = oct,
  journal = {Proceedings of the National Academy of Sciences},
  volume = {118},
  number = {41},
  publisher = {{National Academy of Sciences}},
  issn = {0027-8424, 1091-6490},
  doi = {10.1073/pnas.2021636118},
  urldate = {2021-10-11},
  chapter = {Social Sciences},
  copyright = {Copyright \textcopyright{} 2021 the Author(s). Published by PNAS.. https://creativecommons.org/licenses/by-nc-nd/4.0/This open access article is distributed under Creative Commons Attribution-NonCommercial-NoDerivatives License 4.0 (CC BY-NC-ND).},
  langid = {english},
  pmid = {34607941}
}

@article{Chuyanov2017,
  title = {Modular Fusion Power Plant},
  author = {Chuyanov, V. A. and Gryaznevich, M. P.},
  year = {2017},
  journal = {Fusion engineering and design},
  volume = {122},
  pages = {238--252},
  publisher = {{Elsevier}},
  doi = {10.1016/j.fusengdes.2017.07.017}
}

@article{Cianciosa2017,
  title = {Helical Core Reconstruction of a {{DIII-D}} Hybrid Scenario Tokamak Discharge},
  author = {Cianciosa, M. and Wingen, A. and Hirshman, S. P. and Seal, S. K. and Unterberg, E. A. and Wilcox, R. S. and Piovesan, P. and Lao, L. and Turco, F.},
  year = {2017},
  month = may,
  journal = {Nuclear Fusion},
  volume = {57},
  number = {7},
  pages = {076015},
  publisher = {{IOP Publishing}},
  issn = {0029-5515},
  doi = {10.1088/1741-4326/aa6f82},
  urldate = {2021-06-21},
  langid = {english}
}

@article{Cianciosa2018,
  title = {{{3D}} Equilibrium Reconstruction with Islands},
  author = {Cianciosa, M. and Hirshman, S. P. and Seal, S. K. and Shafer, M. W.},
  year = {2018},
  month = mar,
  journal = {Plasma Physics and Controlled Fusion},
  volume = {60},
  number = {4},
  pages = {044017},
  publisher = {{IOP Publishing}},
  issn = {0741-3335},
  doi = {10.1088/1361-6587/aaaf90},
  urldate = {2021-06-21},
  langid = {english}
}

@article{Ciresan2010,
  title = {Deep {{Big Simple Neural Nets Excel}} on {{Handwritten Digit Recognition}}},
  author = {Ciresan, Dan Claudiu and Meier, Ueli and Gambardella, Luca Maria and Schmidhuber, Juergen},
  year = {2010},
  month = mar,
  journal = {arXiv:1003.0358 [cs]},
  eprint = {1003.0358},
  primaryclass = {cs},
  doi = {10.1162/NECO_a_00052},
  urldate = {2020-12-08},
  archiveprefix = {arxiv}
}

@article{Citrin2015,
  title = {Real-Time Capable First Principle Based Modelling of Tokamak Turbulent Transport},
  author = {Citrin, J. and Breton, S. and Felici, F. and Imbeaux, F. and Aniel, T. and Artaud, J. and Baiocchi, B. and Bourdelle, C. and Camenen, Y. and Garc{\'i}a, J.},
  year = {2015},
  doi = {10.1088/0029-5515/55/9/092001}
}

@article{Citrin2020,
  title = {Neural Network Surrogate Modelling of Tokamak Plasma Turbulence},
  author = {Citrin, Jonathan and Bourdelle, Clarisse and Camenen, Yann and Felici, Federico and Ho, Aaron and Horn, Philipp and Kremers, Bart and {van de Plassche}, Karel},
  year = {2020},
  journal = {Bulletin of the American Physical Society},
  publisher = {{APS}}
}

@article{Clements,
  title = {Linear {{Quantum Optics}}: {{Components}} and {{Applications}}},
  author = {Clements, William R and Hall, Lady Margaret},
  pages = {240},
  langid = {english}
}

@inproceedings{Cohen2016,
  title = {On the Expressive Power of Deep Learning: {{A}} Tensor Analysis},
  shorttitle = {On the Expressive Power of Deep Learning},
  booktitle = {Conference on Learning Theory},
  author = {Cohen, Nadav and Sharir, Or and Shashua, Amnon},
  year = {2016},
  pages = {698--728},
  publisher = {{PMLR}}
}

@article{Cohen2016a,
  title = {Mini-Course on {{Geometric Numerical Integration}}},
  author = {Cohen, David and University, Ume{\aa}},
  year = {2016},
  pages = {45},
  langid = {english}
}

@article{Coleman2019,
  title = {{{BLUEPRINT}}: {{A}} Novel Approach to Fusion Reactor Design},
  shorttitle = {{{BLUEPRINT}}},
  author = {Coleman, M. and McIntosh, S.},
  year = {2019},
  month = feb,
  journal = {Fusion Engineering and Design},
  volume = {139},
  pages = {26--38},
  issn = {0920-3796},
  doi = {10.1016/j.fusengdes.2018.12.036},
  urldate = {2021-11-04},
  langid = {english}
}

@article{Conlin2022,
  title = {The {{DESC Stellarator Code Suite Part II}}: {{Perturbation}} and Continuation Methods},
  shorttitle = {The {{DESC Stellarator Code Suite Part II}}},
  author = {Conlin, Rory and Dudt, Daniel W. and Panici, Dario and Kolemen, Egemen},
  year = {2022},
  month = mar,
  journal = {arXiv:2203.15927 [physics]},
  eprint = {2203.15927},
  primaryclass = {physics},
  urldate = {2022-04-21},
  archiveprefix = {arxiv}
}

@article{Conlon2008,
  title = {The New Engineer: Between Employability and Social Responsibility},
  shorttitle = {The New Engineer},
  author = {Conlon, E.},
  year = {2008},
  month = may,
  journal = {European Journal of Engineering Education},
  volume = {33},
  number = {2},
  pages = {151--159},
  publisher = {{Taylor \& Francis}},
  issn = {0304-3797},
  doi = {10.1080/03043790801996371},
  urldate = {2021-02-28}
}

@book{Coolen2005,
  title = {Theory of Neural Information Processing Systems},
  author = {Coolen, Anthony CC and K{\"u}hn, Reimer and Sollich, Peter},
  year = {2005},
  publisher = {{OUP Oxford}}
}

@article{Cooley1965,
  title = {An Algorithm for the Machine Calculation of Complex {{Fourier}} Series},
  author = {Cooley, James W. and Tukey, John W.},
  year = {1965},
  journal = {Mathematics of Computation},
  volume = {19},
  number = {90},
  pages = {297--301},
  issn = {0025-5718, 1088-6842},
  doi = {10.1090/S0025-5718-1965-0178586-1},
  urldate = {2021-12-16},
  langid = {english}
}

@article{Cooper2011,
  title = {Magnetohydrodynamic Equilibrium and the Stability of Tokamak and Reversed-Field Pinch Systems with {{3D}} Helical Cores},
  author = {Cooper, W A and Graves, J P and Sauter, O and Terranova, D and Gobbin, M and Marrelli, L and Martin, P and Predebon, I},
  year = {2011},
  month = aug,
  journal = {Plasma Physics and Controlled Fusion},
  volume = {53},
  number = {8},
  pages = {084001},
  issn = {0741-3335, 1361-6587},
  doi = {10.1088/0741-3335/53/8/084001},
  urldate = {2020-10-14},
  langid = {english}
}

@article{Costley2021,
  title = {Fusion Performance of Spherical and Conventional Tokamaks: Implications for Compact Pilot Plants and Reactors},
  shorttitle = {Fusion Performance of Spherical and Conventional Tokamaks},
  author = {Costley, A. E. and McNamara, S. A. M.},
  year = {2021},
  month = jan,
  journal = {Plasma Physics and Controlled Fusion},
  volume = {63},
  number = {3},
  pages = {035005},
  publisher = {{IOP Publishing}},
  issn = {0741-3335},
  doi = {10.1088/1361-6587/abcdfc},
  urldate = {2021-03-02},
  langid = {english}
}

@book{Cover1999,
  title = {Elements of Information Theory},
  author = {Cover, Thomas M.},
  year = {1999},
  publisher = {{John Wiley \& Sons}}
}

@article{Crawford1991,
  title = {Introduction to Bifurcation Theory},
  author = {Crawford, John David},
  year = {1991},
  month = oct,
  journal = {Reviews of Modern Physics},
  volume = {63},
  number = {4},
  pages = {991--1037},
  publisher = {{American Physical Society}},
  doi = {10.1103/RevModPhys.63.991},
  urldate = {2021-05-20}
}

@article{Creely2020,
  title = {Overview of the {{SPARC}} Tokamak},
  author = {Creely, A. J. and Greenwald, M. J. and Ballinger, S. B. and Brunner, D. and Canik, J. and Doody, J. and F{\"u}l{\"o}p, T. and Garnier, D. T. and Granetz, R. and Gray, T. K. and Holland, C. and Howard, N. T. and Hughes, J. W. and Irby, J. H. and Izzo, V. A. and Kramer, G. J. and Kuang, A. Q. and LaBombard, B. and Lin, Y. and Lipschultz, B. and Logan, N. C. and Lore, J. D. and Marmar, E. S. and Montes, K. and Mumgaard, R. T. and {Paz-Soldan}, C. and Rea, C. and Reinke, M. L. and {Rodriguez-Fernandez}, P. and S{\"a}rkim{\"a}ki, K. and Sciortino, F. and Scott, S. D. and Snicker, A. and Snyder, P. B. and Sorbom, B. N. and Sweeney, R. and Tinguely, R. A. and Tolman, E. A. and Umansky, M. and Vallhagen, O. and Varje, J. and Whyte, D. G. and Wright, J. C. and Wukitch, S. J. and Zhu, J. and Team, the SPARC},
  year = {2020},
  month = oct,
  journal = {Journal of Plasma Physics},
  volume = {86},
  number = {5},
  publisher = {{Cambridge University Press}},
  issn = {0022-3778, 1469-7807},
  doi = {10.1017/S0022377820001257},
  urldate = {2021-02-25},
  langid = {english}
}

@article{Crichton2020,
  title = {Neural Networks for Open and Closed {{Literature-based Discovery}}},
  author = {Crichton, Gamal and Baker, Simon and Guo, Yufan and Korhonen, Anna},
  year = {2020},
  month = may,
  journal = {PLOS ONE},
  volume = {15},
  number = {5},
  pages = {e0232891},
  publisher = {{Public Library of Science}},
  issn = {1932-6203},
  doi = {10/gjpptn},
  urldate = {2021-04-12},
  langid = {english}
}

@article{Cust2017,
  title = {Evidence for a Presource Curse? Oil Discoveries, Elevated Expectations, and Growth Disappointments},
  shorttitle = {Evidence for a Presource Curse?},
  author = {Cust, James Frederick and Mihalyi, David},
  year = {2017},
  journal = {Oil Discoveries, Elevated Expectations, and Growth Disappointments (July 10, 2017). World Bank Policy Research Working Paper},
  number = {8140}
}

@article{Cybenko1989,
  title = {Approximation by Superpositions of a Sigmoidal Function},
  author = {Cybenko, G.},
  year = {1989},
  month = dec,
  journal = {Mathematics of Control, Signals and Systems},
  volume = {2},
  number = {4},
  pages = {303--314},
  issn = {1435-568X},
  doi = {10.1007/BF02551274},
  urldate = {2020-11-12},
  langid = {english}
}

@article{Dai2017,
  title = {Investigation of Heat Flux Deposition on Divertor Target on the {{Large Helical Device}} with {{EMC3-EIRENE}} Modelling},
  author = {Dai, Shuyu and Kobayashi, M. and Kawamura, G. and Masuzaki, S. and Tanaka, H. and Suzuki, Y. and Feng, Y. and {and}, D. Z. Wang},
  year = {2017},
  month = jun,
  journal = {Plasma Physics and Controlled Fusion},
  volume = {59},
  number = {8},
  pages = {085013},
  publisher = {{IOP Publishing}},
  issn = {0741-3335},
  doi = {10.1088/1361-6587/aa7599},
  urldate = {2021-08-18},
  langid = {english}
}

@article{Dalal2018,
  title = {Safe {{Exploration}} in {{Continuous Action Spaces}}},
  author = {Dalal, Gal and Dvijotham, Krishnamurthy and Vecerik, Matej and Hester, Todd and Paduraru, Cosmin and Tassa, Yuval},
  year = {2018},
  month = jan,
  journal = {arXiv:1801.08757 [cs]},
  eprint = {1801.08757},
  primaryclass = {cs},
  urldate = {2020-10-14},
  archiveprefix = {arxiv},
  langid = {english}
}

@article{Damizia2021,
  title = {Optical Tomography of the Plasma on the {{PROTO-SPHERA}} Experiment},
  author = {Damizia, Yacopo and Iafrati, Matteo and Liuzza, Davide and Boncagni, Luca and Micozzi, Paolo and Alladio, Franco},
  year = {2021},
  journal = {Fusion Engineering and Design},
  volume = {169},
  pages = {112461},
  publisher = {{Elsevier}},
  doi = {10/gjh8c7}
}

@inproceedings{DanForesee1997,
  title = {Gauss-{{Newton}} Approximation to {{Bayesian}} Learning},
  booktitle = {Proceedings of {{International Conference}} on {{Neural Networks}} ({{ICNN}}'97)},
  author = {Dan Foresee, F. and Hagan, M.T.},
  year = {1997},
  month = jun,
  volume = {3},
  pages = {1930-1935 vol.3},
  doi = {10.1109/ICNN.1997.614194}
}

@article{Dao2020,
  title = {Learning {{Fast Algorithms}} for {{Linear Transforms Using Butterfly Factorizations}}},
  author = {Dao, Tri and Gu, Albert and Eichhorn, Matthew and Rudra, Atri and R{\'e}, Christopher},
  year = {2020},
  month = dec,
  journal = {arXiv:1903.05895 [cs, stat]},
  eprint = {1903.05895},
  primaryclass = {cs, stat},
  urldate = {2021-04-30},
  archiveprefix = {arxiv}
}

@article{Das2000,
  title = {Robustness {{Optimization}} for {{Constrained Nonlinear Programming Problems}}},
  author = {Das, Indraneel},
  year = {2000},
  month = jan,
  journal = {Engineering Optimization},
  volume = {32},
  number = {5},
  pages = {585--618},
  publisher = {{Taylor \& Francis}},
  issn = {0305-215X},
  doi = {10.1080/03052150008941314},
  urldate = {2021-12-14}
}

@book{Das2003,
  title = {Introduction to Nuclear and Particle Physics},
  author = {Das, Ashok and Ferbel, Thomas},
  year = {2003},
  publisher = {{World Scientific}}
}

@article{Dattani2019,
  title = {Pegasus: {{The}} Second Connectivity Graph for Large-Scale Quantum Annealing Hardware},
  shorttitle = {Pegasus},
  author = {Dattani, Nike and Szalay, Szilard and Chancellor, Nick},
  year = {2019},
  month = jan,
  journal = {arXiv:1901.07636 [quant-ph]},
  eprint = {1901.07636},
  primaryclass = {quant-ph},
  urldate = {2019-09-01},
  archiveprefix = {arxiv}
}

@article{Dax2021,
  title = {Real-{{Time Gravitational Wave Science}} with {{Neural Posterior Estimation}}},
  author = {Dax, Maximilian and Green, Stephen R. and Gair, Jonathan and Macke, Jakob H. and Buonanno, Alessandra and Sch{\"o}lkopf, Bernhard},
  year = {2021},
  month = dec,
  journal = {Physical Review Letters},
  volume = {127},
  number = {24},
  pages = {241103},
  publisher = {{American Physical Society}},
  doi = {10.1103/PhysRevLett.127.241103},
  urldate = {2022-03-11}
}

@article{Daxberger2021,
  title = {Laplace {{Redux}} -- {{Effortless Bayesian Deep Learning}}},
  author = {Daxberger, Erik and Kristiadi, Agustinus and Immer, Alexander and Eschenhagen, Runa and Bauer, Matthias and Hennig, Philipp},
  year = {2021},
  month = oct,
  journal = {arXiv:2106.14806 [cs, stat]},
  eprint = {2106.14806},
  primaryclass = {cs, stat},
  urldate = {2022-02-28},
  archiveprefix = {arxiv}
}

@article{DeChiara2018,
  title = {Public Data and Value Creation in {{Italy}}. {{The}} Findings from the {{Open Data}} 200 Study},
  author = {De Chiara, Francesca},
  year = {2018},
  journal = {Sociologia del Lavoro},
  publisher = {{FrancoAngeli Editore}},
  doi = {10.3280/SL2018-152004}
}

@article{Degrave2022,
  title = {Magnetic Control of Tokamak Plasmas through Deep Reinforcement Learning},
  author = {Degrave, Jonas and Felici, Federico and Buchli, Jonas and Neunert, Michael and Tracey, Brendan and Carpanese, Francesco and Ewalds, Timo and Hafner, Roland and Abdolmaleki, Abbas and {de las Casas}, Diego and Donner, Craig and Fritz, Leslie and Galperti, Cristian and Huber, Andrea and Keeling, James and Tsimpoukelli, Maria and Kay, Jackie and Merle, Antoine and Moret, Jean-Marc and Noury, Seb and Pesamosca, Federico and Pfau, David and Sauter, Olivier and Sommariva, Cristian and Coda, Stefano and Duval, Basil and Fasoli, Ambrogio and Kohli, Pushmeet and Kavukcuoglu, Koray and Hassabis, Demis and Riedmiller, Martin},
  year = {2022},
  month = feb,
  journal = {Nature},
  volume = {602},
  number = {7897},
  pages = {414--419},
  publisher = {{Nature Publishing Group}},
  issn = {1476-4687},
  doi = {10.1038/s41586-021-04301-9},
  urldate = {2022-02-17},
  copyright = {2022 The Author(s)},
  langid = {english}
}

@article{Delucchi2011,
  title = {Providing All Global Energy with Wind, Water, and Solar Power, {{Part II}}: {{Reliability}}, System and Transmission Costs, and Policies},
  shorttitle = {Providing All Global Energy with Wind, Water, and Solar Power, {{Part II}}},
  author = {Delucchi, Mark A. and Jacobson, Mark Z.},
  year = {2011},
  journal = {Energy policy},
  volume = {39},
  number = {3},
  pages = {1170--1190},
  publisher = {{Elsevier}},
  doi = {10.1016/j.enpol.2010.11.045}
}

@book{Dendy1990,
  title = {Plasma Dynamics},
  author = {Dendy, Richard O.},
  year = {1990},
  publisher = {{Clarendon Press}}
}

@article{Deng2021,
  title = {Convergence Rate of {{DeepONets}} for Learning Operators Arising from Advection-Diffusion Equations},
  author = {Deng, Beichuan and Shin, Yeonjong and Lu, Lu and Zhang, Zhongqiang and Karniadakis, George Em},
  year = {2021},
  month = mar,
  journal = {arXiv:2102.10621 [cs, math]},
  eprint = {2102.10621},
  primaryclass = {cs, math},
  urldate = {2021-05-20},
  archiveprefix = {arxiv}
}

@article{DeOliveira2017,
  title = {Learning {{Particle Physics}} by {{Example}}: {{Location-Aware Generative Adversarial Networks}} for {{Physics Synthesis}}},
  shorttitle = {Learning {{Particle Physics}} by {{Example}}},
  author = {{de Oliveira}, Luke and Paganini, Michela and Nachman, Benjamin},
  year = {2017},
  month = sep,
  journal = {Computing and Software for Big Science},
  volume = {1},
  number = {1},
  pages = {4},
  issn = {2510-2044},
  doi = {10.1007/s41781-017-0004-6},
  urldate = {2021-08-11},
  langid = {english}
}

@article{Deppe2020,
  title = {Nighttime {{Photovoltaic Cells}}: {{Electrical Power Generation}} by {{Optically Coupling}} with {{Deep Space}}},
  shorttitle = {Nighttime {{Photovoltaic Cells}}},
  author = {Deppe, Tristan and Munday, Jeremy N.},
  year = {2020},
  month = jan,
  journal = {ACS Photonics},
  volume = {7},
  number = {1},
  pages = {1--9},
  publisher = {{American Chemical Society}},
  doi = {10.1021/acsphotonics.9b00679},
  urldate = {2021-05-29}
}

@article{Desai2021,
  title = {One-{{Shot Transfer Learning}} of {{Physics-Informed Neural Networks}}},
  author = {Desai, Shaan and Mattheakis, Marios and Joy, Hayden and Protopapas, Pavlos and Roberts, Stephen},
  year = {2021},
  month = oct,
  journal = {arXiv:2110.11286 [physics]},
  eprint = {2110.11286},
  primaryclass = {physics},
  urldate = {2021-11-19},
  archiveprefix = {arxiv}
}

@article{DeutschDavidElieser1995,
  title = {Universality in Quantum Computation},
  author = {{Deutsch David Elieser} and {Barenco Adriano} and {Ekert Artur}},
  year = {1995},
  month = jun,
  journal = {Proceedings of the Royal Society of London. Series A: Mathematical and Physical Sciences},
  volume = {449},
  number = {1937},
  pages = {669--677},
  doi = {10.1098/rspa.1995.0065},
  urldate = {2019-08-29}
}

@article{Dewar2022,
  title = {Relaxed Magnetohydrodynamics with Ideal {{Ohm}}'s Law Constraint},
  author = {Dewar, R. L. and Qu, Z. S.},
  year = {2022},
  month = feb,
  journal = {Journal of Plasma Physics},
  volume = {88},
  number = {1},
  publisher = {{Cambridge University Press}},
  issn = {0022-3778, 1469-7807},
  doi = {10.1017/S0022377821001355},
  urldate = {2022-03-22},
  langid = {english}
}

@book{Dhaeseleer1991,
  title = {Flux {{Coordinates}} and {{Magnetic Field Structure}}},
  author = {D'haeseleer, William Denis and Hitchon, William Nicholas Guy and Callen, James D. and Shohet, J. Leon},
  year = {1991},
  publisher = {{Springer Berlin Heidelberg}},
  address = {{Berlin, Heidelberg}},
  doi = {10.1007/978-3-642-75595-8},
  urldate = {2020-10-14},
  isbn = {978-3-642-75597-2 978-3-642-75595-8},
  langid = {english}
}

@article{DiCiccio1996,
  title = {Bootstrap Confidence Intervals},
  author = {DiCiccio, Thomas J. and Efron, Bradley},
  year = {1996},
  month = sep,
  journal = {Statistical Science},
  volume = {11},
  number = {3},
  pages = {189--228},
  publisher = {{Institute of Mathematical Statistics}},
  issn = {0883-4237, 2168-8745},
  doi = {10.1214/ss/1032280214},
  urldate = {2021-08-13}
}

@inproceedings{DiGiovanni2020,
  title = {Finding Multiple Solutions of Odes with Neural Networks},
  booktitle = {{{AAAI-MLPS}} 2020},
  author = {Di Giovanni, Marco and Sondak, David and Protopapas, Pavlos and Brambilla, Marco},
  year = {2020},
  volume = {2587},
  pages = {1--7},
  publisher = {{CEUR-WS}}
}

@article{Dikovsky2021,
  title = {Multi-Instrument {{Bayesian}} Reconstruction of Plasma Shape Evolution in the {{C-2W}} Experiment},
  author = {Dikovsky, M. and Baltz, E. A. and Von Behren, R. and Geraedts, S. and Kast, A. and Langmore, I. and Madams, T. and Norgaard, P. and Platt, J. C. and Romero, J. and Roche, T. and Smith, R. and Trask, E. and Dettrick, S. and Gota, H. and Titus, J. B. and Magee, R. M.},
  year = {2021},
  month = jun,
  journal = {Physics of Plasmas},
  volume = {28},
  number = {6},
  pages = {062503},
  publisher = {{American Institute of Physics}},
  issn = {1070-664X},
  doi = {10.1063/5.0049530},
  urldate = {2021-06-30}
}

@book{Dinklage2005,
  title = {Plasma {{Physics}}: {{Confinement}}, {{Transport}} and {{Collective Effects}}},
  shorttitle = {Plasma {{Physics}}},
  editor = {Dinklage, Andreas and Klinger, Thomas and Marx, Gerrit and Schweikhard, Lutz and Beig, R. and Beiglb{\"o}ck, W. and Domcke, W. and Englert, B.-G. and Frisch, U. and H{\"a}nggi, P. and Hasinger, G. and Hepp, K. and Hillebrandt, W. and Imboden, D. and Jaffe, R. L. and Lipowsky, R. and v. L{\"o}hneysen, H. and Ojima, I. and Sornette, D. and Theisen, S. and Weise, W. and Wess, J. and Zittartz, J.},
  year = {2005},
  series = {Lecture {{Notes}} in {{Physics}}},
  volume = {670},
  publisher = {{Springer Berlin Heidelberg}},
  address = {{Berlin, Heidelberg}},
  doi = {10.1007/b103882},
  urldate = {2020-10-14},
  isbn = {978-3-540-25274-0 978-3-540-31521-6},
  langid = {english}
}

@article{Dinklage2018,
  title = {Magnetic Configuration Effects on the {{Wendelstein}} 7-{{X}} Stellarator},
  author = {Dinklage, A. and Beidler, C. D. and Helander, P. and Fuchert, G. and Maa{\ss}berg, H. and Rahbarnia, K. and Sunn Pedersen, T. and Turkin, Y. and Wolf, R. C. and Alonso, A. and Andreeva, T. and Blackwell, B. and Bozhenkov, S. and Buttensch{\"o}n, B. and Czarnecka, A. and Effenberg, F. and Feng, Y. and Geiger, J. and Hirsch, M. and H{\"o}fel, U. and Jakubowski, M. and Klinger, T. and Knauer, J. and Kocsis, G. and {Kr{\"a}mer-Flecken}, A. and Kubkowska, M. and Langenberg, A. and Laqua, H. P. and Marushchenko, N. and Moll{\'e}n, A. and Neuner, U. and Niemann, H. and Pasch, E. and Pablant, N. and Rudischhauser, L. and Smith, H. M. and Schmitz, O. and Stange, T. and Szepesi, T. and Weir, G. and Windisch, T. and Wurden, G. A. and Zhang, D.},
  year = {2018},
  month = aug,
  journal = {Nature Physics},
  volume = {14},
  number = {8},
  pages = {855--860},
  publisher = {{Nature Publishing Group}},
  issn = {1745-2481},
  doi = {10.1038/s41567-018-0141-9},
  urldate = {2021-04-26},
  copyright = {2018 The Author(s)},
  langid = {english}
}

@article{DIsa2020,
  title = {Performance Analysis of a 2.45 {{GHz}} Microwave Plasma Torch for {{CO2}} Decomposition in Gas Swirl Configuration},
  author = {D'Isa, F. A. and Carbone, E. A. D. and Hecimovic, A. and Fantz, U.},
  year = {2020},
  month = oct,
  journal = {Plasma Sources Science and Technology},
  volume = {29},
  number = {10},
  pages = {105009},
  publisher = {{IOP Publishing}},
  issn = {0963-0252},
  doi = {10.1088/1361-6595/abaa84},
  urldate = {2021-05-10},
  langid = {english}
}

@article{DiSiena2020,
  title = {Turbulence {{Suppression}} by {{Energetic Particle Effects}} in {{Modern Optimized Stellarators}}},
  author = {Di Siena, A. and Ba{\~n}{\'o}n Navarro, A. and Jenko, F.},
  year = {2020},
  month = sep,
  journal = {Physical Review Letters},
  volume = {125},
  number = {10},
  pages = {105002},
  publisher = {{American Physical Society}},
  doi = {10.1103/PhysRevLett.125.105002},
  urldate = {2021-05-04}
}

@article{Dissanayake1994,
  ids = {Dissanayake1994a},
  title = {Neural-Network-Based Approximations for Solving Partial Differential Equations},
  author = {Dissanayake, M. W. M. G. and Phan-Thien, N.},
  year = {1994},
  journal = {Communications in Numerical Methods in Engineering},
  volume = {10},
  number = {3},
  pages = {195--201},
  issn = {1099-0887},
  doi = {10/c8mzns},
  urldate = {2021-04-07},
  langid = {english}
}

@article{DiVincenzo1995,
  title = {Two-Bit Gates Are Universal for Quantum Computation},
  author = {DiVincenzo, David P.},
  year = {1995},
  month = feb,
  journal = {Physical Review A},
  volume = {51},
  number = {2},
  pages = {1015--1022},
  doi = {10.1103/PhysRevA.51.1015},
  urldate = {2019-08-29}
}

@article{DiVincenzo2000,
  title = {The {{Physical Implementation}} of {{Quantum Computation}}},
  author = {DiVincenzo, David P. and IBM},
  year = {2000},
  month = feb,
  doi = {10.1002/1521-3978(200009)48:9/11<771::AID-PROP771>3.0.CO;2-E},
  urldate = {2019-08-25},
  langid = {english}
}

@book{DoCarmo2016,
  title = {Differential Geometry of Curves and Surfaces: Revised and Updated Second Edition},
  shorttitle = {Differential Geometry of Curves and Surfaces},
  author = {Do Carmo, Manfredo P.},
  year = {2016},
  publisher = {{Courier Dover Publications}}
}

@book{Dolan2013,
  title = {Fusion {{Research}}: {{Principles}}},
  shorttitle = {Fusion {{Research}}},
  author = {Dolan, Thomas James},
  year = {2013},
  publisher = {{Elsevier}}
}

@article{Dong2020,
  title = {A Survey on Ensemble Learning},
  author = {Dong, Xibin and Yu, Zhiwen and Cao, Wenming and Shi, Yifan and Ma, Qianli},
  year = {2020},
  month = apr,
  journal = {Frontiers of Computer Science},
  volume = {14},
  number = {2},
  pages = {241--258},
  issn = {2095-2236},
  doi = {10.1007/s11704-019-8208-z},
  urldate = {2021-08-11},
  langid = {english}
}

@article{Dreval2021,
  title = {Determination of Poloidal Mode Numbers of {{MHD}} Modes and Their Radial Location Using a Soft X-Ray Camera Array in the {{Wendelstein}} 7-{{X}} Stellarator},
  author = {Dreval, M. B. and Brandt, C. and Schilling, J. and Thomsen, H. and Beletskii, A. and K{\"o}nies, A. and Team, the W7-X.},
  year = {2021},
  month = apr,
  journal = {Plasma Physics and Controlled Fusion},
  volume = {63},
  number = {6},
  pages = {065006},
  publisher = {{IOP Publishing}},
  issn = {0741-3335},
  doi = {10.1088/1361-6587/abf449},
  urldate = {2021-04-27},
  langid = {english}
}

@article{Drevlak1998,
  title = {Automated Optimization of Stellarator Coils},
  author = {Drevlak, Michael},
  year = {1998},
  journal = {Fusion Technology},
  volume = {33},
  number = {2},
  pages = {106--117},
  publisher = {{Taylor \& Francis}}
}

@article{Drevlak2005,
  title = {{{PIES}} Free Boundary Stellarator Equilibria with Improved Initial Conditions},
  author = {Drevlak, M and Monticello, D and Reiman, A},
  year = {2005},
  pages = {11},
  doi = {10.1088/0029-5515/45/7/022},
  langid = {english}
}

@article{Drevlak2014,
  title = {Fast Particle Confinement with Optimized Coil Currents in the {{W7-X}} Stellarator},
  author = {Drevlak, M. and Geiger, J. and Helander, P. and Turkin, Y.},
  year = {2014},
  month = apr,
  journal = {Nuclear Fusion},
  volume = {54},
  number = {7},
  pages = {073002},
  publisher = {{IOP Publishing}},
  issn = {0029-5515},
  doi = {10.1088/0029-5515/54/7/073002},
  urldate = {2021-05-12},
  langid = {english}
}

@article{Drevlak2018,
  title = {Optimisation of Stellarator Equilibria with {{ROSE}}},
  author = {Drevlak, M. and Beidler, C. D. and Geiger, J. and Helander, P. and Turkin, Y.},
  year = {2018},
  month = nov,
  journal = {Nuclear Fusion},
  volume = {59},
  number = {1},
  pages = {016010},
  publisher = {{IOP Publishing}},
  issn = {0029-5515},
  doi = {10.1088/1741-4326/aaed50},
  urldate = {2020-11-17},
  langid = {english}
}

@article{Du2021,
  title = {Evolutional {{Deep Neural Network}}},
  author = {Du, Yifan and Zaki, Tamer A.},
  year = {2021},
  month = mar,
  journal = {arXiv:2103.09959 [physics]},
  eprint = {2103.09959},
  primaryclass = {physics},
  urldate = {2021-04-08},
  archiveprefix = {arxiv}
}

@article{Duan2016,
  title = {{{RL}}\$\^2\$: {{Fast Reinforcement Learning}} via {{Slow Reinforcement Learning}}},
  shorttitle = {{{RL}}\$\^2\$},
  author = {Duan, Yan and Schulman, John and Chen, Xi and Bartlett, Peter L. and Sutskever, Ilya and Abbeel, Pieter},
  year = {2016},
  month = nov,
  journal = {arXiv:1611.02779 [cs, stat]},
  eprint = {1611.02779},
  primaryclass = {cs, stat},
  urldate = {2020-10-14},
  archiveprefix = {arxiv},
  langid = {english}
}

@article{Dudt2020,
  ids = {Dudt2020a},
  title = {{{DESC}}: {{A}} Stellarator Equilibrium Solver},
  shorttitle = {{{DESC}}},
  author = {Dudt, D. W. and Kolemen, E.},
  year = {2020},
  month = oct,
  journal = {Physics of Plasmas},
  volume = {27},
  number = {10},
  pages = {102513},
  publisher = {{American Institute of Physics}},
  issn = {1070-664X},
  doi = {10.1063/5.0020743},
  urldate = {2020-11-19}
}

@article{Dudt2022,
  title = {The {{DESC Stellarator Code Suite Part III}}: {{Quasi-symmetry}} Optimization},
  shorttitle = {The {{DESC Stellarator Code Suite Part III}}},
  author = {Dudt, Daniel and Conlin, Rory and Panici, Dario and Kolemen, Egemen},
  year = {2022},
  month = mar,
  journal = {arXiv:2204.00078 [physics]},
  eprint = {2204.00078},
  primaryclass = {physics},
  urldate = {2022-04-21},
  archiveprefix = {arxiv}
}

@article{Dullemond,
  title = {Introduction to {{Tensor Calculus}}},
  author = {Dullemond, Kees and Peeters, Kasper},
  pages = {53},
  langid = {english}
}

@article{E2017,
  title = {Deep {{Learning-Based Numerical Methods}} for {{High-Dimensional Parabolic Partial Differential Equations}} and {{Backward Stochastic Differential Equations}}},
  author = {E, Weinan and Han, Jiequn and Jentzen, Arnulf},
  year = {2017},
  month = dec,
  journal = {Communications in Mathematics and Statistics},
  volume = {5},
  number = {4},
  pages = {349--380},
  issn = {2194-671X},
  doi = {10/gg9zwp},
  urldate = {2021-04-06},
  langid = {english}
}

@article{E2018,
  title = {The {{Deep Ritz Method}}: {{A Deep Learning-Based Numerical Algorithm}} for {{Solving Variational Problems}}},
  shorttitle = {The {{Deep Ritz Method}}},
  author = {E, Weinan and Yu, Bing},
  year = {2018},
  month = mar,
  journal = {Communications in Mathematics and Statistics},
  volume = {6},
  number = {1},
  pages = {1--12},
  issn = {2194-671X},
  doi = {10/gjnmhf},
  urldate = {2021-04-06},
  langid = {english}
}

@article{E2020,
  title = {Integrating {{Machine Learning}} with {{Physics-Based Modeling}}},
  author = {E, Weinan and Han, Jiequn and Zhang, Linfeng},
  year = {2020},
  month = jun,
  journal = {arXiv:2006.02619 [physics]},
  eprint = {2006.02619},
  primaryclass = {physics},
  urldate = {2021-03-09},
  archiveprefix = {arxiv}
}

@article{E2020a,
  title = {Algorithms for {{Solving High Dimensional PDEs}}: {{From Nonlinear Monte Carlo}} to {{Machine Learning}}},
  shorttitle = {Algorithms for {{Solving High Dimensional PDEs}}},
  author = {E, Weinan and Han, Jiequn and Jentzen, Arnulf},
  year = {2020},
  month = sep,
  journal = {arXiv:2008.13333 [cs, math]},
  eprint = {2008.13333},
  primaryclass = {cs, math},
  urldate = {2021-03-09},
  archiveprefix = {arxiv}
}

@book{Edenhofer2011,
  title = {Renewable Energy Sources and Climate Change Mitigation: {{Special}} Report of the Intergovernmental Panel on Climate Change},
  shorttitle = {Renewable Energy Sources and Climate Change Mitigation},
  author = {Edenhofer, Ottmar and {Pichs-Madruga}, Ram{\'o}n and Sokona, Youba and Seyboth, Kristin and Kadner, Susanne and Zwickel, Timm and Eickemeier, Patrick and Hansen, Gerrit and Schl{\"o}mer, Steffen and {von Stechow}, Christoph},
  year = {2011},
  publisher = {{Cambridge University Press}}
}

@article{Editors1999,
  title = {Chapter 1: {{Overview}} and Summary},
  shorttitle = {Chapter 1},
  author = {Editors, ITER Physics Basis and an {Co-Chairs}, ITER Physics Expert Group Chairs and Team, ITER Joint Central and Unit, Physics},
  year = {1999},
  month = dec,
  journal = {Nuclear Fusion},
  volume = {39},
  number = {12},
  pages = {2137--2174},
  publisher = {{IOP Publishing}},
  issn = {0029-5515},
  doi = {10.1088/0029-5515/39/12/301},
  urldate = {2021-05-06},
  langid = {english}
}

@article{Edwards2017,
  title = {Academic Research in the 21st Century: {{Maintaining}} Scientific Integrity in a Climate of Perverse Incentives and Hypercompetition},
  shorttitle = {Academic Research in the 21st Century},
  author = {Edwards, Marc A. and Roy, Siddhartha},
  year = {2017},
  journal = {Environmental engineering science},
  volume = {34},
  number = {1},
  pages = {51--61},
  publisher = {{Mary Ann Liebert, Inc. 140 Huguenot Street, 3rd Floor New Rochelle, NY 10801 USA}}
}

@article{Efron1997,
  title = {Improvements on {{Cross-Validation}}: {{The}} 632+ {{Bootstrap Method}}},
  shorttitle = {Improvements on {{Cross-Validation}}},
  author = {Efron, Bradley and Tibshirani, Robert},
  year = {1997},
  month = jun,
  journal = {Journal of the American Statistical Association},
  volume = {92},
  number = {438},
  pages = {548--560},
  publisher = {{Taylor \& Francis}},
  issn = {0162-1459},
  doi = {10.1080/01621459.1997.10474007},
  urldate = {2021-06-21}
}

@article{Endler2015,
  title = {Engineering Design for the Magnetic Diagnostics of {{Wendelstein}} 7-{{X}}},
  author = {Endler, M. and Brucker, B. and Bykov, V. and Cardella, A. and Carls, A. and Dobmeier, F. and Dudek, A. and Fellinger, J. and Geiger, J. and Grosser, K. and Grulke, O. and Hartmann, D. and Hathiramani, D. and H{\"o}chel, K. and K{\"o}ppen, M. and Laube, R. and Neuner, U. and Peng, X. and Rahbarnia, K. and Rummel, K. and Sieber, T. and Thiel, S. and Vork{\"o}per, A. and Werner, A. and Windisch, T. and Ye, M. Y.},
  year = {2015},
  month = nov,
  journal = {Fusion Engineering and Design},
  volume = {100},
  pages = {468--494},
  issn = {0920-3796},
  doi = {10.1016/j.fusengdes.2015.07.020},
  urldate = {2023-01-23},
  langid = {english}
}

@article{Falcon2019,
  title = {Pytorch {{Lightning}}},
  author = {Falcon, William},
  year = {2019},
  volume = {3},
  number = {6},
  doi = {10.5281/zenodo.3828935}
}

@article{Fuchert2022,
  title = {A Novel Technique for an Alignment-Insensitive Density Calibration of {{Thomson}} Scattering Diagnostics Developed at {{W7-X}}},
  author = {Fuchert, G. and Nelde, P. and Pasch, E. and Beurskens, M. N. A. and Bozhenkov, S. A. and Brunner, K. J. and Meineke, J. and Scott, E. R. and Wolf, R. C. and {team}, W7-X.},
  year = {2022},
  month = mar,
  journal = {Journal of Instrumentation},
  volume = {17},
  number = {03},
  pages = {C03012},
  publisher = {{IOP Publishing}},
  issn = {1748-0221},
  doi = {10.1088/1748-0221/17/03/C03012},
  urldate = {2023-02-21},
  langid = {english}
}

@article{Geiger2010,
  title = {Magnetic {{Diagnostics}} for {{Equiibrium Reconstruction}} at {{W7-X}}},
  author = {Geiger, J. and Endler, M. and Werner, A.},
  year = {2010},
  journal = {Contributions to Plasma Physics},
  volume = {50},
  number = {8},
  pages = {736--740},
  issn = {1521-3986},
  doi = {10.1002/ctpp.200900025},
  urldate = {2023-02-21},
  langid = {english}
}

@article{Geiger2014,
  title = {Physics in the Magnetic Configuration Space of {{W7-X}}},
  author = {Geiger, Joachim and Beidler, Craig D. and Feng, Y. and Maa{\ss}berg, Henning and Marushchenko, Nikolai B. and Turkin, Yuryi},
  year = {2014},
  journal = {Plasma Physics and Controlled Fusion},
  volume = {57},
  number = {1},
  pages = {014004},
  publisher = {{IOP Publishing}},
  doi = {10.1088/0741-3335/57/1/014004}
}

@article{Grieger1993,
  title = {{Das Fusionsexperiment WENDELSTEIN 7-X}},
  author = {Grieger, G. and Milch, I.},
  year = {1993},
  journal = {Physikalische Bl\"atter},
  volume = {49},
  number = {11},
  pages = {1001--1005},
  issn = {1521-3722},
  doi = {10.1002/phbl.19930491106},
  urldate = {2022-07-17},
  langid = {ngerman}
}

@article{Hanson2009,
  title = {{{V3FIT}}: A Code for Three-Dimensional Equilibrium Reconstruction},
  shorttitle = {{{V3FIT}}},
  author = {Hanson, James D. and Hirshman, Steven P. and Knowlton, Stephen F. and Lao, Lang L. and Lazarus, Edward A. and Shields, John M.},
  year = {2009},
  month = jul,
  journal = {Nuclear Fusion},
  volume = {49},
  number = {7},
  pages = {075031},
  issn = {0029-5515, 1741-4326},
  doi = {10.1088/0029-5515/49/7/075031},
  urldate = {2020-10-14},
  langid = {english}
}

@article{Harris2020,
  title = {Array Programming with {{NumPy}}},
  author = {Harris, Charles R. and Millman, K. Jarrod and {van der Walt}, St{\'e}fan J. and Gommers, Ralf and Virtanen, Pauli and Cournapeau, David and Wieser, Eric and Taylor, Julian and Berg, Sebastian and Smith, Nathaniel J. and Kern, Robert and Picus, Matti and Hoyer, Stephan and {van Kerkwijk}, Marten H. and Brett, Matthew and Haldane, Allan and {del R{\'i}o}, Jaime Fern{\'a}ndez and Wiebe, Mark and Peterson, Pearu and {G{\'e}rard-Marchant}, Pierre and Sheppard, Kevin and Reddy, Tyler and Weckesser, Warren and Abbasi, Hameer and Gohlke, Christoph and Oliphant, Travis E.},
  year = {2020},
  month = sep,
  journal = {Nature},
  volume = {585},
  number = {7825},
  pages = {357--362},
  publisher = {{Nature Publishing Group}},
  issn = {1476-4687},
  doi = {10.1038/s41586-020-2649-2},
  urldate = {2022-10-22},
  copyright = {2020 The Author(s)},
  langid = {english}
}

@article{Hastings1970,
  title = {Monte {{Carlo}} Sampling Methods Using {{Markov}} Chains and Their Applications},
  author = {Hastings, W. Keith},
  year = {1970},
  publisher = {{Oxford University Press}},
  doi = {10.1093/biomet/57.1.97}
}

@article{Hirshman1983,
  title = {Steepest-descent Moment Method for Three-dimensional Magnetohydrodynamic Equilibria},
  author = {Hirshman, S P and Whitson, J C},
  year = {1983},
  pages = {17},
  doi = {10.1063/1.864116}
}

@article{Hirshman2004,
  title = {Magnetic Diagnostic Responses for Compact Stellarators},
  author = {Hirshman, Steven P. and Lazarus, Edward A. and Hanson, James D. and Knowlton, Stephen F. and Lao, Lang L.},
  year = {2004},
  month = feb,
  journal = {Physics of Plasmas},
  volume = {11},
  number = {2},
  pages = {595--603},
  publisher = {{American Institute of Physics}},
  issn = {1070-664X},
  doi = {10.1063/1.1637347},
  urldate = {2023-01-24}
}

@article{Hooke1961,
  title = {``{{Direct Search}}''{{Solution}} of {{Numerical}} and {{Statistical Problems}}},
  author = {Hooke, Robert and Jeeves, Terry A.},
  year = {1961},
  journal = {Journal of the ACM (JACM)},
  volume = {8},
  number = {2},
  pages = {212--229},
  publisher = {{ACM New York, NY, USA}},
  doi = {10.1145/321062.321069}
}

@article{Howell2020,
  title = {Development of a Non-Parametric {{Gaussian}} Process Model in the Three-Dimensional Equilibrium Reconstruction Code {{V3FIT}}},
  author = {Howell, Eric C. and Hanson, J. D.},
  year = {2020},
  month = feb,
  journal = {Journal of Plasma Physics},
  volume = {86},
  number = {1},
  pages = {905860102},
  publisher = {{Cambridge University Press}},
  issn = {0022-3778, 1469-7807},
  doi = {10.1017/S0022377819000813},
  urldate = {2023-01-24},
  langid = {english}
}

@article{Hunter2007,
  title = {Matplotlib: {{A 2D}} Graphics Environment},
  shorttitle = {Matplotlib},
  author = {Hunter, John D.},
  year = {2007},
  journal = {Computing in science \& engineering},
  volume = {9},
  number = {03},
  pages = {90--95},
  publisher = {{IEEE Computer Society}},
  doi = {10.1109/MCSE.2007.55}
}

@article{Killer2019a,
  title = {Characterization of the {{W7-X}} Scrape-off Layer Using Reciprocating Probes},
  author = {Killer, Carsten and Grulke, Olaf and Drews, Philipp and Gao, Yu and Jakubowski, Marcin and Knieps, Alexander and Nicolai, Dirk and Niemann, Holger and Sitjes, Aleix Puig and Satheeswaran, Guruparan and Team, W7-X.},
  year = {2019},
  month = jun,
  journal = {Nuclear Fusion},
  volume = {59},
  number = {8},
  pages = {086013},
  publisher = {{IOP Publishing}},
  issn = {0029-5515},
  doi = {10.1088/1741-4326/ab2272},
  urldate = {2023-03-02},
  langid = {english}
}

@article{Klinger2016,
  title = {Performance and Properties of the First Plasmas of {{Wendelstein}} 7-{{X}}},
  author = {Klinger, T. and Alonso, A. and Bozhenkov, S. and Burhenn, R. and Dinklage, A. and Fuchert, G. and Geiger, J. and Grulke, O. and Langenberg, A. and Hirsch, M. and Kocsis, G. and Knauer, J. and {Kr{\"a}mer-Flecken}, A. and Laqua, H. and Lazerson, S. and Landreman, M. and Maa{\ss}berg, H. and Marsen, S. and Otte, M. and Pablant, N. and Pasch, E. and Rahbarnia, K. and Stange, T. and Szepesi, T. and Thomsen, H. and Traverso, P. and Velasco, J. L. and Wauters, T. and Weir, G. and {and}, T. Windisch},
  year = {2016},
  month = oct,
  journal = {Plasma Physics and Controlled Fusion},
  volume = {59},
  number = {1},
  pages = {014018},
  publisher = {{IOP Publishing}},
  issn = {0741-3335},
  doi = {10.1088/0741-3335/59/1/014018},
  urldate = {2021-07-15},
  langid = {english}
}

@inproceedings{Knauer2016,
  title = {A New Dispersion Interferometer for the Stellarator {{Wendelstein}} 7-{{X}}},
  booktitle = {43rd {{EPS Conference}} on {{Plasma Physics}}},
  author = {Knauer, J. and Kornejew, P. and Trimino Mora, H. and Hirsch, M. and Werner, A. and Wolf, R. C.},
  year = {2016},
  publisher = {{European Physical Society}}
}

@article{Langenberg2019,
  title = {Inference of Temperature and Density Profiles via Forward Modeling of an X-Ray Imaging Crystal Spectrometer within the {{Minerva Bayesian}} Analysis Framework},
  author = {Langenberg, A. and Svensson, J. and Marchuk, O. and Fuchert, G. and Bozhenkov, S. and Damm, H. and Pasch, E. and Pavone, A. and Thomsen, H. and Pablant, N. A. and Burhenn, R. and Wolf, R. C.},
  year = {2019},
  month = jun,
  journal = {Review of Scientific Instruments},
  volume = {90},
  number = {6},
  pages = {063505},
  publisher = {{American Institute of Physics}},
  issn = {0034-6748},
  doi = {10.1063/1.5086283},
  urldate = {2020-11-12}
}

@article{Lao1985,
  title = {Reconstruction of Current Profile Parameters and Plasma Shapes in Tokamaks},
  author = {Lao, L. L. and John, H. St and Stambaugh, R. D. and Kellman, A. G. and Pfeiffer, W.},
  year = {1985},
  month = nov,
  journal = {Nuclear Fusion},
  volume = {25},
  number = {11},
  pages = {1611--1622},
  publisher = {{IOP Publishing}},
  issn = {0029-5515},
  doi = {10/b5qc2x},
  urldate = {2021-04-08},
  langid = {english}
}

@article{Lao1990,
  title = {Equilibrium Analysis of Current Profiles in Tokamaks},
  author = {Lao, L. L. and Ferron, J. R. and Groebner, R. J. and Howl, W. and John, H. St and Strait, E. J. and Taylor, T. S.},
  year = {1990},
  month = jun,
  journal = {Nuclear Fusion},
  volume = {30},
  number = {6},
  pages = {1035},
  issn = {0029-5515},
  doi = {10.1088/0029-5515/30/6/006},
  urldate = {2023-04-12},
  langid = {english}
}

@article{Lazerson2015,
  title = {Three-Dimensional Equilibrium Reconstruction on the {{DIII-D}} Device},
  author = {Lazerson, S. A.},
  year = {2015},
  month = jan,
  journal = {Nuclear Fusion},
  volume = {55},
  number = {2},
  pages = {023009},
  publisher = {{IOP Publishing}},
  issn = {0029-5515},
  doi = {10.1088/0029-5515/55/2/023009},
  urldate = {2021-06-21},
  langid = {english}
}

@article{Ma2015,
  title = {Non-Axisymmetric Equilibrium Reconstruction of a Current-Carrying Stellarator Using External Magnetic and Soft x-Ray Inversion Radius Measurements},
  author = {Ma, X. and Maurer, D. A. and Knowlton, S. F. and ArchMiller, M. C. and Cianciosa, M. R. and Ennis, D. A. and Hanson, J. D. and Hartwell, G. J. and Hebert, J. D. and Herfindal, J. L. and Pandya, M. D. and Roberds, N. A. and Traverso, P. J.},
  year = {2015},
  month = dec,
  journal = {Physics of Plasmas},
  volume = {22},
  number = {12},
  pages = {122509},
  publisher = {{American Institute of Physics}},
  issn = {1070-664X},
  doi = {10.1063/1.4938031},
  urldate = {2023-01-24}
}

@article{Merlo2023,
  title = {Physics-Regularized Neural Network of the Ideal-{{MHD}} Solution Operator in {{Wendelstein}} 7-{{X}} Configurations},
  author = {Merlo, Andrea and B{\"o}ckenhoff, Daniel Fabian and Schilling, Jonathan and Lazerson, Samuel A and Sunn Pedersen, Thomas},
  year = {2023},
  journal = {Nuclear Fusion},
  issn = {0029-5515},
  doi = {10.1088/1741-4326/acc852},
  urldate = {2023-04-01},
  langid = {english}
}

@article{Metropolis1953,
  title = {Equation of State Calculations by Fast Computing Machines},
  author = {Metropolis, Nicholas and Rosenbluth, Arianna W. and Rosenbluth, Marshall N. and Teller, Augusta H. and Teller, Edward},
  year = {1953},
  journal = {The journal of chemical physics},
  volume = {21},
  number = {6},
  pages = {1087--1092},
  publisher = {{American Institute of Physics}},
  doi = {10.1063/1.1699114}
}

@article{Nelde2021,
  title = {Quantification of Systematic Errors in the Electron Density and Temperature Measured with {{Thomson}} Scattering at {{W7-X}}},
  author = {Nelde, Philipp and Fuchert, Golo and Pasch, Ekkehard and Beurskens, Marc NA and Bozhenkov, Sergey A. and Brunner, Kai Jakob and H{\"o}fel, Udo and Kwak, Sehyun and Meineke, Jens and Scott, Evan R.},
  year = {2021},
  journal = {arXiv preprint arXiv:2111.03562},
  eprint = {2111.03562},
  archiveprefix = {arxiv}
}

@misc{ONNX2018,
  title = {{{ONNX}} Runtime},
  author = {{developers}, ONNX Runtime},
  year = {2018},
  month = nov,
  copyright = {MIT}
}

@article{Pandya2021,
  title = {The Magnetic Structure of Long-Wavelength Magnetohydrodynamic Modes in Current-Carrying Stellarator Plasmas},
  author = {Pandya, M. D. and Ennis, D. A. and Hanson, J. D. and Hartwell, G. J. and Maurer, D. A.},
  year = {2021},
  month = nov,
  journal = {Physics of Plasmas},
  volume = {28},
  number = {11},
  pages = {112503},
  publisher = {{American Institute of Physics}},
  issn = {1070-664X},
  doi = {10.1063/5.0061806},
  urldate = {2023-01-24}
}

@article{Pasch2016,
  title = {The {{Thomson}} Scattering System at {{Wendelstein}} 7-{{X}}},
  author = {Pasch, E. and Beurskens, M. N. A. and Bozhenkov, S. A. and Fuchert, G. and Knauer, J. and Wolf, R. C.},
  year = {2016},
  month = nov,
  journal = {Review of Scientific Instruments},
  volume = {87},
  number = {11},
  pages = {11E729},
  publisher = {{American Institute of Physics}},
  issn = {0034-6748},
  doi = {10.1063/1.4962248},
  urldate = {2023-02-20}
}

@inproceedings{Paszke2019,
  title = {{{PyTorch}}: {{An Imperative Style}}, {{High-Performance Deep Learning Library}}},
  shorttitle = {{{PyTorch}}},
  booktitle = {Advances in {{Neural Information Processing Systems}}},
  author = {Paszke, Adam and Gross, Sam and Massa, Francisco and Lerer, Adam and Bradbury, James and Chanan, Gregory and Killeen, Trevor and Lin, Zeming and Gimelshein, Natalia and Antiga, Luca and Desmaison, Alban and Kopf, Andreas and Yang, Edward and DeVito, Zachary and Raison, Martin and Tejani, Alykhan and Chilamkurthy, Sasank and Steiner, Benoit and Fang, Lu and Bai, Junjie and Chintala, Soumith},
  year = {2019},
  volume = {32},
  publisher = {{Curran Associates, Inc.}},
  urldate = {2022-10-22}
}

@article{Pavone2019a,
  title = {Measurements of Visible Bremsstrahlung and Automatic {{Bayesian}} Inference of the Effective Plasma Charge {{Zeff}} at {{W7-X}}},
  author = {Pavone, A. and Hergenhahn, U. and Krychowiak, M. and Hoefel, U. and Kwak, S. and Svensson, J. and Kornejew, P. and Winters, V. and Koenig, R. and Hirsch, M. and Brunner, K.-J. and Pasch, E. and Knauer, J. and Fuchert, G. and Scott, E. R. and Beurskens, M. and Effenberg, F. and Zhang, D. and Ford, O. and Van{\'o}, L. and Wolf, R. C.},
  year = {2019},
  month = oct,
  journal = {Journal of Instrumentation},
  volume = {14},
  number = {10},
  pages = {C10003},
  publisher = {{IOP Publishing}},
  issn = {1748-0221},
  doi = {2019100214130200},
  urldate = {2021-04-27},
  langid = {english}
}

@techreport{Pomphrey2005,
  title = {Plasma Control for {{NCSX}} and Development of Equilibrium Reconstruction for Stellarators},
  author = {Pomphrey, N. and Lao, L. L. and Lazarus, E. A. and Hirshman, S. P. and Zarnstorff, M. C. and Hudson, S. R. and Ku, L.-P. and McCune, D. C. and Mikkelsen, D. and Monticello, D. A.},
  year = {2005}
}

@article{Salvatier2016,
  title = {Probabilistic Programming in {{Python}} Using {{PyMC3}}},
  author = {Salvatier, John and Wiecki, Thomas V. and Fonnesbeck, Christopher},
  year = {2016},
  journal = {PeerJ Computer Science},
  volume = {2},
  pages = {e55},
  publisher = {{PeerJ Inc.}},
  doi = {10.7717/peerj-cs.55}
}

@techreport{Schilling2018,
  type = {Master {{Thesis}}},
  title = {Experimental {{MHD Equilibrium Analysis}} of {{Magnetic Configurations}} in the {{Wendelstein}} 7-{{X Stellarator}}},
  author = {Schilling, J.},
  year = {2018},
  number = {IPP 2018-20},
  institution = {{Max-Planck-Institute for Plasma Physics}}
}

@article{Schmitt2013,
  title = {Measurement of a Helical {{Pfirsch}}\textendash{{Schl\"uter}} Current with Reduced Magnitude in {{HSX}}},
  author = {Schmitt, J. C. and Talmadge, J. N. and Anderson, D. T.},
  year = {2013},
  journal = {Nuclear Fusion},
  volume = {53},
  number = {8},
  pages = {082001},
  publisher = {{IOP Publishing}},
  doi = {10.1088/0029-5515/53/8/082001}
}

@article{Schmitt2014,
  title = {Modeling, Measurement, and 3-{{D}} Equilibrium Reconstruction of the Bootstrap Current in the {{Helically Symmetric Experiment}}},
  author = {Schmitt, J. C. and Talmadge, J. N. and Anderson, D. T. and Hanson, J. D.},
  year = {2014},
  journal = {Physics of Plasmas},
  volume = {21},
  number = {9},
  pages = {092518},
  publisher = {{AIP Publishing LLC}},
  doi = {10.1063/1.4895899}
}

@article{Sengupta2004,
  title = {Fast Recovery of Vacuum Magnetic Configuration of the {{W7-X}} Stellarator Using Function Parametrization and Artificial Neural Networks},
  author = {Sengupta, A and McCarthy, P.J and Geiger, J and Werner, A},
  year = {2004},
  month = nov,
  journal = {Nuclear Fusion},
  volume = {44},
  number = {11},
  pages = {1176--1188},
  issn = {0029-5515, 1741-4326},
  doi = {10.1088/0029-5515/44/11/003},
  urldate = {2020-10-14},
  langid = {english}
}

@article{Sengupta2007,
  title = {Statistical Analysis of the Equilibrium Configurations of the {{W7-X}} Stellarator},
  author = {Sengupta, A. and Geiger, J. and Carthy, P. J. Mc},
  year = {2007},
  month = apr,
  journal = {Plasma Physics and Controlled Fusion},
  volume = {49},
  number = {5},
  pages = {649--673},
  publisher = {{IOP Publishing}},
  issn = {0741-3335},
  doi = {10.1088/0741-3335/49/5/007},
  urldate = {2020-10-19},
  langid = {english}
}

@inproceedings{Svensson2010,
  title = {Connecting Physics Models and Diagnostic Data Using {{Bayesian Graphical Models}}},
  booktitle = {37th {{EPS Conference}} on {{Plasma Physics}}},
  author = {Svensson, Jakob and Ford, O. and Werner, A. and Von Nessi, Gregory and Hole, Matthew and McDonald, D. C. and Appel, L. and Beurskens, M. and Boboc, A. and Brix, M.},
  year = {2010},
  publisher = {{European Physical Society}},
  address = {{Dublin, Ireland}}
}

@article{Virtanen2020,
  title = {{{SciPy}} 1.0: Fundamental Algorithms for Scientific Computing in {{Python}}},
  shorttitle = {{{SciPy}} 1.0},
  author = {Virtanen, Pauli and Gommers, Ralf and Oliphant, Travis E. and Haberland, Matt and Reddy, Tyler and Cournapeau, David and Burovski, Evgeni and Peterson, Pearu and Weckesser, Warren and Bright, Jonathan and {van der Walt}, St{\'e}fan J. and Brett, Matthew and Wilson, Joshua and Millman, K. Jarrod and Mayorov, Nikolay and Nelson, Andrew R. J. and Jones, Eric and Kern, Robert and Larson, Eric and Carey, C. J. and Polat, {\.I}lhan and Feng, Yu and Moore, Eric W. and VanderPlas, Jake and Laxalde, Denis and Perktold, Josef and Cimrman, Robert and Henriksen, Ian and Quintero, E. A. and Harris, Charles R. and Archibald, Anne M. and Ribeiro, Ant{\^o}nio H. and Pedregosa, Fabian and {van Mulbregt}, Paul},
  year = {2020},
  month = mar,
  journal = {Nature Methods},
  volume = {17},
  number = {3},
  pages = {261--272},
  publisher = {{Nature Publishing Group}},
  issn = {1548-7105},
  doi = {10.1038/s41592-019-0686-2},
  urldate = {2022-09-15},
  copyright = {2020 The Author(s)},
  langid = {english}
}

@article{Wolf2008,
  title = {A Stellarator Reactor Based on the Optimization Criteria of {{Wendelstein}} 7-{{X}}},
  author = {Wolf, R. C.},
  year = {2008},
  month = dec,
  journal = {Fusion Engineering and Design},
  series = {Proceedings of the {{Eight International Symposium}} of {{Fusion Nuclear Technology}}},
  volume = {83},
  number = {7},
  pages = {990--996},
  issn = {0920-3796},
  doi = {10.1016/j.fusengdes.2008.05.008},
  urldate = {2023-03-14},
  langid = {english}
}

@inproceedings{Wurden2022,
  title = {A {{Special Case}} of {{Long-Pulse High Performance Operation}} in {{W7-X}}},
  booktitle = {48th {{EPS Conference}} on {{Plasma Physics}}},
  author = {Wurden, Glen Anthony and Bozhenkov, S. and Fuchert, G. and Zhang, D. and Jablonski, S. and Kubkowska, M. and von Stechow, A. and Beurskens, M. and Brandt, C. and Brunner, K.},
  year = {2022},
  publisher = {{European Physical Society}}
}

@misc{Yadan2019,
  title = {Hydra},
  author = {Yadan, Omry},
  year = {2022},
  month = oct,
  urldate = {2022-10-22},
  copyright = {MIT},
  howpublished = {Meta Research}
}
\end{refsection}

}{}

\ifthenelseproperty{compilation}{acknowledgement}{%
    \chapter{Acknowledgements}\label{sec:acknowledgement}
This work was not possible without the help of many people.
\TODO{Thanksgiving}
\par
\TODO{Maybe the following applies.}
This work has been carried out within the framework of the EUROfusion Consortium and has received funding from the Euratom research and training programme 2014-2018 and 2019-2020 under grant agreement No 633053. 
The views and opinions expressed herein do not necessarily reflect those of the European Commission.

}{}

\ifthenelseproperty{compilation}{affidavit}{%
    \thispagestyle{empty}
\ifthenelseproperty{compilation}{clsdefineschapter}{%
	\ifKOMA
		\addchap[Statutory declaration]{Statutory declaration}
	\else
    	\chapter[Statutory declaration]{Statutory declaration}
    \fi
}{%
	\ifKOMA
		\addsec[Statutory declaration]{Statutory declaration}
	\else
    	\section[Statutory declaration]{Statutory declaration}
    \fi
}
I hereby declare in accordance with the examination regulations that I myself have written this document, that no other sources as those indicated were used and all direct and indirect citations are properly designated, that the document handed in was neither fully nor partly subject to another examination procedure or published and that the content of the electronic exemplar is identical to the printing copy.

\Signature{\getproperty{document}{location}}{\textsc{%
    \ifluatex
        \IfSubStr{\getproperty{author}{firstname}}{TODO}{%
            \getproperty{author}{firstname}
        }{%
            \FirstWord{\getproperty{author}{firstname}}
        }
    \else
        \getproperty{author}{firstname}
    \fi
    \getproperty{author}{familyname}}}

}{}

    \typeout{----- END OF DOCUMENT -----}

\end{document}
\typeout{----- END OF MAIN -----}